\documentclass[final,1p,times]{elsarticle}

\usepackage{graphicx,amsmath,amssymb,array,multirow}

\biboptions{numbers,square,comma,sort&compress}

\journal{Nuclear Physics B}

\def\beq{\begin{equation}}
\def\eeq{\end{equation}}
\def\bea{\begin{eqnarray}}
\def\eea{\end{eqnarray}}
\def\bq{\begin{quote}}
\def\eq{\end{quote}}
\def\ie{{\em i.e.}~}
\def\eg{{\em e.g.}~}
\def\ala{{\it \`{a} la}~}
\def\vev#1{\langle #1\rangle}
\newlength{\plotwidth}
\begin{document}

\begin{frontmatter}

\title{Elements of $\mathcal{F}$-ast Proton Decay}

\author[affil1,affil2]{Tianjun~Li}
\ead{junlt@physics.tamu.edu}
\author[affil1,affil3,affil4]{Dimitri~V.~Nanopoulos}
\ead{dimitri@physics.tamu.edu}
\author[affil5]{Joel~W.~Walker\corref{cor1}}
\ead{jwalker@shsu.edu}
\address[affil1]{George P. and Cynthia W. Mitchell Institute for Fundamental Physics,\\ Texas A$\&$M University, College Station, TX 77843, USA}
\address[affil2]{Key Laboratory of Frontiers in Theoretical Physics, Institute of Theoretical Physics,\\ Chinese Academy of Sciences, Beijing 100190, P. R. China}
\address[affil3]{Astroparticle Physics Group, Houston Advanced Research Center (HARC),\\ Mitchell Campus, Woodlands, TX 77381, USA}
\address[affil4]{Academy of Athens, Division of Natural Sciences,\\ 28 Panepistimiou Avenue, Athens 10679, Greece}
\address[affil5]{Department of Physics, Sam Houston State University,\\ Huntsville, TX 77341, USA}
\cortext[cor1]{Corresponding author}

\begin{abstract}
Gauge coupling unification in the Minimal Supersymmetric Standard Model (MSSM)
strongly suggests the existence of a Grand Unified Theory (GUT),
which could be probed by the observation of proton decay.
Proton lifetime in the $p \!\rightarrow\! {(e\vert\mu)}^{\!+}\! \pi^0$
dimension six mode is proportional in the fourth power to the GUT mass scale,
and inversely proportional in the fourth power to the GUT coupling.
Flipped $SU(5)$ with strict MSSM field content is known to 
survive existing null detection limits for proton decay approaching $10^{34}$ years, 
and indeed, the lifetime predicted by prior studies can be so
long that successful detection is not currently plausible.
We provide an updated dictionary of solutions for the relevant
flipped unification parameters with generic $\beta$-function coefficients,
significantly upgrading the level of detail with which
second order effects are treated, and correcting subtle published errors.

Recently studied classes of $\mathcal{F}$-theory
derived GUT models postulate additional vector-like multiplets at the TeV scale
which modify the renormalization group to yield a substantial
increase in the $SU(3)_C\times SU(2)_L$ unified coupling.
In conjunction with the naturally depressed
$\mathcal{F}$-lipped $SU(5)$ partial unification mass $M_{32}$,
the $\mathcal{F}$-resh analysis undertaken predicts
comparatively $\mathcal{F}$-ast proton decay which only narrowly evades existing
detection limits, and likely falls within the observable range of
proposed next generation detectors such as DUSEL and Hyper-Kamiokande.
The TeV-scale vector multiplets are themselves suitable for
cross correlation by the Large Hadron Collider.
Their presence moreover magnifies the gap between the dual mass scales
of Flipped $SU(5)$, allowing for an elongated second stage renormalization,
pushing the $\mathcal{F}$-inal grand unification to the doorstep
of the reduced Planck mass.
\end{abstract}


\end{frontmatter}

%%%%%%%%%%%%%%%%%%%%%%%%%%%%%%%%%%%%%%%%%%%%%%%%%%%%%%%%%%%%

\section{$\mathcal{F}$-undamentals\label{main:fundamentals}}

%%%%%%%%%%%%%%%%%%%%%%%%%%%%%%%%%%%%%%%%%%%%%%%%%%%%%%%%%%%%

\subsection{Introduction}

The major problem in the Standard Model (SM) is that of the gauge hierarchy,
{\it i.e.}, the Higgs boson mass square is not stable against quantum
corrections and has quadratic divergences. Supersymmetry (SUSY) provides
a natural solution to gauge hierarchy problem due to the opposing spin
statistics for bosons and fermions. In the Minimal Supersymmetric Standard
Model (MSSM) with $R$ parity, under which the SM particles are even
while their supersymmetric partners are odd, gauge coupling unification
can be achieved~\cite{Ellis:1990zq, Ellis:1990wk, Amaldi:1991cn, Langacker:1991an},
the lightest supersymmetric particle (LSP), typically the neutralino,
can be a cold dark matter candidate~\cite{Ellis:1983wd, Ellis:1983ew, Goldberg:1983nd},
and the precision electroweak constraints can be satisfied. Also, the electroweak
gauge symmetry can be broken radiatively due to the large top quark Yukawa
coupling~\cite{Ellis:1983bp}, and smallness of the neutrino masses can be explained via the seesaw
mechanism~\cite{gellmann,yanagida,Georgi:1979dq}. In particular, gauge coupling unification strongly suggests
the existence of a Grand Unified Theory (GUT), which can further explain
charge quantization.  If this scenario is realized in nature, there is a further
unique and necessary implication which is of great and pressing interest. 

As the lightest baryon, the proton, having no viable decay products within the Standard Model,
seems to enjoy a life without age or end.  It has long been suspected however,
that protons are not actually forever~\cite{Nanopoulos:1978kz}. 
Instability of the proton is indeed a ubiquitous signature of Grand Unification,
the merger of fundamental forces necessarily linking quarks to leptons,
and providing a narrow channel $p \!\rightarrow\! {(e\vert\mu)}^{\!+}\! \pi^0$
of dimension six decay via heavy gauge boson exchange.
Following the direct verification of coupling unification, neutrino masses, and the third quark generation,
and convincing secondary evidence for supersymmetry and the Higgs mechanism, proton decay is the
last great unresolved prediction of the particle physics generation. 

We will consider the problem of proton decay within both the
Georgi-Glashow $SU(5)$~\cite{Georgi:1974sy} GUT and also the 
Flipped $SU(5)$~\cite{Barr:1981qv, Derendinger:1983aj, Antoniadis:1987dx,
Ellis:1988tx,Nanopoulos:2002qk} variation, both as pure particle theory and supplemented by string
theoretic considerations from free fermionic constructions~\cite{Antoniadis:1989zy,Lopez:1992kg} and
Cumrun Vafa's F-theory~\cite{Vafa:1996xn, Donagi:2008ca, Beasley:2008dc, Beasley:2008kw,
Donagi:2008kj, Heckman:2008rb, Jiang:2009zza, Jiang:2009za}.
Flipped $SU(5)$ with strict MSSM field content is
known to survive existing null detection limits for proton decay
approaching $10^{34}$ years~\cite{Ellis:1988tx,Ellis:1995at,Ellis:2002vk,Li:2009fq},
while Standard $SU(5)$, via dimension five decay, could be in trouble~\cite{Murayama:2001ur,Ellis:2002vk}.
Flipped $SU(5)$, in which the `ignoble' dimension five operators are automatically suppressed,
is moreover phenomenologically preferred by the authors for a host of additional benefits~\cite{Nanopoulos:2002qk}
which include natural doublet-triplet splitting, avoidance of unwieldy adjoint Higgs multiplets,
a mechanism of baryogenesis, consistency with precision electroweak data,
a more inclusive accommodation of right-handed neutrinos,
and an improved two-step seesaw mechanism~\cite{Georgi:1979dq} for generation of their eV scale mass.

It is an amusing game we have seen played out with the proton lifetime,
in that the first success of a sufficiently suppressed decay rate can turn next
into a `failure' to predict results which are potentially observable.
Of course, nature can do as she wishes.  But there remains a comic absurdity to the notion
that we could be right on the cusp of a detection at $10^{35-36}$ years,
at just the scale where the macroscopic size of the experiment becomes unviable.
The enormity of Avogadro's number has run most of the relay and our
engineering limitations have let him down at the finish line!

A proper treatment of the problem necessarily touches upon an astonishingly broad set of interconnected topics.
As input, we call on the precision measurements of couplings at the $Z$ boson mass boundary,
and numerically evaluated proton wavefunction matrix elements from lattice Quantum Chromodynamics.
Updates to the rate of coupling renormalization from the detailed mass spectrum of light supersymmetric particles
play a decisive role in the determination of the unified mass scale,
on which proton lifetime is very strongly dependent.
The preferred position within the supersymmetric parameter space is in turn constrained by
the ability to realize electrically neutral cold dark matter and electroweak symmetry breaking,
cosmological limits on dark matter relic density, the anomalous magnetic moment of the muon,
and search limits for the Higgs boson.
Detection and direct measurement of the light superpartners themselves constitutes one of the
key reasons of entry for the Large Hadron Collider (LHC). 

The subject of proton decay is highly differentially sensitive,
in both preferred mode and predicted rate of decay, to the unified representation.
The lifetime in the dimension six decay mode is proportional in the fourth power to the GUT mass scale,
and inversely proportional in the fourth power to the GUT coupling.
This extraordinary sensitivity to the unification point enhances even subtle distinctions between
competing scenarios, and argues for great care and precision in the evaluation of the renormalization group
running and of any model dependent peculiarities.

Flipped $SU(5)$ could be unique among proposed GUT scenarios for its
simultaneously consistency with existing proton decay limits and all low energy phenomenology.
It is a key feature of Flipped $SU(5)$, or more properly $SU(5) \times U(1)_{\rm X}$,
that the $SU(3)_{\rm C}\times SU(2)_{\rm L}$ gauge couplings $g_{3}$ and $g_{2}$
first merge into the common value $g_5(M_{32}) \equiv g_{32}$
at an intermediate mass $M_{32}$, proximal to the usual GUT scale and
prior to a point of later full unification with the hypercharge-like remixed coupling $g_{\rm X}$.
In fact, it is exactly this additional compensatory freedom in the position of $M_{32}$
which has allowed the flipped scenario to match the precision electro-weak measurements of LEP without
unreasonable invocation of heavy threshold effects.

The current study represents, in our opinion, the most careful, comprehensive and accurate survey to date of
proton lifetime predictions in the context of Flipped $SU(5)$,
correcting certain mistakes of prior analysis, and significantly reducing the extent to which
simplifying approximations produce errors in the calculation of critical second order effects.
The results obtained are both surprising and noteworthy, yielding where one might expect only inconsequential variation
within the noise, a highly significant downward shift of the central predicted lifetime.
The net motion of more than a full order of magnitude toward the region of experimental detectability is of clear
relevance to the great experimental proton decay searches, both ongoing
(Super Kamiokande) and proposed (DUSEL, Hyper Kamiokande, LAGUNA).

Superstring phenomenology may be considered as a natural partner to particle physics, providing theoretical impetus
and an ultimate connection to energies approaching the Planck scale.
It is a generic prediction of string model building that extra heavy multiplets beyond the MSSM may exist.
We have focused here on extensions to GUT phenomenology from the formulation called F-theory, which is highly favorable
in its own right for the decoupling of gravitational physics, allowing the construction of local models.
Such models naturally yield very specific predictions for new particle representations near the TeV scale,
only a small number of such configurations successfully avoiding a Landau pole in a physically reasonable manner.
The presence of such fields, themselves potentially verifiable at the LHC, has direct bearing on gauge
renormalization and proton lifetime.  Specifically, The modifications to the renormalization group from these fields
yield a significant increase in $\alpha_{32}$, which enhances the rate of proton decay.

It is the reduced scale $M_{32}$ of Flipped $SU(5)$ relative to its Standard counterpart
which is directly relevant to dimension six decay predictions, also acting to shorten the proton lifetime.
Traditional analysis with MSSM field content has tended to indicate a rather mild splitting~\cite{Ellis:1995at,Ellis:2002vk},
with $M_{32}/M^{\rm max}_{32}$ in the neighborhood of one or two parts out of three.
However, inclusion here of extra matter at the TeV scale has the effect of making the splitting substantially more dramatic,
due precisely to the enhanced gap between the $SU(5)$ and $U(1)_{\rm X}$ couplings,
realizing an `extreme' Flipped $SU(5)$ scenario.
Although the location of $M_{32}$ itself experiences no great modification, and has actually been seen to increase mildly,
the true grand unification point tends to push farther upward by two orders of magnitude or more,
tantalizingly close to the string scale.
The F-theoretic field content is in this sense just what the doctor ordered for
Flipped $SU(5)$, tangibly justifying the expected separation between the GUT and String/Planck scales.

%%%%%%%%%%%%%%%%%%%%%%%%%%%%%%%%%%%%%%%%%%%%%%%%%%%%%%%%%%%%

\subsection{Outline of Topics}

Continuing in Section~(\ref{main:fundamentals}), {\it $\mathcal{F}$-undamentals}~,
we proceed first with a general review of proton decay considerations,
including a survey of existing experimental decay limits, and a look forward
to the realistic proposals for next generation experiments.  We also
provide a broad background on the available proton decay channels, including some
discussion of the fundamental amplitudes, and focusing in particular on the differential
decay signatures of Standard vs. Flipped $SU(5)$ grand unification.  Simple numerically
parameterized formulae for the dimension six
$p \!\rightarrow\! {(e\vert\mu)}^{\!+}\! \pi^0$
decay are provided for each scenario, including the dominant dependencies on the unification
mass scale and coupling.

Also provided in Section~(\ref{main:fundamentals}) is a detailed presentation of the
appropriate manner in which to analytically correct for the step-wise entrance of new
particles into the gauge renormalization as their respective mass thresholds
are crossed.  The distinction in treatment between light and heavy thresholds is made clear.
The section concludes with a detailed breakdown of the $\beta$-function coefficients
for each gauge coupling, for each MSSM field, as is
required for implementation of the previously described threshold factors.  For handy
reference, we also tabulate the $\beta$-coefficients for the field groupings which
are typically consolidated in the output files of the standard
particle physics computational analysis libraries.

Section~(\ref{main:flipped}), {\it $\mathcal{F}$-lipped $SU(5)$}~, broadly reviews
the phenomenology of our preferred GUT framework, Flipped $SU(5) \times U(1)_{\rm X}$,
including a discussion of field content and quantum numbers, neutrino masses,
doublet-triplet splitting, compliance with precision low energy measurements,
and the related implications for proton decay.

Section~(\ref{main:ftheory}), {\it $\mathcal{F}$-theory}~, consist of an
introduction to Cumrun Vafa's F-theory, and its related model building techniques.
In particular, we focus on the possible extensions to the TeV
scale field content which F-theory can naturally
provide.  It is these fields which are directly responsible for magnifying 
the coupling separation at $M_{32}$, allowing for an elongated
second stage renormalization, pushing grand unification to the doorstep of the
reduced Planck mass.  An immediate consequence of the enhanced $SU(5)$ gauge
coupling is substantial speeding of the dimension six proton decay.
We also supplement the previously published descriptions of possible heavy
vector content in Flipped $SU(5)$ GUT models with the corresponding
explicit constructions which extend Georgi-Glashow $SU(5)$.

Section~(\ref{main:fresh}), {\it $\mathcal{F}$-resh Analysis}~, documents the
majority of the original efforts undertaken in this study.  Beginning with the
renormalization group equations (RGEs) of the MSSM, we
systematically establish a dictionary of solutions for the unification parameters
of Standard and Flipped $SU(5)$, including threshold and second loop effects
individually for each of the three couplings.  We both distinguish and relate the
two model classes, while carefully identifying and isolating the set of dependent
variables in use at any moment, demonstrating how lack of caution in this regard
has sometimes in the past led to a subtle pollution of resulting formulae.

Subsequently presented in Section~(\ref{main:fresh}), is a detailed approximate
solution of the second loop contributions to the renormalization group.
While applications are readily available for the numerical evaluation of second order effects,
they do not generally enforce the flipped unification paradigm which we favor, and they
moreover suffer from the lack of visual transparency which is endemic to any processed solution.
The methods presented offer either an advanced launching position from which to initiate
a numerical analysis, or a full, albeit more greatly simplified, closed form solution.
Both approaches are parameterized to quickly and efficiently interface with the main results
on gauge unification, with the formalism of the second loop corrections placed on equal
footing with the companion correction factors from the thresholds.
The updated treatment of these two factors is found to speed the baseline predictions
for proton decay even without inclusion of the exotic F-theory field content.

Section~(\ref{main:fresh}) concludes with an elaboration on the methods used for
evaluation of the GUT scale (heavy) mass thresholds.  We reproduce an established formula
for the single parameter approximation of the consequences of heavy thresholds to Standard $SU(5)$,
and demonstrate that the historical cross-application of this formula to Flipped $SU(5)$ is
inappropriate, proposing a more satisfactory alternative.  We also make a correction to a
previously published formula for the shift in the unification scale $M_{32}$ 
which accompanies the activation of heavy thresholds in Flipped $SU(5)$.

Section~(\ref{main:fast}), {\it $\mathcal{F}$-ast Proton Decay}~, contains the
principle results of the present study, in the form of description, tables and figures.
We summarize the updates to data, field content, and methodology which have
culminated in the present reductions in the predicted proton lifetime, distinguishing
between the portion of the overall change which is to be attributed to each modification.
Applying the techniques established in the prior section, we go on to undertake a thorough
survey of the possible heavy threshold contributions to both Flipped and Standard $SU(5)$.
Our study supports the previously advertised conclusion that heavy thresholds cannot save
Standard $SU(5)$ from overly rapid dimension five proton decay, and makes the new suggestion
that heavy effects in Flipped $SU(5)$ act only to suppress,
and never to enhance, dimension six proton decay.

Section~(\ref{main:finale}), {\it $\mathcal{F}$-inale}~, opens with the final stage of
Flipped $SU(5)$ grand unification, namely the second stage running of $\alpha_5$ and
$\alpha_{\rm X}$.  We observe that the `weakness' of the two step unification
becomes here a great strength, without which the rationale for the splitting of the
traditional single GUT scale from the Planck mass becomes more strained.  The F-theory
fields play a decisive role in this picture by stretching the natural Flipped $SU(5)$
scale separation across the required two-plus orders of magnitude.

We close the final section with a brief summary and conclusions, emphasizing the cohesive
plot line that the cumulative effect of Flipped $SU(5)$ grand unification in the
F-theory model building context, applying the freshly detailed methods of analysis presented
herein, is surprisingly fast proton decay.

In an appendix, we offer a brief review of del Pezzo surfaces.

%%%%%%%%%%%%%%%%%%%%%%%%%%%%%%%%%%%%%%%%%%%%%%%%%%%%%%%%%%%%

\subsection{Proton Lifetime Considerations\label{sct:plife}}

Let us first briefly review the
existing and proposed proton decay experiments.
Despite the incredible contribution of knowledge made by Super-Kamiokande to
the physics of neutrino oscillations, this is in fact not even the principal
task of that experiment.
Faced with the incomprehensibly long time scales on which proton decay is expected
to be manifest, some $23-26$ orders of magnitude older than the universe itself,
we can only hope to observe this process within some reasonably finite interval
by leveraging Avogadro's number to our benefit and watching some very large number
of nuclei simultaneously.
Indeed, that is precisely the role of the extravagant size allowed to this detector.
This 50-kiloton (kt) water \v{C}erenkov detector
has set the current 
lower bounds of $8.2\times 10^{33}$ and $6.6\times 10^{33}$ years
at the $90\%$ confidence level for the partial lifetimes in the
$p\rightarrow e^+ \pi^0$ and $p\rightarrow \mu^+ \pi^0$ modes~\cite{:2009gd}.

Hyper-Kamiokande is a proposed 1-Megaton detector, about 20 times larger
volumetrically than Super-K~\cite{Nakamura:2003hk},
which we can expect to explore partial lifetimes 
up to a level near $2\times 10^{35}$ years for
$p \!\rightarrow\! {(e\vert\mu)}^{\!+}\! \pi^0$ across a decade long run.
The proposal for the DUSEL
experiment~\cite{cline,kearns,Raby:2008pd}
features both water \v{C}erenkov and liquid Argon (which is around five times
more sensitive per kilogram to $p\rightarrow K^+ {\bar \nu}_{\mu}$ than water)
detectors, in the neighborhood of 500 and 100 kt respectively,
with the stated goal of probing partial lifetimes into the
order of $10^{35}$ years for both the neutral pion and $K^+$ channels.

For the $p \rightarrow K^+ \bar{\nu}$ partial lifetime as mediated by
the triplet Higgsinos of $SU(5)$, a lower bound of $6.7 \times 10^{32}$
years has been established at the $90 \%$ confidence level.
This is a so-called dimension-five decay, summing the mass level of the
two-boson, two-fermion effective vertex \ala Fermi.
The upper bound on its rate translates directly to a minimal mass for the
color-triplet Higgs of around $10^{17}~{\rm [GeV]}$~\cite{Murayama:2001ur}.
Conversely though, compatibility of a strict unification with the precision
LEP measurements of SM parameters at $M_Z$ places an upper limit on this mass
of order $10^{15}~{\rm [GeV]}$.

Flipped $SU(5)$ evades this incongruity by means of the `missing-partner
mechanism'~\cite{Antoniadis:1987dx} which naturally splits the heavy triplets $H_\mathbf{3}$
within the five of Higgs ($h$) away from the light electroweak components
$H_\mathbf{2}$.
Specifically, since the flipped $\mathbf{10}$ now contains a neutral element
it is possible to allow vacuum expectation values for the breaking of $SU(5)$
to arise within a Higgs decaplet $H$ from this representation.
The GUT superpotential elements $HHh$ and $\bar{H}\bar{H}\bar{h}$ then provide
for the mass terms $\vev{\nu^c_H} d^c_H H_\mathbf{3}$ (and conjugate), while
$H_\mathbf{2}$ is left light, having no partner in $H$ with which to make
a neutral pairing.
So then is adroitly bypassed all insinuation of a hand-built term
$Mh_\mathbf{5}h_\mathbf{\bar{5}}$ to finely tune against the
putative adjoint GUT Higgs $h_\mathbf{5}h_\mathbf{\bar{5}}\Sigma_\mathbf{24}$
for fulfillment of this same goal.
With that term goes also the undesirable triplet mixing, the dangerously
fast proton decay channel and the fatal limits\footnote{The
$d=5$ mode is not {\it entirely} abandoned, as there does remain
the supersymmetric term $\mu h \bar{h}$ for suppression of electroweak axions.
It is slower though by a ratio $(\frac{\mu}{M_{H_\mathbf{3}}})^2$,
where $\mu \sim$~TeV.}.
on the mass of $H_\mathbf{3}$
As for Standard $SU(5)$ however, this is just another splash of cold water
from our friends at Super-K.
And as we have mentioned, they have no shortage of cold water.

This diversion put aside, the dimension six decay
$p \!\rightarrow\! {(e\vert\mu)}^{\!+}\! \pi^0$
may now regain our attention~\cite{Ellis:1988tx,Ellis:1995at,Ellis:2002vk}.
With aid of the SUSY extension, neither theory is under any fear from the
current lower bound for this mode.
That is not to say though that interesting differences do not exist between
the pictures.
In Standard $SU(5)$ there are two effective operators which contribute in
sum to this rate.
The first vertex arises from the term $\mathbf{10\,\bar{5}\,10^*\,\bar{5}^*}$
and the second from $\mathbf{10\,10\,10^*\,10^*}$ with a relative strength of
$(1 + |V_{ud}|^2)^2$.
However, in Flipped $SU(5)$, $e^c_L$ no longer resides within the $\mathbf{10}$,
so the positronic channel makes use of only $e^c_R$ decays utilizing the
operator which contains the representation $\mathbf{\bar{5}}$.
Taking the central value of $.9738 (5)$ for the Cabibbo quark-mixing phase
$V_{ud}$ leads to a suppression of the total rate by a factor of
about five after dividing out the correction $(1 + (1 + |V_{ud}|^2)^2)$.
In opposition to this effect is a tendency toward more rapid decay
due to dependence on the intermediate partial unification scale
$M_{32}$ rather than the traditional GUT value.
In fact, we will see that this second distinction generally overwhelms the
first, leading on balance to a net shorter prediction of the proton lifetime
in Flipped $SU(5)$.

The effective dimension six operator for proton decay~\cite{Buras:1977yy,Ellis:1979hy}
in Flipped $SU(5)$ is given following, suppressing the hermitian conjugate terms,
where $g_{5}$ is the $SU(3)_C\times SU(2)_L$ unified gauge coupling,
$\theta_c$ is the Cabibbo angle, and $u$, $d$, $s$
are the up, down and strange quarks, respectively.
\bea
{\bar {\cal L}}_{\Delta B \ne 0} & =  & \frac{g_5^2}{2 M_{32}^2}
 \left[ (\epsilon^{ijk}
{\bar d^c}_{k} e^{2 i \eta_{11}} \gamma^\mu P_L d_{j}) (u_{i}
 \gamma_\mu P_L \nu_L)  \right. \nonumber \\
 & + & \left. (\epsilon^{ijk}
({\bar d^c}_{k} e^{2 i \eta_{11}} \cos \theta_c
 + {\bar s^c}_{k} e^{2 i \eta_{21}} \sin \theta_c)
\gamma^\mu P_L u_{j}) (u_{i}
 \gamma_\mu P_L \ell_L) \right]
\label{leff}
\eea

The decay amplitude is proportional to the overall normalization of the 
proton wave function at the origin.
The relevant matrix elements 
$\alpha$ and $\beta$ are defined as: 
\begin{eqnarray}
\langle 0 | \epsilon_{ijk} (u^i d^j)_R u^k_L | p ({\mathbf k}) \rangle & \equiv & 
\alpha \, {\rm u}_L ({\mathbf k}) \nonumber \\ 
\langle 0 | \epsilon_{ijk} (u^i d^j)_L u^k_L | p ({\mathbf k}) \rangle & \equiv & 
\beta \, {\rm u}_L ({\mathbf k})
\label{alphabeta}
\end{eqnarray}

The reduced matrix elements $\alpha$ and $\beta$ have been
calculated in a lattice approach~\cite{Kuramashi:2000hw}, with central values
$\alpha = \beta = 0.015~[{\rm GeV}^3]$ reported.
Quoted errors are below 10\%, corresponding to an
uncertainty of less than 20\% in the proton partial lifetime,
negligible compared to other uncertainties present in our calculation.

Following the results of~\cite{Ellis:1993ks,Murayama:2001ur} and references therein, we present
a numerically parameterized expression for the desired lifetime in both Standard and Flipped $SU(5)$,
with coefficients appropriate to the flipped specialization already absorbed in the latter case.
\begin{subequations}
\label{eqs:plife}
\bea
{\textrm{\Large $\tau$ }}_{p \rightarrow {(e\vert\mu)}^{\!+}\! \pi^0}^{\textrm{\scriptsize SU(5)}} &=& 0.8 \times
{\left( \frac{M_{32}^{\rm max} }{10^{16}~{\textrm{[GeV]}}} \right)}^4
\times {\left( \frac{ 0.0412 }{ \alpha_5^{\rm max }} \right)}^2 \times 10^{35}~{\textrm{[Y]}} \label{plife_std} \\
{\textrm{\Large $\tau$ }}_{p \rightarrow {(e\vert\mu)}^{\!+}\! \pi^0}^{\mathcal{F}\textrm{\scriptsize -SU(5)}} &=& 3.8 \times
{\left( \frac{M_{32} }{10^{16}~{\textrm{[GeV]}}} \right)}^4
\times {\left( \frac{0.0412 }{ \alpha_5} \right)}^2 \times 10^{35}~{\textrm{[Y]}} \label{plife_flipped}
\eea
\end{subequations}

The proton lifetime scales as a fourth power
of the $SU(5)$ unification scale $M_{32}$, and inversely, again in the fourth
power, to the coupling $g_{32} \equiv \sqrt{4\pi\alpha_5}$ evaluated at that scale.
This extreme sensitivity argues for great care in the
selection and study of a unification scenario.
Lower bounds can only ever exclude a model, never truly supporting any one
competing suggestion.
The real goal of course is to constrain this number from both directions.
Assisted by the shorter net Flipped $SU(5)$ lifetime, the increased reach of next-generation
experiments offers the tantalizing prospect of probing the most relevant parameter space.

We close this section with a survey of some characteristic predictions which have
been made for Flipped $SU(5)$ proton decay based on the baryon-number violating
effective potential of Eq.~(\ref{leff}).
Unknown parameters in this expression are the the CP-violating phases $\eta_{11,21}$
and lepton flavor eigenstates $\nu_L$ and $\ell_L$, related to the mass
diagonal mixtures as:
\beq
\nu_L =\nu_F U_\nu \qquad;\qquad \ell_L =\ell_F U_\ell
\label{33}
\eeq

These mixing matrices $U_{(\nu\vert\ell)}$ take on added currency in the age of
neutrino oscillations.
Having seen there evidence for near-maximal mixing, it seems reasonable
to suspect that at least some $(e\vert\mu)$ entries are also ${\cal O}(1)$
in $U_\ell$.
From this point it will indeed be assumed that $|U_{\ell_{11,12}}|^2$
are ${\cal O}(1)$, thus avoiding further large numerical suppressions
of both the $p \!\rightarrow\! {(e\vert\mu)}^{\!+}\! \pi^0$ rates\footnote{Note
that there is no corresponding suppression of the
$p \rightarrow {\bar \nu} \pi^+$ and $n \rightarrow {\bar \nu} \pi^0$ modes,
since all neutrino flavors are summed over.}.
No more can be said though regarding the ratio of $p \rightarrow e^+ X$ and
$p \rightarrow \mu^+ X$ decays, and as such it would be good that any
next-generation detector be equally adept at the exposure of either mode.
Throughout this report we thus diligently emphasize the electronic-muonic
product ambiguity as an essential Flipped $SU(5)$ characteristic. 

Despite the discussed points of ignorance,
it can still be robustly stated that~\cite{Ellis:1988tx}:
\bea
&
\Gamma(p \rightarrow e^+ \pi^o) \quad=\quad \frac{\cos ^2 \theta_c}{2}
 |U_{\ell_{11}}|^2
\Gamma(p \rightarrow {\bar \nu} \pi^+) \quad=\quad \cos ^2 \theta_c
|U_{\ell_{11}}|^2
\Gamma(n \rightarrow {\bar \nu} \pi^o) \nonumber
&
\\[4pt]
&
\Gamma(n \rightarrow e^+ \pi^-) \,=\, 2
\Gamma(p \rightarrow e^+ \pi^o) \quad;\quad
\Gamma(n \rightarrow \mu^+ \pi^-) \,=\, 2
\Gamma(p \rightarrow \mu^+ \pi^o)
&
\label{gammas} \\[4pt]
&
\Gamma(p \rightarrow \mu^+ \pi^o) \quad=\quad \frac{\cos ^2 \theta_c}{2}
 |U_{\ell_{12}}|^2
\Gamma(p \rightarrow {\bar \nu} \pi^+) \quad=\quad \cos ^2 \theta_c
|U_{\ell_{12}}|^2
\Gamma(n \rightarrow {\bar \nu} \pi^o)
&
\nonumber
\eea

We note~\cite{Ellis:1988tx,Ellis:1993ks} that the Flipped $SU(5)$ predictions for decay
ratios involving strange particles, neutrinos and charged leptons differ
substantially from those of conventional $SU(5)$.
Comparison of such characteristic signals then constitutes a potentially
powerful tool for establishing mixing patterns and differentiating between
GUT proposals.

%%%%%%%%%%%%%%%%%%%%%%%%%%%%%%%%%%%%%%%%%%%%%%%%%%%%%%%%%%%%

\subsection{Multiple Threshold Renormalization\label{sct:multiplethresholds}}

To determine the effects of embedding TeV scale vector multiplets within a GUT scenario,
it is useful to start by studying generally, and from first principles, the way in which renormalization
kinks at intermediate thresholds influence the relationship between coupling strengths
at the bounding scales.

\begin{figure}[htp]
\begin{center}
\includegraphics[width=.8\plotwidth,angle=0]{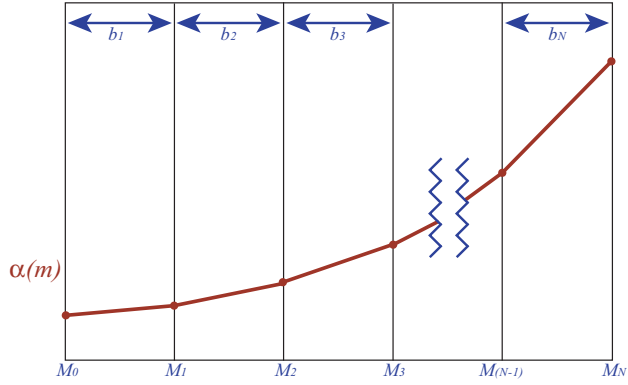}
\end{center}
\caption{\label{fig:kink}
Kinking is depicted at multiple threshold points labeled $(0 \ldots N)$
in the extended running of some coupling $\alpha$.
The naming convention shown is used in related equations and calculations.}
\end{figure}

The transition between any two such intermediate scales (containing no threshold transition)
is given by the standard one-loop renormalization group equation\footnote{The
index $i$ here labels the threshold {\it segment}
of the running being considered.  It does not label the group to which $\alpha$ belongs.}
(cf. Figure~\ref{fig:kink}).
\beq
\frac{2\pi}{\alpha_{i-1}}-\frac{2\pi}{\alpha_{i}} \:=\:
{b_i} \left(\ln{M_i} - \ln {M_{i-1}} \right) \:\equiv\:
\Delta_i
\label{RGEE}\\
\eeq

Summing over all steps, the total shift in the coupling is obtained.  To facilitate comparison
to the case with no intermediate thresholds, we will define a factor $\xi_N$ to encapsulate all
dependence on internal masses $M_i$ and $\beta$-coefficients $b_i$.
\beq
\sum_{i=1}^N \Delta_i \:=\:
\frac{2\pi}{\alpha_{0}}-\frac{2\pi}{\alpha_{N}} \:=\:
b_N \ln{\frac{M_N}{M_0}} - \xi_N
\label{eq:onestep}
\eeq
\bea
(-)\, \xi_N &\equiv&
b_N \ln{M_0} - b_1 \ln{M_0} + b_1 \ln{M_1} - b_2 \ln{M_1} \nonumber \\
&\phantom{=}&
{} + b_2 \ln{M_2} - b_3 \ln{M_2} + b_3 \ln{M_3} + \cdots - b_N \ln{M_{N-1}}
\label{eq:internal} \\
&=&
\ln{\left[
{\left(\frac{M_1}{M_0}\right)}^{b_1}
{\left(\frac{M_2}{M_1}\right)}^{b_2}
{\left(\frac{M_3}{M_2}\right)}^{b_3}
\times \cdots \times
{\left(\frac{M_0}{M_{N-1}}\right)}^{b_N}
\right]} \nonumber
\eea

For concreteness in evaluation of Eq.~(\ref{eq:internal}), we will will specialize to the example $(N=3)$.
\bea
\xi_3
&=& (-)\,
\ln{\left[
{\left(\frac{M_1}{M_0}\right)}^{b_1}
{\left(\frac{M_2}{M_1}\right)}^{b_2}
{\left(\frac{M_0}{M_2}\right)}^{b_3}
\right]} \nonumber \\
&=& (-)\,
\ln{\left[
{\left(\frac{M_1}{M_0}\right)}^{(b_1 - b_2)}
{\left(\frac{M_2}{M_0}\right)}^{b_2}
{\left(\frac{M_0}{M_2}\right)}^{b_3}
\right]} \label{eq:neq3} \\
&=&
\ln{\left[
{\left(\frac{M_1}{M_0}\right)}^{(b_2 - b_1)}
{\left(\frac{M_2}{M_0}\right)}^{(b_3 - b_2)}
\right]} \nonumber
\eea

This compact form readily regeneralizes to any choice of $N$.
Picking back up from Eq.~(\ref{eq:internal}), we group together repeated occurrences
of each $M_i$, separating out the special case $M_0$.
This term is then reabsorbed into the sum over intermediate masses by use of the identity
$\left(b_N - b_1 \right) = \sum_{i=1}^{N-1} \left( b_{i+1} - b_i \right)$.
The notation $\delta b_i \equiv \left( b_{i+1} - b_i \right)$ is adopted.
\beq
\xi_N \,=\, - \ln{M_0^{(b_N-b_1)}} + \sum_{i=1}^{N-1} \ln{M_i^{(b_{i+1}-b_i)}}
\,=\, \sum_{i=1}^{N-1} \delta b_i \ln{\left( \frac{M_i}{M_0} \right)}
\label{eq:alln}
\eeq

The final result for running which crosses internal threshold points is then:
\beq
\left\{ \frac{2\pi}{\alpha_{0}}-\frac{2\pi}{\alpha_{N}} \right\} +
\left\{ \xi_N \equiv \sum_{\rm internal}\!\!\delta b \ln{\left( \frac{M}{M_0} \right)} \right\} \:=\:
b_N \ln{\frac{M_N}{M_0}}
\label{eq:allsteps}
\eeq
This expression has the appealing feature that ordering of the kinks at each $M_i$
is fully decoupled, depending only on the {\it change} induced in the $\beta$-coefficient
at each given mass.  

The underlying symmetry between the bounding energy scales suggests that it should likewise be possible
to achieve an analogous formula which references each threshold against the heavy mass $M_N$.
Of course, our general low energy perspective on the unification mass as an output prediction of the experimentally determined
electroweak parameters would seem to preclude this view.
However, it is exactly the form which would be useful for consideration of the ultra heavy threshold terms which
arise from masses in direct associated proximity to the unification scale.
Guided by this observation, we may attempt to split our perspective by defining an intermediate index $I$, $(1 \le I \le N)$,
such that the coefficient $b_I$ prevails after action of what is considered to be the final light threshold,
and until action of the subsequent, and by definition initial heavy, threshold.
The principal analysis then refactors as follows, starting from Eqs.~(\ref{eq:onestep},\ref{eq:alln}).
\bea
\frac{2\pi}{\alpha_{0}}-\frac{2\pi}{\alpha_{N}}
&=&
b_N \ln{\frac{M_N}{M_0}} + \ln{M_0^{(b_N-b_1)}} - \sum_{i=1}^{N-1} \ln{M_i^{(b_{i+1}-b_i)}} \nonumber \\
&=&
b_I \ln{\frac{M_N}{M_0}} + \ln{M_0^{(b_I-b_1)}} + \ln{M_N^{(b_N-b_I)}} - \sum_{i=1}^{N-1} \ln{M_i^{(b_{i+1}-b_i)}} \label{eq:splitthr} \\
&=&
b_I \ln{\frac{M_N}{M_0}} - \sum_{i=1}^{I-1} \, (b_{i+1}-b_i) \times \ln{\left( \frac{M_i}{M_0} \right)}
- \sum_{i=I}^{N-1} \, (b_{i+1}-b_i) \times \ln{\left( \frac{M_i}{M_N} \right)} \nonumber
\eea
We have then the following alternate prescription of Eq.~(\ref{eq:allsteps})
for the efficient distinction between light and heavy thresholds. 
It will be particularly useful for the work of subsequent sections that {\it only} the better known
light fields contribute to the main $\beta$-coefficient $b_I$. 
\beq
\left\{ \frac{2\pi}{\alpha_{0}}-\frac{2\pi}{\alpha_{N}} \right\} +
\left\{
\xi_N \equiv
\sum_{\rm light}\,\delta b \ln{\left( \frac{M}{M_0} \right)} -
\sum_{\rm heavy}\delta b \ln{\left( \frac{M_N}{M} \right)}
\right\}
\:=\:
b_I \ln{\frac{M_N}{M_0}}
\label{eq:lightandheavy}
\eeq
As a note in passing, we remark that prior study~\cite{Ellis:1991ri} suggests the requisite conversion factor
between the minimal subtraction and dimensional regularization
renormalization schemes to be a comparatively insignificant contribution.

%%%%%%%%%%%%%%%%%%%%%%%%%%%%%%%%%%%%%%%%%%%%%%%%%%%%%%%%%%%%

\subsection{Computation of MSSM Thresholds\label{sct:mssmthresh}}

We review~\cite{masiero} in this section the determination of Standard Model quantum numbers and $\beta$-function coefficients.
The Standard Model (plus right-handed neutrinos) electric- and hyper-charge strengths
for the sixteen matter fields of each generation are given in Table~(\ref{tb:smf}).

\begin{table}[htbp]
\begin{center}
\begin{footnotesize}
\setlength{\extrarowheight}{12pt}
$
\begin{array}{c|c|c|c|c|c|c}
& {
\setlength{\extrarowheight}{0pt}
\begin{pmatrix}
u\cr d
\end{pmatrix}}_L
\setlength{\extrarowheight}{12pt}
\otimes 3 & u^c_R \otimes 3 & d^c_R \otimes 3 & 
{
\setlength{\extrarowheight}{0pt}
\begin{pmatrix}
\nu_e\cr e
\end{pmatrix}}_L
\setlength{\extrarowheight}{12pt}
& e^c_R & \nu^c_R \\[12pt]
\hline
\{{\rm Q},{\rm Y}'\}&
\left\{
\setlength{\extrarowheight}{0pt}
\begin{pmatrix}
+2/3\cr-1/3
\end{pmatrix}
\setlength{\extrarowheight}{12pt}
, +\frac{1}{6} \right\} &
\{-\frac{2}{3},-\frac{2}{3}\} &
\{+\frac{1}{3},+\frac{1}{3}\} &
\left\{
\setlength{\extrarowheight}{0pt}
\begin{pmatrix}
0\cr-1
\end{pmatrix}
\setlength{\extrarowheight}{12pt}
, -\frac{1}{2} \right\} &
\{+1,+1\} &
\{0,0\} \\[12pt]
\end{array}
$
\setlength{\extrarowheight}{0pt}
\end{footnotesize}
\end{center}
\caption{\label{tb:smf}
Standard Model (plus right-handed neutrino) field content and quantum numbers.}
\end{table}

The diagonal eigenvalues of ${\rm T}_3$ onto each doublet of $SU(2)_L$ are $\pm 1/2$ for up/down elements respectively,
and the hypercharge ${\rm Y}'$ is assigned so as to satisfy the defining relation ${\rm Q} = {\rm T}_3 + {\rm Y}'$.
However, in a unification scenario, $U(1)_{\rm Y}$ and $SU(2)_L$ must have a common normalization,
meaning that $\sum ({\rm T}_3)^2 = \sum ({\rm Y})^2$ for each family.
There are four doublets, times two fields each, times $((\pm 1/2)^2 = 1/4)$, for a total sum over ${\rm T}_3$ of $2$.
For hypercharge ${\rm Y}'$:
\beq
3 \times (+1/6)^2 \times 2
\enspace+\enspace
3 \times (-2/3)^2
\enspace+\enspace
3 \times (+1/3)^2
\enspace+\enspace
(-1/2)^2 \times 2
\enspace+\enspace
1
\enspace=\enspace
10/3
\eeq
This necessitates the scaling ${\rm Y} = \sqrt{5/3} \,{\rm Y}'$, where absence of the prime denotes the 
GUT normalized hypercharge\footnote{The defining relation $\tan {\theta}_{\rm W} \equiv {g'}^{}_{{\rm Y}'} / {g}^{}_{{\rm L}}$
for the Weinberg angle references the original primed hypercharge coupling.}.  Specifically, the physical
product ${g'}^{}_{{\rm Y}'}\times {\rm Y}' \equiv g^{}_{{\rm Y}}\times {\rm Y}$ of coupling times assigned charge must be
invariant, so to {\it reduce} an overly large sum of numerical hypercharge counts, we must {\it increase} the coupling
by the same proportionality.

The $\beta$-function coefficients have renormalization effects from self interactions of the gauge bosons, fermion loops, and scalar loops.
\beq
b_i = \frac{-11}{3} \, C_2(G_i) + \sum_f \frac{4}{3} \frac{d(R) C_2(R)}{d(G_i)} + \sum_\Phi \frac{1}{6}
\label{eq:betas}
\eeq
$C_2$ is the second Casimir invariant of the representation, and $d$ is the dimension of the representation.
For $SU(N \ge 2)$, $C_2(G_i) = N$, and for $U(1)$, which lacks self interactions, $C_2 = 0$.
Matter fields of the Standard Model are exclusively housed in fundamental representations, for which the dimension $d(R)$ is just $N$.
The dimension $d(G_i)$ of the $SU(N)$ adjoint is $N^2-1$, the number of generators.
The second Casimir eigenvalue of the fundamental is $C_2(R) = \frac{N^2-1}{2N}$.
The fermion loop thus reduces to $\sum_f \frac{4}{3} \times \frac{1}{2}$ for a $L/R$ vector pair with $N\ge 2$.
For the $U(1)$ special case however, it becomes $\sum_f \frac{4}{3} \times \frac{1}{2} \times ({\rm Y})^2$, leaving just a factor of $4/3$ per family after making use of the previously summed normalization.
For $SU(2)_L$, there is an overall factor of $1/2$ since only one chirality is active. 
There are four fundamental doublets, for a net of $4/3 \times 1/2 \times 1/2 \times 4 = 4/3$ per generation.
For color $SU(3)$, there are two vector triplets per family, for a total again of $4/3 \times 1/2 \times 2 = 4/3$.  
There is a single scalar doublet Higgs in the Standard Model, and thus a single term in the sum for $U(1)$ and $SU(2)$ of the scalar $\Phi$ variety.
As before, the normalization on ${\rm Y}'$ must be included, for a total of $1/6 \times 3/5 = 1/10$.
All together, with totals for $N_G = 3$ generations and $N_H = 1$ Higgs doublet:
\setlength{\extrarowheight}{8pt}
\beq
\begin{array}{ccccccccc}
b_{{\rm Y}} &=& 0 &+& \frac{4}{3} N_G &+& \frac{1}{10} N_H &\Rightarrow& \frac{41}{10} \\[4pt]
b_{2} &=& \frac{-22}{3} &+& \frac{4}{3} N_G &+& \frac{1}{6} N_H &\Rightarrow& -\frac{19}{6} \\[4pt]
b_{3} &=& -11 &+& \frac{4}{3} N_G &+& 0 &\Rightarrow& -7
\end{array}
\label{eq:sm_betas}
\eeq
\setlength{\extrarowheight}{0pt}

In the MSSM, there is an extra Higgs doublet, required to avoid complex conjugation in the holomorphic superpotential while providing both up- and down-type quark
masses\footnote{Or alternatively to avoid a chiral anomaly for the superpartner $\tilde{H}$.}.
Each of the $8+3+1 = 12$ gauge bosons, and each of the 16 matter fields also receive a superpartner shifted in spin by a half increment.
The `gauginos' are fermionic, spin-$1/2$, as are the `Higgsinos'.
The matter fields shift down to spin-$0$.
There are no new spacetime vectors, which would be necessity be gauged, thus further enlarging the spectrum and moreover creating a representational mismatch between the partners. 
For $SU(3)$, the only new spin-$1/2$ content are the adjoint of gluinos, $\tilde{g}$, for which $d(R) = d(G_i) = N^2-1$.
This leaves just $C_2(R)$, which, for the adjoint, equals $N$.
However, there is also a factor of $\frac{1}{2}$, similar to that seen previously in the $SU(2)_L$ case.
One set of gluons/gluinos service all generations, so there is no $N_G$ type factor.
The sum is over a single $8$-plet: $4/3 \times (N=3) \times 1/2 = 2$.
On the scalar side, there are generational sets of $\begin{pmatrix} \tilde{u} \cr \tilde{d} \end{pmatrix} \otimes 3$,
$\tilde{u}^c \otimes 3$, and $\tilde{d}^c \otimes 3$.
This represents four $3$-plets of complex scalars for each family, each contributing a factor of $1/6$ to each $b_i$, for a total of $1/6 \times 4 \times N_G = 2/3 N_G$
The net shift in $b_3$ is $\delta b_3 = 2 + 2/3 N_G$.

Focusing on $SU(2)$, the wino triplet contributes $4/3 \times (N=2) \times 1/2 = 4/3$.
{\em Each} of the Higgs doublets adds $1/6$, and their fermionic partners are in for $4/3 \times 1/2 \times N_H \times 1/2$, with appearance again of the halving term.
The total Higgs portion of $b_2$ is $N_H \times (1/6 + 1/3) = N_H/2$, which replaces the prior $N_H/6$ factor.
The generational scalar $SU(2)$ contributions come from the super partners of the left handed quarks and leptons,
$\begin{pmatrix} \tilde{u} \cr \tilde{d} \end{pmatrix} \otimes 3$, and  $\begin{pmatrix} \tilde{\nu} \cr \tilde{e}\end{pmatrix}$.
The effect is tallied for four complex doublets at a rate of $1/6$ per family, for a total $2/3 N_G$.
The net shift is $\delta b_2 = 4/3 + 2/3 N_G + (1/2 - 1/6) N_H$.

Finally, we consider the normalized hypercharge ${\rm Y}$.
The $W$'s carry $Q$, but are hypercharge neutral.
The $U(1)_{\rm Y}$ generator $B^0$ does not self-interact.
The sole source for new spin-$1/2$ contributions is thus the Higgsinos.
$H_u$ and $\tilde{H}_u$ carry ${{\rm Y}'}=-1/2$, while the value is $+1/2$ for ($H_d,\tilde{H}_d$).
The sum of ${{\rm Y}'}^2$ over two Higgsino doublets is thus $4 \times (1/2)^2 = 1$, giving a normalized hypercharge of $\sum ({\rm Y})^2 = 3/5$,
and a $\beta$-function contribution of $\sum 4/3 \times 1/2 \times ({\rm Y})^2 = 2/5$.
This is for $N_H = 2$, and we may more generally write the term as $N_H/5$.
The original scalar $H_d$ and the new $H_u$ contribute $1/6 \times 3/5 = 1/10$ each, or $N_H/10$ together.
The combined Higgs sector contribution to $b_{{\rm Y}}$ is thus $\frac{3}{10} N_H$.
Finally, we have the scalar matter partners for each generation.
Note that the Higgs contribution which was written as $1/6 \times 3/5$ is alternatively equivalent to
$1/6 \times 2 \sum ({\rm Y})^2$.
Taking instead the known value of $\sum ({\rm Y})^2 = 2$ for a full matter-partner generation, we apparently get a total
$1/6 \times 2 \times N_G \times 2 = \frac{2}{3} N_G$.
The net shift is $\delta b_{{\rm Y}} = \frac{2}{3} N_G + \frac{3-1}{10} N_H$.
All together, the MSSM coefficients, with ($N_G = 3, N_H = 2$) in the totals, are:
\setlength{\extrarowheight}{8pt}
\beq
\begin{array}{ccccccccc}
b_{{\rm Y}} &=& 0 &+& 2 N_G &+& \frac{3}{10} N_H &\Rightarrow& \frac{33}{5} \\[4pt]
b_{2} &=& -6 &+& 2 N_G &+& \frac{1}{2} N_H &\Rightarrow& 1 \\[4pt]
b_{3} &=& -9 &+& 2 N_G &+& 0 &\Rightarrow& -3 
\end{array}
\label{eq:mssm_betas}
\eeq
\setlength{\extrarowheight}{0pt}

The need for such measured accounting exists because we must ultimately assign individual shifts to each of the $\beta$-coefficients at the threshold mass where each MSSM field enters the renormalization.
We provide in Table~(\ref{tb:detailed_betas}) the resulting detailed breakdown, noting respect for the general principle that the fermionic member of a spin ($0,\frac{1}{2}$) super field doubles the $\beta$-contribution of its scalar partner.
Multiplicities ($\otimes$) refer to the group theoretic representational degeneracy.
The numbers shown are per generation where applicable. 

\begin{table}[htbp]
\begin{center}
\setlength{\extrarowheight}{8pt}
$
\begin{array}{|c|c|c|c||c|c|c|c|}
\hline
\textrm{Field} & \delta b_{{\rm Y}} & \delta b_2 & \delta b_3 &
\textrm{{\it S}-Field} & \delta b_{{\rm Y}} & \delta b_2 & \delta b_3 \\[4pt]
\hline
\hline
B^0 & 0 & 0 & 0 & 
\tilde{B}^0 & 0 & 0 & 0 \\[4pt] 
\hline
(W^{\pm,0}) & 0 & \frac{-22}{3} & 0 & 
(\tilde{W}^{\pm,0}) & 0 & \frac{4}{3} & 0 \\[4pt]
\hline
g \otimes 8 & 0 & 0 & -11 & 
\tilde{g} \otimes 8 & 0 & 0 & 2 \\[4pt]
\hline
\hline
u_L \otimes 3 & \frac{1}{30} & \frac{1}{2} & \frac{1}{3} & 
\tilde{u}_L \otimes 3 & \frac{1}{60} & \frac{1}{4} & \frac{1}{6} \\[4pt]
\hline
d_L \otimes 3 & \frac{1}{30} & \frac{1}{2} & \frac{1}{3} & 
\tilde{d}_L \otimes 3 & \frac{1}{60} & \frac{1}{4} & \frac{1}{6} \\[4pt]
\hline
u^c_R \otimes 3 & \frac{8}{15} & 0 & \frac{1}{3} & 
\tilde{u}^c_R \otimes 3 & \frac{4}{15} & 0 & \frac{1}{6} \\[4pt]
\hline
d^c_R \otimes 3 & \frac{2}{15} & 0 & \frac{1}{3} & 
\tilde{d}^c_R \otimes 3 & \frac{1}{15} & 0 & \frac{1}{6} \\[4pt]
\hline
\hline
\nu_L & \frac{1}{10} & \frac{1}{6} & 0 & 
\tilde{\nu}_L & \frac{1}{20} & \frac{1}{12} & 0 \\[4pt]
\hline
e_L & \frac{1}{10} & \frac{1}{6} & 0 & 
\tilde{e}_L & \frac{1}{20} & \frac{1}{12} & 0 \\[4pt]
\hline
\nu^c_R & 0 & 0 & 0 & 
\tilde{\nu}^c_R & 0 & 0 & 0 \\[4pt]
\hline
e^c_R & \frac{2}{5} & 0 & 0 & 
\tilde{e}^c_R & \frac{1}{5} & 0 & 0 \\[4pt]
\hline
\hline
H_u \otimes 2 & \frac{1}{10} & \frac{1}{6} & 0 & 
\tilde{H}_u \otimes 2 & \frac{1}{5} & \frac{1}{3} & 0 \\[4pt]
\hline
H_d \otimes 2 & \frac{1}{10} & \frac{1}{6} & 0 & 
\tilde{H}_d \otimes 2 & \frac{1}{5} & \frac{1}{3} & 0 \\[4pt]
\hline
\end{array}
$
\setlength{\extrarowheight}{0pt}
\end{center}
\caption{\label{tb:detailed_betas}
Detailed breakdown of contributions to the one-loop MSSM $\beta$-function coefficients.}
\end{table}

As a practical matter, output of the standard tools such as SSARD, FeynHiggs and ISASUGRA used to determine
sparticle masses is not in one-to-one correspondence with the fields listed in Table~(\ref{tb:detailed_betas}).
Table~(\ref{tb:isasugra_betas}) provides $\beta$-contributions for all mass content above $M_{\rm Z}$, tallied for the specific spectral output of these program.
Print style symbols and text style field designations are given, as well as the full corresponding MSSM representation.
Multiplicities ($\times$) refer to the number of included generations.
For values spanning multiple rows, the predicted masses will merged as a geometric mean for purposes of computation.
The fields ($h^0,H^0,A^0$) represent mixtures of the three surviving neutral Higgs elements from the complex
doublets $\left(H_u \equiv \begin{pmatrix} 0\cr-\end{pmatrix},H_d \equiv \begin{pmatrix} +\cr0\end{pmatrix}\right)$ after $Z^0$ becomes massive in the symmetry breaking.
Likewise, $H^\pm$ are the two charged Higgs elements surviving after generation of the $W^\pm$ masses.

\begin{table}[htbp]
\begin{center}
\setlength{\extrarowheight}{3pt}
$
\begin{array}{|c|c|c|c|c|c|}
\hline
\textrm{Print} & \textrm{Text} & \textrm{MSSM Content} & \delta b_{{\rm Y}} & \delta b_2 & \delta b_3 \\[2pt]
\hline
\hline
t & \textrm{t} & (u_L,u^c_R) \times 1 & \frac{17}{30} & \frac{1}{2} & \frac{2}{3} \\[2pt]
\hline
\hline
h^0 & \textrm{HL} & \multirow{3}{*}{($H^0_u,H^0_d$)} & \frac{1}{40} & \frac{1}{24} & 0 \\[2pt]
\cline{1-2}
\cline{4-6}
H^0 & \textrm{HH} & & \frac{1}{40} & \frac{1}{24} & 0 \\[2pt]
\cline{1-2}
\cline{4-6}
A^0 & \textrm{HA} & & \frac{1}{40} & \frac{1}{24} & 0 \\[2pt]
\hline
H^\pm & \textrm{H+} & (H^-_u,H^+_d)
 & \frac{1}{20} & \frac{1}{12} & 0 \\[2pt]
\hline
\hline
\tilde{x}^0_1 & \textrm{Z1} & \multirow{4}{*}{($\tilde{B}^0,\tilde{W}^0,\tilde{H}^0_u,\tilde{H}^0_d$)} &
\multirow{4}{*}{$\frac{1}{5}$} & \multirow{4}{*}{$\frac{7}{9}$} & \multirow{4}{*}{$0$} \\[2pt]
\cline{1-2}
\tilde{x}^0_2 & \textrm{Z2} &&&& \\[2pt]
\cline{1-2}
\tilde{x}^0_3 & \textrm{Z3} &&&& \\[2pt]
\cline{1-2}
\tilde{x}^0_4 & \textrm{Z4} &&&& \\[2pt]
\hline
\tilde{x}^\pm_1 & \textrm{W1} & \multirow{2}{*}{($\tilde{W}^-,\tilde{W}^+,\tilde{H}^-_u,\tilde{H}^+_d$)} &
\multirow{2}{*}{$\frac{1}{5}$} & \multirow{2}{*}{$\frac{11}{9}$} & \multirow{2}{*}{$0$} \\[2pt]
\cline{1-2}
\tilde{x}^\pm_2 & \textrm{W2} &&&& \\[2pt]
\hline
\tilde{g} & \textrm{GL} & \tilde{g} & 0 & 0 & 2 \\[2pt]
\hline
\hline
\tilde{e}^\pm_L & \textrm{EL} & \tilde{e}_L \times 2 & \frac{1}{10} & \frac{1}{6} & 0 \\[2pt]
\hline
\tilde{e}^\pm_R & \textrm{ER} & \tilde{e}^c_R \times 2 & \frac{2}{5} & 0 & 0 \\[2pt]
\hline
\tilde{\nu} & \textrm{SN} & (\tilde{\nu}_L,\tilde{\nu}^c_R) \times 2 & \frac{1}{10} & \frac{1}{6} & 0 \\[2pt]
\hline
\tilde{\tau}^\pm_1 & \textrm{TAU1} & \multirow{2}{*}{$(\tilde{e}_L,\tilde{e}^c_R) \times 1$} &
\multirow{2}{*}{$\frac{1}{4}$} & \multirow{2}{*}{$\frac{1}{12}$} & \multirow{2}{*}{$0$} \\[2pt]
\cline{1-2}
\tilde{\tau}^\pm_2 & \textrm{TAU2} &&&& \\[2pt]
\hline
\tilde{\nu}_\tau & \textrm{NTAU} & (\tilde{\nu}_L,\tilde{\nu}^c_R) \times 1 & \frac{1}{20} & \frac{1}{12} & 0 \\[2pt]
\hline
\hline
\tilde{u}_L & \textrm{UL} & \tilde{u}_L \times 2 & \frac{1}{30} & \frac{1}{2} & \frac{1}{3} \\[2pt]
\hline
\tilde{u}_R & \textrm{UR} & \tilde{u}^c_R \times 2 & \frac{8}{15} & 0 & \frac{1}{3} \\[2pt]
\hline
\tilde{d}_L & \textrm{DL} & \tilde{d}_L \times 2 & \frac{1}{30} & \frac{1}{2} & \frac{1}{3} \\[2pt]
\hline
\tilde{d}_R & \textrm{DR} & \tilde{d}^c_R \times 2 & \frac{2}{15} & 0 & \frac{1}{3} \\[2pt]
\hline
\tilde{t}_1 & \textrm{T1} & \multirow{2}{*}{$(\tilde{u}_L,\tilde{u}^c_R) \times 1$} &
\multirow{2}{*}{$\frac{17}{60}$} & \multirow{2}{*}{$\frac{1}{4}$} & \multirow{2}{*}{$\frac{1}{3}$} \\[2pt]
\cline{1-2}
\tilde{t}_2 & \textrm{T2} &&&& \\[2pt]
\hline
\tilde{b}_1 & \textrm{B1} & \multirow{2}{*}{$(\tilde{d}_L,\tilde{d}^c_R) \times 1$} &
\multirow{2}{*}{$\frac{1}{12}$} & \multirow{2}{*}{$\frac{1}{4}$} & \multirow{2}{*}{$\frac{1}{3}$} \\[2pt]
\cline{1-2}
\tilde{b}_2 & \textrm{B2} &&&& \\[2pt]
\hline
\end{array}
$
\setlength{\extrarowheight}{0pt}
\end{center}
\caption{\label{tb:isasugra_betas}
Beta function contributions for the full MSSM field content above $M_{\rm Z}$ are broken down according to the output groupings of
standard sparticle program libraries.}
\end{table}

We may now proceed with a computation of the sum from Eq.~(\ref{eq:alln}) for each of the three gauge couplings as soon
as an appropriate sparticle mass spectrum has been specified.
We will generally defer here to the suggested post-WMAP CMSSM benchmark scenarios\footnote{A
partial update (excepting the focus point region and $\mu < 0$)
from single to double-primed benchmarks was made in 2005~\cite{DeRoeck:2005bw},
but the distinction is not significant to our purposes here.}
of Ref.~\cite{Battaglia:2003ab}.
Specifically, we will default to the point $B'$ as characterized by the values following due to its proximity to
the region of parameter space favored by a least-squares analysis~\cite{Ellis:2004tc} of predictions for the $W$ boson mass,
the effective Weinberg angle, $(g-2)$ of the muon, and the branching ratio for $(b \rightarrow s\gamma)$.
Recognizing that improvements both in experimental constraint and in the refinement of certain program libraries
have transpired since publication of these references, we nevertheless suggest that the present benchmark set sufficiently
serves the purpose of collectively establishing an approximate border on variation within plausible parameter
space bounds while facilitating ease of comparison with other work.
\beq
\begin{array}{ccccc}
A_0 = 0 &;&  \mu > 0 &;& \tan \beta = 10 \\[8pt]
&m_{0} = 60 &;& m_{1/2} = 250&
\end{array}
\label{bestfitIN}
\eeq

We will accept at face value the supersymmetric mass textures suggested in~\cite{Battaglia:2003ab},
which were computed using the SSARD and FeynHiggs library
routines\footnote{The gluino mass scale appears to be generically larger than that produced by ISASUGRA, but the spectra are otherwise in reasonable agreement.}.
Recognizing also that we will shortly diverge from certain assumptions made there as regards flipped unification and the addition
of extra TeV scale multiplets, we again consider the spectrum to be a sufficiently plausible one for our purposes,
and judge that any potential discrepancy represents a correction to the correction.
For the top quark mass, we adopt the printed `benchmark' value only for its own threshold factor,
rejecting it for all other calculations\footnote{We likewise
prefer the PDG value of the strong coupling $\alpha_3$ for all calculations.},
\eg the Yukawa boundary on the second loop, in favor of the Tevatron Electroweak Working Group world average
$m_t = 173.1 \pm 1.3~{\rm [GeV]}$, courtesy of CDF and D\O~\cite{:2009ec}. We refer back to Eq. (\ref{lepdg}) for $M_{\rm Z}$.
Our baseline threshold corrections for the MSSM are then calculated to take the values shown following.
\beq
\xi_{\rm Y} \approx 3.20 \quad;\quad
\xi_2 \approx 5.48 \quad;\quad
\xi_3 \approx 8.05
\label{mssmthresh}
\eeq

We have recently studied~\cite{Li:2010mr} the implications of gravity mediated F-theoretic supersymmetry
breaking for the electroweak scale gaugino mass relations, presenting two scenarios which are consistent
with the latest experimental constraints, including CDMS~II.  The interplay of these considerations
with the topic of proton decay will be taken up in a forthcoming letter. 

\section{$\mathcal{F}$-lipped $SU(5)$\label{main:flipped}}

If the notion of a GUT is to be feasible, then one must necessarily inquire
as to candidates for the unified group structure and what representational
form the known interactions and fields would take within that group.
To contain the SM, $SU(3)_C \times SU(2)_L \times U(1)_Y$, which carry respectively
two, one and one diagonal generators, a group of minimal rank four is required.
The natural starting point is then $SU(5)$, whose lowest order representations
are the singlet $\mathbf{1}$, the fundamental $\mathbf{5}$, the (anti) symmetric
tensors $\mathbf{10}$ and $\mathbf{15}$, the adjoint $\mathbf{24}$ which transforms
as the generators, and their related conjugates.

Within each family, all SM states must be assigned a residence which is compatible
with their existing quantum numbers, as shown in Table~(\ref{tb:smcontent}).
A charge-parity involution is understood where needed such that all grouped
fields will carry a consistent handedness.
The six quark states of the left-doublet, color-triplet must then fit in at least
an antisymmetric $\mathbf{10}$.
This leaves four spots open, tidily filled by one right-handed color triplet
and a right handed singlet.
The remaining color-triplet partners neatly with the electron-neutrino left-doublet
as a $\mathbf{\bar{5}}$, canceling the non-Abelian anomaly of the $\mathbf{10}$.
Not only are the fifteen Standard Model states thus uniquely and compactly
represented, but we are gifted additional `wisdom' in the process.
By assignment each of the $\mathbf{10}$ and $\mathbf{\bar{5}}$ are electrically neutral,
thus correlating charge quantization to the number of color degrees of freedom.
Furthermore, the apparent masslessness of the neutrinos finds a pleasing justification:
there is simply no room remaining in which to house a right-handed component.

\begin{table}[htbp]
\begin{center}
$
\begin{array}{ccccc}
{\begin{pmatrix} u \cr d \end{pmatrix}}_L &;& u^c_L &;& d^c_L \\
\vspace{1mm}\\
{\begin{pmatrix} \nu_e \cr e \end{pmatrix}}_L &;& e^c_L &;& \fbox{$\nu^c_L$} \\
\end{array}
\quad
\begin{matrix}\textrm{\small{JUST}} \cr \longrightarrow \cr \textrm{\small{RIGHT}} \end{matrix}
\quad
\begin{array}{ccccc}
\setlength{\extrarowheight}{3pt}
{\begin{pmatrix}
d^c_1 \cr d^c_2 \cr d^c_3
\cr e \cr \nu_e \end{pmatrix}}_L
\setlength{\extrarowheight}{0pt}
&;&
{\begin{pmatrix} {\begin{pmatrix} u \cr d \end{pmatrix}}_L&\!\!u^c_L&\!e^c_L \end{pmatrix}} &;& \fbox{$\nu^c_L$} \\
\vspace{-.5mm}\\
\bar{\mathbf{5}} &\phantom{;}& \mathbf{10} &\phantom{;}& \fbox{$\mathbf{1}$} \\
\end{array}
$
\end{center}
\caption{\label{tb:smcontent}
The Standard $SU(5)$ charge assignments are `just right' to compactly
house a full generation within a fundamental five-plet and an antisymmetric ten-plet.
However, this applies only if if one is willing to either neglect
the right-handed neutrino or exile it to a singlet representation.}
\end{table}

But experimental evidence cannot any longer allow an agnostic position on
neutrino masses.
The Super-Kamiokande facility in Japan, which houses $50,000$ metric tonnes of
ultra-pure water inside a $40$-meter high by $40$-meter diameter cylindrical
tank faced on all sides by a collection of $13,000$ photomultiplier
{\v C}erenkov detectors and shielded beneath $2,700$ meters of earth within
the cavity of an old mine, has been diligently studying the problem for many years.
By comparing the careful observation of neutrino fluxes with atmospheric origination
against the expected detection ratios from known interaction cascades they have
borne convincing witness to oscillation between the $\nu_\mu$ and $\nu_\tau$ sectors\footnote{Likewise,
$\nu_e \leftrightarrow \nu_\mu$ oscillation has been induced from
the observation of solar neutrinos by the SNO collaboration.}.
This phenomenon may only occur when the related states carry non-equivalent masses,
of which at least one must then be non-zero.
However, chirality can only be an invariant quantum number for massless states,
and we are thus compelled to introduce a sixteenth element for accommodation of the
right-handed neutrino\footnote{There is a loophole to this consideration for
Standard $SU(5)$ formulations with pure Majorana neutrino masses}.
It is certainly possible to imagine the new state as a simple singlet outside the
main representations already laid down.
Surely though it is presumptive to suppose that this right-handed neutrino,
while arriving `last' must also be seated as the `least'.
Every existing position must instead be subject to reassignment.
There is indeed then another way, if one is willing to sacrifice charge quantization.
We can choose to `flip'\,\footnote{For a comprehensive
review of Flipped $SU(5)$, please consult~\cite{Nanopoulos:2002qk}}
the right-handed quarks placed inside the $\mathbf{\bar{5}}$
and $\mathbf{10}$ while also swapping $e^c_L$ for $\nu^c_L$.

\begin{table}[htbp]
\begin{center}
$
f_{\mathbf{\bar{5}}} =
\setlength{\extrarowheight}{3pt}
{\begin{pmatrix}u^c_1\cr u^c_2\cr u^c_3\cr e\cr \nu_e \end{pmatrix}}_L
\setlength{\extrarowheight}{0pt}
\quad;\quad
F_{\mathbf{10}} =
{\begin{pmatrix}{\begin{pmatrix} u \cr d \end{pmatrix}}_L&\!\!d^c_L&\!\nu^c_L \end{pmatrix}}
\quad;\quad
l_{\mathbf{1}} = e^c_L
$
\end{center}
\caption{\label{tb:flippedcontent}
The Flipped $SU(5)$ assignment exchanges the roles of the two right-handed quark triplets and
also the two right-handed lepton singlets.  The resulting structure is no longer a simple group.}
\end{table}

\begin{figure}[htp]
\begin{center}
\includegraphics[width=\plotwidth,angle=0]{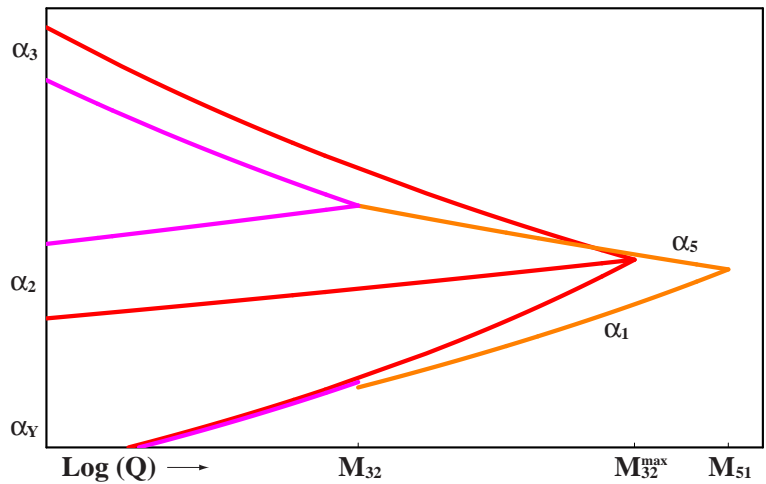}
\end{center}
\caption{\label{fig:heuristic}
An exaggerated heuristic demonstration of the prominent features of flipped coupling unification.
Notice the discontinuity in the (purple) line of $U(1)_Y$, as it remixes between the `grand unified'
$U(1)_{\rm X}$, and that which emerges out of broken $SU(5)$ at the scale $M_{32}$.
Proceeding upward from this interim stage in orange, $SU(5) \times U(1)_{\rm X}$
is itself unified at some higher scale $M_{51}$.
For comparison, the Standard $SU(5)$ scenario is shown in red with a single unification
at $M^{\rm max}_{32} \geq M_{32}$, and `predicting' a larger value for $\alpha_s(M_Z)$.}
\end{figure}

The cost of this flipping is nothing less than the loss of grand unification, as
the resulting symmetry group is enlarged to the non-simple variant
$SU(5) \times U(1)_X$.
As demonstrated in Figure~(\ref{fig:heuristic}),
the hypercharge does not descend out of $SU(5)$ together with the
nuclear forces, but is instead of an admixture of $U(1)_X$ together with the additional
U(1) factor which is emergent from that breaking.
If this tune sounds familiar though, it is simply a reprise of the theme played out
some fourteen orders of magnitude below in the Glashow-Weinberg-Salam $SU(2)_L \times U(1)_Y$
electroweak `unification'.
Just as it was there no loss to save a true convergence for the future inclusion of
color perhaps it is here folly to imagine a full unification which occurs on the border
of the Planck mass without waiting on gravity. 
Just as only experiment could there reject the truly unified but dysfunctional
$SU(2)$ model of Georgi and Glashow, between (or against!) these contenders we can
again only let phenomenology decide.
And here there is no finer judge than the consideration of proton decay.

One of the major predictions of GUTs is that the proton 
becomes destabilized due to the quark and lepton unification. 
Pairs of quarks may transform into a lepton and an anti-quark
via dimension six operators for 
the exchange of heavy gauge bosons, and thus
the proton may decay into a lepton plus meson final state.  
Because the masses of heavy gauge bosons are near to the GUT scale, 
such processes are expected to be very rare. Indeed, 
proton decay has not yet been seen in the expansive
Super-Kamiokande experiment, which currently places a 
lower bound on the dimension six partial lifetimes around
$6-8\times 10^{33}$ years~\cite{:2009gd}.

In Standard $SU(5)$~\cite{Georgi:1974sy}, there exists the 
Higgs doublet-triplet splitting problem, and the additional threat of proton 
decay via dimension five operators from the 
colored Higgsino field (supersymmetric partners of the
colored triplet Higgs fields) 
exchange~\cite{Murayama:2001ur}.
These difficulties are solved elegantly in Flipped
$SU(5)\times U(1)_X$~\cite{Barr:1981qv, Derendinger:1983aj, Antoniadis:1987dx},
due to the missing partner 
mechanism~\cite{Antoniadis:1987dx},
and we thus only need consider dimension six proton decay.
This initial salvation from the overly rapid `ignoble' sparticle channel
can however turn subsequently into frustration that allowed portions of the
parameter space of minimal Flipped $SU(5)$ models appears to predict lifetimes
so long as to be unobservable by even hypothetical proposals for future experiments
~\cite{Ellis:1995at,Murayama:2001ur,Ellis:2002vk}.

In this paper, we consider the testable Flipped $SU(5)\times U(1)_X$
models of Ref.~\cite{Jiang:2006hf} with inclusion of TeV-scale vector-like particles.
Such models can be realized within free fermionic string
constructions~\cite{Lopez:1992kg} and also F-theory~\cite{Vafa:1996xn, Donagi:2008ca, Beasley:2008dc, Beasley:2008kw,
Donagi:2008kj, Heckman:2008rb, Jiang:2009zza, Jiang:2009za}.
Interestingly, we can solve the little hierarchy problem
between the string scale and the GUT scale in the free
fermionic constructions~\cite{Jiang:2006hf}, and we can
explain the decoupling scenario in F-theory model
building~\cite{Jiang:2009zza, Jiang:2009za}.

The minimal Flipped $SU(5)$ models~\cite{Barr:1981qv, Derendinger:1983aj, Antoniadis:1987dx}. 
contain three families of SM fermions as in Table~(\ref{tb:flippedcontent}),
whose quantum numbers under $SU(5)\times U(1)_{X}$ are
as follows, with $i=(1, 2, 3)$. 
\bea
F_i={\mathbf{(10, 1)}} \quad;\quad
{\bar f}_i={\mathbf{(\bar 5, -3)}} \quad;\quad
{\bar l}_i={\mathbf{(1, 5)}}
\label{smfermions}
\eea

To break the GUT and electroweak gauge symmetries, we 
introduce two pairs of Higgs fields, and also add a SM singlet field $\Phi$.
\bea
H={\mathbf{(10, 1)}} \quad;\quad
{\overline{H}}={\mathbf{({\overline{10}}, -1)}} \quad;\quad
h={\mathbf{(5, -2)}} \quad;\quad
{\overline h}={\mathbf{({\bar {5}}, 2)}}
\label{Higgse1}
\eea
The particle assignments of Higgs fields are given following, where
$H_d$ and $H_u$ are one pair of Higgs doublets in the supersymmetric SM.
\bea
H=(Q_H, D_H^c, N_H^c) \quad&;&\quad
{\overline{H}} = ({\overline{Q}}_{\overline{H}},
{\overline{D}}^c_{\overline{H}},
{\overline {N}}^c_{\overline H})
\label{Higgse3} \\[4pt]
h=(D_h, D_h, D_h, H_d) \quad&;&\quad
{\overline h}=({\overline {D}}_{\overline h}, {\overline {D}}_{\overline h},
{\overline {D}}_{\overline h}, H_u)
\nonumber
\eea

To break the $SU(5)\times U(1)_{X}$ gauge symmetry down to the Standard Model,
we introduce the following Higgs superpotential at the GUT scale.
\bea
{\it W}_{\rm GUT}=\lambda_1 H H h + \lambda_2 {\overline H} {\overline H} {\overline
h} + \Phi ({\overline H} H-M_{\rm H}^2)
\label{spgut} 
\eea 
There is only
one F-flat and D-flat direction, which can always be rotated into orientation with 
$N^c_H$ and ${\overline {N}}^c_{\overline H}$, yielding 
$\langle N^c_H\rangle=\langle {\overline {N}}^c_{\overline H}\rangle=M_{\rm H}$. In addition, the
superfields $H$ and ${\overline H}$ are absorbed, acquiring large masses via
the supersymmetric Higgs mechanism, except for $D_H^c$ and 
${\overline {D}}^c_{\overline H}$. The superpotential terms 
$ \lambda_1 H H h$ and
$ \lambda_2 {\overline H} {\overline H} {\overline h}$ couple the $D_H^c$ and
${\overline {D}}^c_{\overline H}$ with the $D_h$ and ${\overline {D}}_{\overline h}$,
respectively, to form the massive eigenstates with masses
$2 \lambda_1 \langle N_H^c \rangle$ and $2 \lambda_2 \langle {\overline {N}}^c_{\overline H}\rangle$. So then, we
naturally achieve doublet-triplet splitting due to the missing
partner mechanism~\cite{Antoniadis:1987dx}. Because
the triplets in $h$ and ${\overline h}$ only have
small mixing through the $\mu h {\overline h}$ term with
$\mu$ around the TeV scale, we also solve the dimension five
proton decay problem from colored Higgsino exchange.

In Flipped $SU(5)\times U(1)_X$, the 
$SU(3)_C\times SU(2)_L$ gauge couplings are first joined 
at the scale $M_{32}$, and the $SU(5)$ and $U(1)_X$ gauge
couplings are subsequently unified at the higher scale $M_{51}$.
To separate the $M_{32}$ and $M_{51}$ scales
and obtain true string-scale gauge coupling unification in 
free fermionic models~\cite{Jiang:2006hf} or
the decoupling scenario of F-theory~\cite{Jiang:2008yf},
we introduce, as elaborated in the following section,
vector-like particles which form complete
Flipped $SU(5)\times U(1)_X$ multiplets.
In order to avoid the Landau pole
problem for the strong coupling constant, we can only introduce the
following two sets of vector-like particles around the TeV 
scale~\cite{Jiang:2006hf}.

\begin{subequations}
\label{eqs:z1z2}
\bea
Z1&:& XF={\mathbf{(10, 1)}} ~,~ {\overline{XF}}={\mathbf{({\overline{10}}, -1)}} \\[4pt]
Z2&:& XF ~,~ {\overline{XF}} ~,~ Xl={\mathbf{(1, -5)}} ~,~ {\overline{Xl}}={\mathbf{(1, 5)}}
\eea
\end{subequations}

For notational simplicity, we define the Flipped
$SU(5)\times U(1)_X$ models with $Z1$ and
$Z2$ type sets of vector-like particles as 
Type~I and Type~II Flipped
$SU(5)\times U(1)_X$ models, respectively.
Although we focus in this paper on Type~II,
results for proton decay are not be found to differ significantly
between the Type~I and Type~II scenarios. 
The behavior of the two models above $M_{32}$ is however quite distinct,
and as shown in Section~(\ref{sct:superunification}), the Type~II variation
is in this regard preferable.

%%%%%%%%%%%%%%%%%%%%%%%%%%%%%%%%%%%%%%%%%%%%%%%%%%%%%%%%%%%%

\section{$\mathcal{F}$-theory\label{main:ftheory}}

%%%%%%%%%%%%%%%%%%%%%%%%%%%%%%%%%%%%%%%%%%%%%%%%%%%%%%%%%%%%

\subsection{Elements of $\mathcal{F}$-Theory\label{sct:ftheory}}

Recently, semi-realistic Grand Unified Theories (GUTs) 
have been constructed locally in 
the F-theory with seven-branes, which can be considered as the
strongly coupled formulation of ten-dimensional Type~IIB string 
theory~\cite{Vafa:1996xn, Donagi:2008ca,
Beasley:2008dc, Beasley:2008kw, Donagi:2008kj}.
Model building and phenomenological consequences have been studied 
extensively~\cite{Heckman:2008es, Marsano:2008jq, Font:2008id,
Heckman:2008qa, Jiang:2009zza, Blumenhagen:2008aw, Bourjaily:2009vf,
Hayashi:2009ge, Chen:2009me, Heckman:2009bi, Donagi:2009ra,
Jiang:2009za, Li:2009cy, Cecotti:2009zf, Li:2009fq, Marsano:2009wr}. 
The known GUTs without additional chiral 
exotic particles are asymptotically free, and the asymptotic 
freedom can be translated into the existence of a consistent 
decompactification limit. Interestingly,
the GUT scale $M_{\rm GUT}$ is around
$2\times 10^{16}~{\rm [GeV]}$ while the reduced Planck scale $M_{\rm Pl}$ 
is around $2\times 10^{18}~{\rm [GeV]}$, so $M_{\rm GUT}/M_{\rm Pl}$ is 
indeed a small number around $10^{-2}$.
Thus, it is natural to assume that
$M_{\rm GUT}/M_{\rm Pl}$ is small from the effective
field theory point of view in the bottom-up approach, and
then gravity can be decoupled.
In the decoupling limit where $M_{\rm Pl} \rightarrow \infty$
while $M_{\rm GUT}$ remains finite, semi-realistic
$SU(5)$ models and $SO(10)$ models without chiral exotic particles
have been constructed locally.

To decouple gravity and 
avoid the bulk matter fields on the observable 
seven-branes, we can show that the observable seven-branes should 
wrap a del Pezzo $n$ surface $dP_n$ with $n \ge 2$ for the internal 
space dimensions~\cite{Beasley:2008dc, Beasley:2008kw}.
A review of del Pezzo $n$ surfaces is provided here as an Appendix.
The GUT gauge fields are on the world-volume of the
observable seven-branes, and  the SM fermions and 
Higgs fields are localized on the codimension-one curves 
in $dP_n$. All the SM fermion Yukawa couplings can be
obtained from the triple
intersections of the SM fermion and Higgs curves.
A brand new feature is that the $SU(5)$ gauge symmetry
can be broken down to the SM gauge symmetry by turning
on the $U(1)_Y$ flux~\cite{Beasley:2008dc, Beasley:2008kw, Chen:2009me, Li:2009cy}, 
and the $SO(10)$ gauge symmetry
can be broken down to $SU(5)\times U(1)_X$ or 
$SU(3)\times SU(2)_L \times SU(2)_R \times U(1)_{B-L}$
by turning on 
the $U(1)_X$ and $U(1)_{B-L}$ fluxes, 
respectively~\cite{Beasley:2008dc, Beasley:2008kw, Font:2008id,
Jiang:2009zza, Jiang:2009za, Li:2009cy}.
In particular,
in the $SO(10)$ models, to eliminate the zero modes of
the chiral exotic particles,  we must break the $SO(10)$ gauge
symmetry down to the Flipped $SU(5)\times U(1)_X$ gauge 
symmetry~\cite{Beasley:2008kw}.

In Flipped $SU(5)\times U(1)_X$ models of
$SO(10)$ origin, there are two unification
scales: the $SU(2)_L \times SU(3)_C$ unification scale $M_{32}$
and the $SU(5)\times U(1)_X$ unification scale $M_{51}$,
where $M_{32}$ is in the vicinity the usual GUT scale around
$2\times 10^{16}~{\rm [GeV]}$.  To address the little hierarchy problem between
the GUT scale and string scale, one may introduce extra
vector-like particles, achieving models with string-scale
$SU(5)\times U(1)_X$ gauge coupling unification~\cite{Lopez:1995cs, Jiang:2006hf}.
These fields may be naturally associated with a highly restricted class of
Flipped $SU(5)\times U(1)_{\rm X}$ F-theory model building constructions~\cite{Jiang:2009za},
as is the perspective taken in the present work.
%Similarly, for the Flipped $SU(5)\times U(1)_X$ models
%from F-theory, we can naturally obtain the decoupling scenario 
%where $M_{32}/M_{51}$ or $M_{32}/M_{\rm Pl}$ can be made small
%by the introduction of additional vector-like particles.

In the F-Theory GUTs proposed in Refs.~\cite{Donagi:2008ca,Beasley:2008dc}, the decoupling scenario under
consideration implies that $M_{\rm GUT}/M_{\rm Pl}$ is a small number, say $\mathcal{O}~(0.1)$ or less.
In our Flipped $SU(5)$ models with F-theory derived vectorlike particle content, the analogous statement
is rather that the ratio $M_{32}/M_{51}$ is a small number.  Satisfaction of this requirement allows for a solution
of the monopole problem and a corresponding realization of string scale gauge unification in a free-fermionic
model building context.  However, because $M_{51}$ is elevated close to the Planck scale, $M_{\rm S}/M_{\rm Pl}$ is
not a small number in the present constructions, where $M_{\rm S}$ is a scale set by the inverse radius
of the K\"ahler surface wrapped by the GUT seven-branes.  Although we therefore cannot decouple
gravity, and there may also be corresponding corrections to the gauge kinetic terms, the picture of gauge
coupling unification which we present will be undisturbed, since there is an $SO(10)$ gauge symmetry
around $M_{\rm S}$ which implies that any corrections will be applied universally.  Moreover,
our immediately central topic of proton decay is likewise protected, as its dominant physics is unearthed
some two orders of magnitude below, at $M_{32}$.  Any shift in the final unified coupling $g_{51}$ will thus
be irrelevant to this discussion, and any correction terms applied around the scale $M_{51}$ should
be expected to experience a relative suppression of around eight orders of magnitude.

We remark that our construction can indeed be made consistent with both perspectives of decoupling,
if one invokes the presence of heavy thresholds to reduce $M_{51}$ to $\mathcal{O}~(10^{17})~{\rm [GeV]}$, keeping
$M_{32} \sim \mathcal{O}~(10^{16})~{\rm [GeV]}$, with $M_{\rm Pl} \sim \mathcal{O}~(10^{18})~{\rm [GeV]}$.  Indeed, this is not
merely a hypothetical consideration, as two concrete examples (models Type IA1 and IA2) satisfying this
triple hierarchy have been presented in Tables~(IX,X) of Ref.~\cite{Jiang:2009za}.  These scenarios thus
allow for perturbation analysis in the small expansion parameter $M_{51}/M_{\rm Pl}$.  However, since
the dimension six proton decay results are unaffected, we will not consider this possibility any further
at the present time.

%%%%%%%%%%%%%%%%%%%%%%%%%%%%%%%%%%%%%%%%%%%%%%%%%%%%%%%%%%%%

\subsection{$\mathcal{F}$-Theory Model Building\label{sct:fmodels}}

In this section we briefly review F-theory model 
building~\cite{Vafa:1996xn, Donagi:2008ca,
Beasley:2008dc, Beasley:2008kw, Donagi:2008kj}.
The twelve-dimensional F theory is a convenient way to describe
Type~IIB vacua with varying
axion ($a$)-dilaton ($\phi$) field 
$\tau=a+ie^{-\phi}$.
We compactify F-theory on a
Calabi-Yau fourfold, which is elliptically fibered $\pi: Y_4 \rightarrow B_3$
with a section $\sigma: B_3 \rightarrow Y_4$. The base $B_3$ is the internal
spatial dimensions of Type~IIB string theory, and the complex structure
of the $T^2$ fibre encodes $\tau$ at each point of $B_3$. The SM or GUT
gauge theories are on the world-volume of the observable
seven-branes that wrap
a complex codimension-one surface in $B_3$. Denoting the complex
coordinate transverse to these seven-branes in $B_3$ as $z$, we can
write the elliptic fibration in  Weierstrass form
\begin{eqnarray}
 y^2=x^3+f(z)x+g(z)~,~\,
\end{eqnarray}
where $f(z)$  and $g(z)$ are sections of $K_{B_3}^{-4}$ and
$K_{B_3}^{-6}$, respectively. The complex structure of the fibre is: 
\begin{eqnarray}
 j(\tau)~=~ \frac{4(24f)^3}{\Delta} \quad;\quad
\Delta~=~ 4 f^3 + 27 g^2
\end{eqnarray}

At the discriminant locus $\{\Delta=0\} \subset B_3$,
the torus $T^2$ degenerates by pinching one of its
cycles and becomes singular. For a generic pinching one-cycle 
 $(p, q)=p\alpha+q\beta$ where $\alpha$ and $\beta$
are one-cycles for the torus $T^2$, we obtain a $(p,q)$ seven-brane in 
the locus where the $(p,q)$ string can end.
The singularity types of the elliptically fibres fall into the 
familiar $ADE$ classifications, and we identify the corresponding 
$ADE$ gauge groups on the seven-brane world-volume. 
This is one of the most important advantages for F-theory model building: 
the exceptional gauge groups appear rather naturally, in contrast
perturbative Type~II string theory. Subsequently all the SM fermion Yukawa
couplings can be generated.

We assume that the observable seven-branes with GUT models
on its world-volume wrap a complex codimension-one 
surface $S$ in $B_3$, and the observable gauge symmetry
is $G_S$. When $h^{1,0}(S)\not=0$, the low energy
spectrum may contain the extra states obtained
by reduction of the bulk supergravity modes of
compactification, so we require that 
$\pi_1(S)$ be a finite group. In order to decouple
gravity and construct models locally, the extension 
of the local metric on $S$ to
a local Calabi-Yau fourfold must have a limit where
the surface $S$ can be shrunk to zero size. This implies
that the anti-canonical bundle on $S$ must be ample. 
Therefore, $S$ is a del Pezzo $n$ surface $dP_n$ with $n \ge 2$
in which $h^{2,0}(S)=0$.
The Hirzebruch surfaces with degree larger than 2
satisfy $h^{2,0}(S)=0$ but do not define the fully 
consistent decoupled models~\cite{Beasley:2008dc, Beasley:2008kw}.

To describe the spectrum, we have to study the gauge theory 
of the world-volume on the seven-branes.  We 
start from the maximal supersymmetric gauge theory on
$\mathbb{R}^{3,1}\times \mathbb{C}^{2}$ and then replace
$\mathbb{C}^{2}$ with the K\"ahler surface $S$. In order to have
four-dimensional ${\cal N}=1$ supersymmetry, the
maximal supersymmetric gauge theory on $\mathbb{R}^{3,1}\times
\mathbb{C}^{2}$ should be twisted. It was shown that there exists a
unique twist preserving ${\cal N}=1$ supersymmetry in four
dimensions, and that chiral matter can arise from the bulk $S$ or the
codimension-one curve $\Sigma$ in $S$ which is the intersection
between the observable seven-branes and non-observable 
seven-brane(s)~\cite{Beasley:2008dc, Beasley:2008kw}.

In order to have matter fields on $S$,
we consider a non-trivial vector bundle with 
structure group $H_S$ which is a subgroup of $G_S$. Then the gauge
group $G_S$ is broken down to $\Gamma_S\times H_S$, and the adjoint
representation ${\rm ad}(G_S)$ of the $G_S$ is decomposed as: 
\begin{equation}
{\rm ad}(G_S)\rightarrow
{\rm ad}(\Gamma_S)\oplus {\rm ad}(H_S)\oplus_j(\tau_j,T_j)
\end{equation}

Employing the vanishing theorem of the del Pezzo surfaces,
we obtain the numbers of the generations and anti-generations 
by calculating the zero modes of the Dirac operator on $S$
\begin{eqnarray}
 n_{\tau_j}~=~ -\chi (S, \mathbf{T_j})~,~~~ 
n_{\tau_j^*}~=~ -\chi (S, \mathbf{T_j}^*)~,~\,
\end{eqnarray}
where $\mathbf{T_j}$ is the vector bundle on $S$ whose 
sections transform in the representation $T_j$ of $H_S$,
and $\mathbf{T_j}^*$ is the dual bundle of $\mathbf{T_j}$.
In particular, when the $H_S$ bundle is a line bundle $L$,
we have:
\begin{eqnarray}
n_{\tau_j}~=~-\chi (S, L^j)~=~
-\Big[1+\frac{1}{2}\big(\int_{S}c_{1}({L}^{j})c_{1}(S)+
\int_{S}c_{1}({L}^{j})^2\big)\Big]
\label{EulerChar}
\end{eqnarray}

In order to preserve supersymmetry, the line bundle $L$ should satisfy
 the BPS equation~\cite{Beasley:2008dc}
\begin{equation}
J_{S}\wedge c_{1}(L)=0,\label{BPS}
\end{equation}
where $J_{S}$ is the K\"ahler form on $S$. Moreover,
the admissible supersymmetric line bundles on del Pezzo surfaces must
satisfy $c_{1}(L)c_{1}(S)=0$, thus,
 $n_{\tau_j}=n_{\tau_j^*}$ and only the vector-like particles
can be obtained. In short, we can not have the chiral matter fields
on the world-volume of the observable seven-branes.

Let us consider a stack of seven-branes with gauge group
$G_{S'}$ that wrap a codimension-one surface $S'$ in $B_3$.
The intersection of $S$ and $S'$ is a codimension-one curve
(Riemann surface) $\Sigma$ in $S$ and $S'$, 
and the gauge symmetry on $\Sigma$
will be enhanced to $G_{\Sigma}$ 
where $G_{\Sigma}\supset G_{S}\times G_{S'}$.
On this curve, there exists chiral matter from 
the decomposition of the adjoint representation
${\rm ad}(G_{\Sigma})$ of $G_{\Sigma}$ as follows:
\begin{equation}
{\rm ad}(G_{\Sigma})={\rm ad}(G_{S})\oplus {\rm ad}(G_{S'})\oplus_{k}
({ U}_{k}\otimes { U'}_{k})
\end{equation}

Turning on the non-trivial gauge bundles on $S$ and $S'$, respectively
with structure groups $H_S$ and $H_{S'}$, we break the gauge 
group $G_S\times G_{S'}$ down to the commutant subgroup 
$\Gamma_{S}\times\Gamma_{S'}$. Defining 
$\Gamma \equiv \Gamma_{S}\times\Gamma_{S'}$ and 
$H \equiv H_{S}\times H_{S'}$,
we can decompose ${ U}\otimes { U'}$ into the irreducible
representations
\begin{equation}
{ U}\otimes { U'}={\oplus}_{k}(r_{k}, {V}_{k}),
\end{equation}
where $r_{k}$ and ${ V}_{k}$ are the representations of $\Gamma$
and $H$, respectively. The light chiral fermions in the
representation $r_{k}$ are determined by the zero modes of the
Dirac operator on $\Sigma$. The net number of chiral superfields
 is given by
\begin{eqnarray}
N_{r_{k}}-N_{r^{*}_{k}}=\chi(\Sigma,K^{1/2}_{\Sigma}\otimes
{\mathbf{V}_{k}}),
\end{eqnarray}
where $K_{\Sigma}$ is the  restriction of
canonical bundle on the curve $\Sigma$, and
$\mathbf{V}_{k}$ is the vector bundle whose sections 
transform in the representation ${ V}_{k}$ of 
the structure group $H$. 

In F-theory model building, we are interested in the 
models where $G_{S'}$ is $U(1)'$, and
$H_S$ and $H_{S'}$ are respectively $U(1)$
and $U(1)'$. Then the vector bundles on $S$ and $S'$ 
are line bundles $L$ and $L'$. The adjoint representation
${\rm ad}(G_{\Sigma})$ of $G_{\Sigma}$ is 
decomposed into a direct sum
of the irreducible representations under the group
$\Gamma_S \times U(1) \times U(1)'$ that can be
denoted as $\mathbf{(r_j, q_j, q'_j)}$.
\begin{equation}
{\rm ad}(G_{\Sigma})={\rm ad}(\Gamma_S)
\oplus {\rm ad}(G_{S'})\oplus_{j}
\mathbf{(r_j, q_j, q_j')}~.~\,
\end{equation}
The numbers of chiral superfields in the representation 
$\mathbf{(r_j, q_j, q'_j)}$ and their Hermitian conjugates
on the curve $\Sigma$ are given by 
\begin{eqnarray}
N_{\mathbf{(r_j, q_j, q'_j)}} ~=~ h^0 (\Sigma, \mathbf{V}_j) ~,~~~
N_{\mathbf{({\bar r}_j, -q_j, -q'_j)}} ~=~ h^1(\Sigma, \mathbf{V}_j)~,~\,
\end{eqnarray}
where 
\begin{eqnarray}
\mathbf{V}_j~=~ K^{1/2}_{\Sigma} \otimes
{L}_{\Sigma}^{q_{j}}\otimes {L'}_{\Sigma}^{q'_{j}} ~,~\,
\end{eqnarray}
where $K^{1/2}_{\Sigma}$, ${L}_{\Sigma}^{r_{j}}$ and
${L'}_{\Sigma}^{q'_{j}}$ are the restrictions of
canonical bundle $K_S$, line bundles $L$ and $L'$ on the curve
$\Sigma$, respectively. In particular, if the
volume of $S'$ is infinite, $G_{S'}=U(1)'$ is decoupled,
and then the index $\mathbf{q'_j}$ can be ignored.

Using the Riemann-Roch theorem, we obtain the net number of 
chiral superfields in the representation $\mathbf{(r_j, q_j, q'_j)}$ as
\begin{eqnarray}
N_{\mathbf{(r_j, q_j, q'_j)}}-
N_{\mathbf{({\bar r}_j, -q_j, -q'_j)}}~=~ 1-g+c_1(\mathbf{V}_j) ~,~\,
\end{eqnarray}
where $g$ is the genus of the curve $\Sigma$, and $c_1$ means the first
Chern class.

Moreover, we can obtain the Yukawa couplings 
at the triple intersection of
three curves $\Sigma_i$, $\Sigma_j$ and $\Sigma_k$ where
 the gauge group or the singularity type is enhanced further.
To have triple intersections, the corresponding
homology classes  $[\Sigma_i]$, $[\Sigma_j]$ and $[\Sigma_k]$
of the curves $\Sigma_i$, $\Sigma_j$ and $\Sigma_k$ must satisfy
the following conditions:
\begin{eqnarray}
[\Sigma_i] \cdot [\Sigma_j] > 0 \quad;\quad
[\Sigma_i] \cdot [\Sigma_k] > 0 \quad;\quad
[\Sigma_j] \cdot [\Sigma_k] > 0
\label{FTYK-Con}
\end{eqnarray}

%%%%%%%%%%%%%%%%%%%%%%%%%%%%%%%%%%%%%%%%%%%%%%%%%%%%%%%%%%%%

\subsection{F-Theory GUTs}

Flipped $SU(5)\times U(1)_X$ models with additional vector-like 
particles have been studied 
systematically~\cite{Beasley:2008dc, Beasley:2008kw, 
Jiang:2009zza, Jiang:2009za}.
In this paper, we supplement those efforts by generic construction
of the Georgi-Glashow $SU(5)$ models
with additional vector-like particles.
In such $SU(5)$
models, we introduce the  vector-like particles $YF$
and ${\overline{YF}}$, and $Yf_i$
and ${\overline{Yf}_i}$, whose quantum numbers under $SU(5)$ are:
\beq
YF={\mathbf{10}}
\quad;\quad
\overline{YF}={\mathbf{\overline{10}}}
\quad;\quad
Yf_i={\mathbf{5}}
\quad;\quad
\overline{Yf}_i={\mathbf{\overline{5}}}
\eeq
The SM field content from the decomposition of
$YF$, ${\overline{YF}}$, $Yf_i$, and ${\overline{Yf}}_i$ is:
\bea
YF = (XQ, XU^c, XE^c) \quad&;&\quad {\overline{YF}}=(XQ^c, XU, XE)
\\[4pt]
Yf_i=(XD, XL^c) \quad&;&\quad {\overline{Yf}}_i= (XD^c, XL)
\nonumber
\eea
Under the $SU(3)_C \times SU(2)_L \times U(1)_Y$ gauge
symmetry, the quantum numbers for the extra vector-like 
particles are:
\bea
XQ={\mathbf{(3, 2, \frac{1}{6})}}
\quad&;&\quad
XQ^c={\mathbf{({\bar 3}, 2,-\frac{1}{6})}}
\nonumber \\[4pt]
XU={\mathbf{({3},1, \frac{2}{3})}}
\quad&;&\quad
XU^c={\mathbf{({\bar 3},  1, -\frac{2}{3})}}
\nonumber \\[4pt]
XD={\mathbf{({3},1, -\frac{1}{3})}}
\quad&;&\quad
XD^c={\mathbf{({\bar 3},  1, \frac{1}{3})}}
\\[4pt]
XL={\mathbf{({1},  2,-\frac{1}{2})}}
\quad&;&\quad
XL^c={\mathbf{(1,  2, \frac{1}{2})}}
\nonumber \\[4pt]
XE={\mathbf{({1},  1, {-1})}}
\quad&;&\quad
XE^c={\mathbf{({1},  1, {1})}}
\nonumber
\eea

We consider two $SU(5)$ models. The Type~I $SU(5)$ model
has one pair of vector-like particles  $YF$
and ${\overline{YF}}$, and the Type~II $SU(5)$ model
has three pairs of vector-like particles $Yf_i$
and ${\overline{Yf}_i}$. 
We assume that the observable gauge group on the $dP_8$ surface $S$ 
is $SU(5)$. On codimension one curves that are intersections 
of the observable seven-branes with other non-observable
seven-branes, we obtain the
SM fermions, Higgs fields, and extra vector-like particles. To break the 
$SU(5)$ gauge symmetry down to
the $SU(3)\times SU(2)_L \times  U(1)_Y$ gauge symmetry, 
we turn on the $U(1)_Y$ flux on $S$ specified by the line bundle $L$. 
To obtain the SM fermions, Higgs fields and vector-like particles, 
we also turn on the $U(1)$ fluxes on the other seven-branes that intersect
with the observable seven-branes, and we specify these fluxes
by the line bundle $L^{\prime n}$. 

We take the line bundle $L=\mathcal{O}_{S}(E_{1}-E_{2})^{1/5}$.
Note that $\chi(S, L^5)=0$, \ie we do not have vector-like particles
on the bulk $S$.
The curves and homology classes for the matter fields, Higgs fields 
and vector-like particles, and the gauge bundle assignments for 
each curve in the Type~I and Type~II $SU(5)$ models are given 
in Table~(\ref{tb:FSU5}). From this table, we see that 
all the SM fermions are localized on the matter curves
$\Sigma_{F}$ and $\Sigma_{\overline{f}}$,
the Higgs fields $H_u$ and $H_d$ are localized on the
curves $\Sigma_{Hu}$ and $\Sigma_{Hd}$ respectively, 
and the vector-like particles
$YF$, $\overline{YF}$, $Yf_i$, and $\overline{Yf}_i$ 
are localized on the curves 
$\Sigma_{F}$, $\Sigma_{\overline{F}}$, $\Sigma_{f} $, 
and $\Sigma_{\overline{f}}$.
In the Type~I $SU(5)$ model, we choose $n=1$ and $m=0$,
while in the Type~II $SU(5)$ model, we choose
$n=0$ and $m=3$.
In addition, there exist singlets
from the intersections of the other seven-branes.
It is easy to check that we can realize the SM fermion
Yukawa coupling terms in the superpotential. 
All the vector-like particles
can obtain masses by giving vacuum expectation values
to the SM singlets at the intersections of 
the other seven-branes.

\begin{table}[htbp]
\begin{center}
\setlength{\extrarowheight}{3pt}
$
\begin{array}{|c|c|c|c|c|c|}
\hline
\textrm{Fields}
&
\textrm{Curves}
&
\textrm{Class}
&
g_{\Sigma}
&
L_{\Sigma}
&
L_{\Sigma}^{\prime n}
\\[2pt]
\hline
\hline
H_u & \Sigma_{Hu} & 2H-E_{1}-E_3 & 0 &
\mathcal{O}(1)^{1/5} & \mathcal{O}(1)^{2/5}\\[2pt]
\hline
H_d & \Sigma_{Hd} & 2H-E_{2}-E_3 & 0 &
\mathcal{O}(-1)^{1/5} & \mathcal{O}(-1)^{2/5}\\[2pt]
\hline
10_i+n\times XF & \Sigma_{F} & 2H-E_{4}-E_6 & 0 
& \mathcal{O}(0)& 
\mathcal{O}(3+n)\\[2pt]
\hline
n\times \overline{XF} & \Sigma_{\overline{F}} & 2H-E_{5}-E_6 & 0 
& \mathcal{O}(0) & 
\mathcal{O}(-n)\\[2pt]
\hline
\overline{5}_i+m\times \overline{Xf}_i & 
\Sigma_{\overline{f}} & -E_7 & 0 
& \mathcal{O}(0) & 
\mathcal{O}(-3-m)\\[2pt]
\hline
m\times Xf_i & 
\Sigma_{f} & H-E_8 & 0 
& \mathcal{O}(0) & 
\mathcal{O}(m)\\[2pt]
\hline
\end{array}
$
\setlength{\extrarowheight}{0pt}
\end{center}
\caption{\label{tb:FSU5}
The particle curves and 
gauge bundle assignments for each curve in the 
$SU(5)$ models from F-theory. The index 
 $i=1,~2,~3$. In the Type~I $SU(5)$ model, we choose $n=1$ and $m=0$.
In the Type~II $SU(5)$ model, we choose
$n=0$ and $m=3$.}
\end{table}

%%%%%%%%%%%%%%%%%%%%%%%%%%%%%%%%%%%%%%%%%%%%%%%%%%%%%%%%%%%%

\subsection{Thresholds From Extra Vector-Like Multiplets\label{sct:vectors}}

Following~\cite{Jiang:2006hf, Barger:2007qb} and references therein, we present the supplementary
step-wise contributions to the MSSM one-
and two-loop $\beta$-function coefficients from the vector-like particles.
$\Delta b \equiv (\Delta b_{\rm Y}, \Delta b_2, \Delta b_3)$, and likewise $\Delta B$ are given as
complete supermultiplets, including the conjugate representation in each case. 
\bea
\Delta b^{XQ + XQ^c} = (\frac{1}{5}, 3, 2) \quad&;&\quad 
\Delta b^{XU + XU^c} = (\frac{8}{5}, 0, 1) \nonumber \\[4pt]
\Delta b^{XD + XD^c} = (\frac{2}{5}, 0, 1) \quad&;&\quad
\Delta b^{XL + XL^c} = (\frac{3}{5}, 1, 0) \label{vectorb} \\[4pt]
\Delta b^{XE + XE^c} = (\frac{6}{5}, 0, 0) \quad&;&\quad
\Delta b^{XN + XN^c} = (0, 0, 0)          \nonumber \\[4pt]
\Delta b^{XY + XY^c} = (5, 3, 2)          \quad&;&\quad
\Delta b^{XT_i + \overline{XT}_i} = (\frac{6}{5}, 0, 0) \nonumber
\eea

\setlength{\extrarowheight}{4pt}
\bea
\Delta B^{XQ + XQ^c} =
\begin{pmatrix}
\frac{1}{75}&
\frac{3}{5}&\frac{16}{15}\cr \frac{1}{5} & 21 & 16 \cr
\frac{2}{15}& 6 & \frac{68}{3}
\end{pmatrix}
\quad&;&\quad
\Delta B^{XU + XU^c} =
\begin{pmatrix}
\frac{128}{75}&
0 &\frac{128}{15}\cr 0 & 0 & 0 \cr
\frac{16}{15} & 0 & \frac{34}{3}
\end{pmatrix}
\nonumber \\[4pt]
\Delta B^{XD + XD^c} =
\begin{pmatrix}
\frac{8}{75}&
0 &\frac{32}{15}\cr 0 & 0 & 0 \cr
\frac{4}{15} & 0 & \frac{34}{3}
\end{pmatrix}
\quad&;&\quad
\Delta B^{XL + XL^c}=\begin{pmatrix}
\frac{9}{25}&
\frac{9}{5}& 0 \cr \frac{3}{5} & 7 & 0 \cr
0 & 0 & 0
\end{pmatrix}
\label{vectorB} \\[4pt]
\Delta B^{XE + XE^c} =
\begin{pmatrix}
\frac{72}{25}&
0 & 0 \cr 0 & 0 & 0 \cr
0 & 0 & 0
\end{pmatrix}
\quad&;&\quad
\Delta B^{XN + XN^c} =
\begin{pmatrix}
0&
0 & 0 \cr 0 & 0 & 0 \cr
0 & 0 & 0
\end{pmatrix}
\nonumber \\[4pt]
\Delta B^{XY + {{XY^c}}} =
\begin{pmatrix}
\frac{25}{3}&
15 &\frac{80}{3}\cr 5 & 21 & 16 \cr
\frac{10}{3}& 6 & \frac{68}{3}
\end{pmatrix}
\quad&;&\quad
\Delta B^{XT_i + \overline{XT}_i} =
\begin{pmatrix}
\frac{18}{25}&
0 & 0 \cr 0 & 0 & 0 \cr
0 & 0 & 0
\end{pmatrix}
\nonumber
\eea
\setlength{\extrarowheight}{0pt}

\begin{table}[htbp]
\begin{center}
\begin{footnotesize}
\setlength{\extrarowheight}{3pt}
$
\begin{array}{|c||c||c|c|c|}
\hline
\textrm{Scenario} & \textrm{Vector Super-Multiplets} &
b_{\rm Y} & b_2 & b_3
\\[2pt]
\hline
\hline
{\textrm{SU(5)}}_0 & 
\textrm{None}
& 6.6 & 1 & -3
\\[2pt]
\hline
{\textrm{SU(5)}}_{\rm I} &
\left( {YF} \equiv {\left\{ XQ,{XU}^c,{XE}^c \right\}}_{\bf 10},\, {\overline{YF}} \right)
& 9.6 & 4 & 0
\\[2pt]
\hline
{\textrm{SU(5)}}_{\rm II} &
3 \:\otimes\: \left( {Yf} \equiv \left\{ {XD}^c, XL \right\}_{\bf 5},\, {\overline{Yf}} \right)
& 9.6 & 4 & 0
\\[2pt]
\hline
\hline
{\mathcal{F}\textrm{-SU(5)}}_0 &
\textrm{None}
& 6.6 & 1 & -3
\\[2pt]
\hline
{\mathcal{F}\textrm{-SU(5)}}_{\rm I} &
\left( {XF} \equiv {\left\{ XQ,{XD}^c,{XN}^c \right\}}_{\bf 10},\, {\overline{XF}} \right)
& 7.2 & 4 & 0
\\[2pt]
\hline
{\mathcal{F}\textrm{-SU(5)}}_{\rm II} &
\left( {XF} \equiv {\left\{ XQ,{XD}^c,{XN}^c \right\}}_{\bf 10},\, {\overline{XF}} \right)
\:\oplus\:
\left( {Xl} \equiv \left\{ XE \right\}_{\bf 1},\, {\overline{Xl}} \right)
& 8.4 & 4 & 0
\\[2pt]
\hline
\end{array}
$
\setlength{\extrarowheight}{0pt}
\end{footnotesize}
\end{center}
\caption{\label{tb:modelcontent}
Light field content and $\beta$-function coefficients for the
Standard $SU(5)$ and Flipped $SU(5) \times U(1)_{\rm X}$ models considered in this report,
including TeV scale vector-like multiplets from F-theory.}
\end{table}

In Table~(\ref{tb:modelcontent}), we present the specific field content of the F-theory
models which are considered in this report.  For both Flipped and Standard $SU(5)$, we begin
with a `Type 0' model representing the bare MSSM.  As discussed in Section~(\ref{sct:ftheory}),
avoidance of a Landau pole in the renormalization places very strict limits on the combinations
of vector multiplets which may be considered.  We study two variations
for each GUT category, labeled sequentially as (I,II), adopting the notation of Ref.~\cite{Li:2009fq}.
There is no immediate connection however between the flipped and standard models which share
the designation of Type~I or Type~II.  Moreover, since the possible vector multiplets
which we may to choose to populate in Flipped and Standard $SU(5)$ are distinct, as required
by the different charge assignments in the two theories, there is no continuous field theoretic
transition between the two broad model classes once the heavy fields are implemented.
The $\beta$-coefficients of the first loop are tallied, starting with the MSSM as presented in
Eq.~{\ref{eq:sm_betas}}, and adding the relevant elements from Eq.~(\ref{vectorb}).
The matrices for the second loop are summed in the identical manner using Eq.~(\ref{vectorB}),
although we suppress a direct printing of the totals.

%%%%%%%%%%%%%%%%%%%%%%%%%%%%%%%%%%%%%%%%%%%%%%%%%%%%%%%%%%%%

\section{$\mathcal{F}$-resh Analysis\label{main:fresh}}

%%%%%%%%%%%%%%%%%%%%%%%%%%%%%%%%%%%%%%%%%%%%%%%%%%%%%%%%%%%%

\subsection{Grand Unification With Thresholds\label{sct:threshunif}}

We will now consider the question of how the presence of a sum over internal scales
effects the picture of unification, particularly in Flipped $SU(5)$. 
We start here with the standard set of 3 one-loop equations,
including thresholds via the factors\footnote{The subscripts on $b$ and $\xi$ now revert to traditional usage,
labeling the given gauge interaction.}
$\xi_i$ introduced in Section~(\ref{sct:multiplethresholds}).
\begin{subequations}
\label{RGE_STD}
\bea
\frac{1}{\alpha_{\rm Y}}-\frac{1}{\alpha_1} \:+ \frac{\xi_{\rm Y}}{2\pi} &=&
\frac{b_{\rm Y}}{2\pi}\, \ln{\frac{M_{32}}{M_{\rm Z}}}
\\*
\frac{1}{\alpha_2}-\frac{1}{\alpha_5} \:+ \frac{\xi_2}{2\pi} &=&
\frac{b_2}{2\pi}\, \ln{\frac{M_{32}}{M_{\rm Z}}}
\\*
\frac{1}{\alpha_3}-\frac{1}{\alpha_5} \:+ \frac{\xi_3}{2\pi} &=&
\frac{b_3}{2\pi}\, \ln{\frac{M_{32}}{M_{\rm Z}}}
\eea
\end{subequations}

The mass scale $M_{32}$ of Flipped $SU(5)$ allows that
the weak and strong couplings may partially unify to $\alpha_5$ in isolation,
earlier than the full triple unification of Standard $SU(5)$ would occur, and $\alpha_1$ defines
the normalized hypercharge evaluated at that point.
As usual:
\begin{subequations}
\label{defcoups}
\bea
\alpha_{\rm Y} &=&
\frac{5\, \alpha_{\rm em}(M_{\rm Z})}{3\,(1-\sin^2\theta_{\rm W})} \\[4pt]
\alpha_2 &=& \frac{\alpha_{\rm em}(M_{\rm Z})}{\sin^2\theta_{\rm W}} \\[4pt]
\alpha_3 &\equiv& \alpha_{\rm s}(M_{\rm Z})
\eea
\end{subequations}

Whenever numbers are required, we will turn to the precision electroweak
measurements compiled by the Particle Data Group~\cite{Amsler:2008zzb}.
\bea
\alpha_{\rm em}(M_{\rm Z}) = \frac{1}{127.925 \pm .016} \quad&;&\quad
\alpha_{\rm s}(M_{\rm Z}) = .1176 \pm .0020
\label{lepdg} \\[0pt]
\sin^2 \theta_{\rm W}^{\overline{\rm MS}}(M_{\rm Z}) = .23119 \pm .00014
\quad&;&\quad M_{\rm Z} = 91.1876 \pm .0021~{\rm [GeV]} 
\nonumber
\eea
Interestingly, the strict unification limit $M_{32} \rightarrow M^{\rm max}_{32}$
is validated to a surprising accuracy in the first loop using the field content of the MSSM
($b_Y= 33/5$; $b_2=+1$; $b_3=-3$) and neglecting thresholds~\cite{Ellis:1990zq}.

We may now attempt to absorb the factors of $\xi_i$ in a convenient way,
concealing explicit references.
It has been a common practice of earlier analysis~\cite{Ellis:1990wk,Ellis:1991ri,Ellis:1995at,Ellis:2002vk}
to account for the thresholds and calculable two-loop
corrections simultaneously by shifting $\sin^2 \theta_{\rm W}$ to an effective value.
\beq
\sin^2\theta_{\rm W}
\Rightarrow
\sin^2(\theta_{\rm W}^{\rm\, eff})
\equiv
\sin^2\theta_{\rm W}-\delta_{\rm 2loop}
-\delta_{\rm light}-\delta_{\rm heavy}\
\label{NLO}
\eeq

However, it appears that this approach treats the factors of
$\sin^2 \theta_{\rm W} \equiv (1-\cos^2 \theta_{\rm W})$ from the running of
the weak and hypercharge couplings consistently {\it only if}
$\:(\xi_{\rm Y} = - 3/5\, \xi_2)$ and $(\xi_3 = 0)$.
There is no general protection which applies to these conditions, nor are 
they found to be regularly satisfied in example.
Nevertheless, it is possible to let this idea guide a notation\footnote{These
definitions will be extended in Section~(\ref{sct:2ndloop}) to include the correction from the second loop.},
which will ease comparison with prior results.
\begin{subequations}
\label{eq:sseff}
\bea
\Xi_{\rm Y} &\equiv& 1 + \frac{\alpha_{\rm em}}{2\pi} \left\{ \frac{5}{3}\:\xi_{\rm Y} + \xi_2 \right\} \\[4pt]
\Theta_{\rm W} &\equiv& \sin^2\theta_{\rm W} + \alpha_{\rm em} \:\xi_2 / 2\pi \\[8pt]
\Xi_3 &\equiv& 1 + \alpha_3 \:\xi_3 / 2\pi
\eea
\end{subequations}

The RGEs with threshold effects accounted now take the form shown following\footnote{It
is emphasized that the $b_i$ are to be tallied {\it inclusive} of all {\it light}
fields with $M_{\rm Z} < M_i \ll M_{32}$.}.
A reduction fully consistent with prior implicit assumptions is achieved
in the limit $(\Xi_{\rm Y} = \Xi_3 = 1)$.
\begin{subequations}
\label{eqs:RGES}
\bea
\frac{3}{5} \: \frac{\left( \Xi_{\rm Y} - \Theta_{\rm W} \right)}{\alpha_{\rm em}}
- \frac{1}{\alpha_1} &=&
\frac{b_{\rm Y}}{2\pi}\, \ln{\frac{M_{32}}{M_{\rm Z}}}
\label{RGEY} \\*
\frac{\Theta_{\rm W}}{\alpha_{\rm em}}-\frac{1}{\alpha_5} &=&
\frac{b_2}{2\pi}\, \ln{\frac{M_{32}}{M_{\rm Z}}}
\label{RGE2} \\*
\frac{\Xi_3}{\alpha_3}-\frac{1}{\alpha_5} &=&
\frac{b_3}{2\pi}\, \ln{\frac{M_{32}}{M_{\rm Z}}}
\label{RGE3}
\eea
\end{subequations}
We note that Eqs. (\ref{eqs:RGES}) are intended formally to bridge the $M_{\rm Z}$ and $M_{32}$ scales,
and as such the referenced factors of Eqs.~(\ref{eq:sseff}) include the totality of second order effects
over this domain.  For example, in Eq.(\ref{RGE3}), if $\alpha_5$ is backed all the way down to $M_{\rm Z}$,
we find $\alpha_5(M_{\rm Z}) = \alpha_3(M_{\rm Z}) \div \Xi_3$, which effectively downshifts
the origin value of the coupling at $M_{\rm Z}$.  We emphasize that this is not an actual departure
from the physically measured value of the coupling, but rather a `visual' artifact of the chosen
calculation methodology.  Specifically, this methodology is adopted to deliver reliable results at
the unknown {\it upper} boundary of the renormalization range terminated at $M_{32}$, and having
absorbed correction factors at the outset for this complete transit, one should be cautious
reinterpreting the results at arbitrarily reduced mass scales.

These three independent algebraic equations allow us to solve for three independent quantities.
We will choose $\alpha_5$, $\alpha_1$, and depending on application, one of either $M_{32}$ or $\Theta_{\rm W}$.
The scale $M_{\rm Z}$ and couplings $\alpha_{({\rm em},3)}(M_{\rm Z})$ are experimentally input,
and the $\beta$-coefficients $b_{{\rm Y},2,3}$ and light thresholds $\delta b_{{\rm Y},2,3}$ will be considered as fixed input
for the context of a given model.
If $M_{32}$ is further input, then the effective shift of (the squared-sine of) the Weinberg angle is
determined as output, and any departure from the value stipulated in Eq.~(\ref{eq:sseff}) will be taken as a sign
of ignorance regarding unaccounted heavy thresholds and higher loop corrections.

It is important to recognize that although the Weinberg angle itself is certainly an
experimental parameter, $\Theta_{\rm W}$ is undetermined to the extent that the
second order factors which it must have absorbed in order to achieve the mandated unification are undetermined.
The key point is that every other variable under
consideration carries {\it implicit} dependence on $M_{32}$, and we must be cautious to
avoid substitutions of a given variable between equations evaluated at disparate scales. 
Indeed, subtle errors of precisely this type have plagued earlier reports with which the
present authors have been associated~\cite{Ellis:1995at,Ellis:2002vk}.
In particular, Eqs.~(\ref{alsmz},\ref{al5gen_min}) now supersede their previously published analogues,
as first advertised in~\cite{Walker:2005dr}.
Our current perspective imparts additional caution to the contextual interpretation and use even of expressions,
such as the $\left. M^{\rm max}_{32} \right|^{\rm MSSM}_{\Xi_i=1}$ limit of Eq.~(\ref{eq:tripleofsstind_3}),
which are formally identical to their predecessors.

If one takes as concrete a specification of the post-$M_{\rm Z}$ spectrum, and diligence is
paid to corrections in the second loop, then it is reasonable to extract in converse a
prediction for the flipped unification scale $M_{32}$. 
In practice, our approach will most often be along this second line,
attempting to leverage the enforcement of phenomenological (including cosmological!) restrictions
on the parameter space of the constrained MSSM against a prediction
for the unification scale $M_{32}$ and related proton lifetime.

To proceed, we will also need to consider the Standard $SU(5)$
scenario of strict unification, \ie $\alpha^{\rm max}_1 = \alpha^{\rm max}_5$,
at an energy scale designated\footnote{The superscript `max' applied to any parameter indicates
evaluation at the extremal scale $M^{\rm max}_{32}$. It does not necessarily imply that the labeled quantity itself
experiences a maximum.}
$M^{\rm max}_{32}$.
This specialization yields an {\it additional} set of three independent equations, which 
are exactly sufficient to solve for the set of variables remaining:
($M^{\rm max}_{32}$,~$\alpha^{\rm max}_5$,~$\Theta^{\rm max}_{\rm W}$).
If the restriction of strict triple unification is relaxed, and $M_{32}$
is allowed to slide downward from its maximal position, the burden
which unification imposes on the thresholds within $\Theta_{\rm W}$ can be dispersed.

We begin by taking, in the triple unification limit, the linear combination of
Eqs.~(\ref{eqs:RGES}) which eliminates the two traditional
variables $\alpha^{\rm max}_5$ and $\Theta^{\rm max}_{\rm W}$.
The result, inverted to solve for the limiting value of $M_{32}$ is:
\beq
M^{\rm max}_{32} = M_{\rm Z} \times \exp \left\{ \frac{2\pi}{\alpha_{\rm em} \alpha_3} \:
\left( \frac{3 \,\alpha_3\,\Xi_{\rm Y} - 8 \,\alpha_{\rm em}\,\Xi_3}
{5 b_{\rm Y} + 3 b_2 - 8 b_3} \right) \right\}
\label{m32max}
\eeq
For the minimal scenario of $(\Xi_{\rm Y} = \Xi_3 = 1)$ and MSSM values for the $b_i$,
this reduces:
\beq
\left. M^{\rm max}_{32} \right|^{\rm MSSM}_{\Xi_i=1} = M_{\rm Z} \times \exp \left\{ \frac{\pi(3\,\alpha_3 - 8\,\alpha_{\rm em})}
{30\,\alpha_{\rm em}\alpha_3} \right\}
\simeq 2.089 \times 10^{16}~{\rm [GeV]}
\label{m32max_min}
\eeq
Similarly excluding $\alpha_5$ between just (\ref{RGE2},\ref{RGE3})
at {\it generic} values for $M_{32}$, and eliminating $M_{\rm Z}$ by use of the expansion
\beq
\ln\frac{M_{32}}{M_{\rm Z}} \,=\,
\ln\frac{M^{\rm max}_{32}}{M_{\rm Z}}
\,-\, \ln\frac{M^{\rm max}_{32}}{M_{32}}
\quad ,
\label{logexp}
\eeq
we arrive at the following expression, using also Eq.~(\ref{m32max}) to reach the second form.
\bea
\ln\frac{M^{\rm max}_{32}}{M_{32}}
&=& \frac{2\pi}{\alpha_{\rm em}\alpha_3}
\:\left( \frac{\alpha_{\rm em}\,\Xi_3 - \alpha_3 \Theta_{\rm W}}{b_2 -b_3} \right)
+ \ln\frac{M^{\rm max}_{32}}{M_{\rm Z}}
\label{lnmaxbym32} \\
&=& \frac{2\pi}{\alpha_{\rm em}\alpha_3}
\:\left(
\frac{3\, \alpha_3 \Xi_{\rm Y}}{(5 b_{\rm Y} + 3 b_2 - 8 b_3)} -
\frac{\alpha_3 \Theta_{\rm W}}{(b_2-b_3)} +
\frac{5\, \alpha_{\rm em} \Xi_3 (b_{\rm Y}-b_2)}{(b_2-b_3)(5 b_{\rm Y} + 3 b_2 - 8 b_3)}
\right)
\nonumber
\eea
This inverts to a `solution' for the strong coupling at $M_{\rm Z}$.
\beq
\frac{1}{\alpha_3}
= \frac{(5 b_{\rm Y} + 3 b_2 - 8 b_3) \times
\left\{ \Theta_{\rm W} + \frac{\alpha_{\rm em}(b_2-b_3)}{2\pi} \,\ln\frac{M^{\rm max}_{32}}{M_{32}} \right\}
- 3\, \Xi_{\rm Y} (b_2-b_3)
}{
5\, \alpha_{\rm em}\, \Xi_3\, (b_{\rm Y} - b_2)}
\label{alsmz_full}
\eeq

We note that it is not truly consistent to interpret this formula
as an independently floating prediction for the strong coupling\footnote{It
is certainly possible to rigorously isolate $\alpha_3$ as a dependent variable.  Although it is not considered profitable
here to do so for Flipped $SU(5)$, Eqs.~(\ref{eqs:tripleofsstind}) 
take exactly this view of the triple unification.},
since other solutions in our current set are dependent upon the experimentally fixed value of $\alpha_3$.
The present is nevertheless useful for visualizing the related slice of parameterizations in $M_{32}$ and
the effective Weinberg angle which are compatible with the experimental bounds on $\alpha_3$.
Going again to the minimal limit, Eqs.~(\ref{lnmaxbym32},\ref{alsmz_full})
compactify significantly.  In particular, Eq.~(\ref{alsmz}) has
been applied historically\footnote{The
coefficient $\frac{10}{\pi}$ in Eq.~(\ref{alsmz}) corrects
the erroneous value $\frac{11}{2\pi}$ which appears in some earlier publications.}
for essentially the purpose just described.
\bea
\ln\frac{M^{\rm max}_{32}}{M_{32}}
\left. \hspace{-2pt} \right|^{\rm MSSM}_{\Xi_i=1}
&=&
2\pi \times \left(
\frac{1- 5 \sin^2(\theta_{\rm W}^{\rm\, eff})}{20\,\alpha_{\rm em}} +
\frac{7}{60\,\alpha_3}
\right)
\label{eq:maxto32_min} \\
\left. \alpha_3(M_{\rm Z}) \right|^{\rm MSSM}_{\Xi_i=1}
&=& \frac{\frac{7}{3}\,\alpha_{\rm em}}
{5 \sin^2(\theta_{\rm W}^{\rm\, eff}) -1
+\frac{10}{\pi}\,\alpha_{\rm em}\ln\left(M^{\rm max}_{32}/M_{32}\right)}
\label{alsmz} 
\eea
Next, we solve Eq.~(\ref{RGE3}) in the triple unification for $\alpha^{\rm max}_5$, using Eq.~(\ref{m32max})
in the logarithm.
\beq
\alpha^{\rm max}_5 \equiv \alpha_5(M^{\rm max}_{32}) =
\left[
\frac{ \alpha_{\rm em} \Xi_3 (5b_{\rm Y} + 3b_2)
- 3 \alpha_3 \Xi_{\rm Y} b_3
}{
\alpha_{\rm em} \alpha_3 (5 b_{\rm Y} + 3 b_2 - 8 b_3)}
\right]^{-1}
\label{al5max}
\eeq

The shift term for generic $M_{32}$ scales can be read off of Eq.~({\ref{RGE3}}) with the aid
again of the logarithmic relation from Eq.~(\ref {logexp}). 
\beq
\frac{1}{\alpha_5} = \frac{1}{\alpha^{\rm max}_5} + \frac{b_3}{2\pi} \,\ln\frac{M^{\rm max}_{32}}{M_{32}}
\label{al5gen}
\eeq
There is a similar expression for the splitting off of $\alpha_1$ which can be obtained by
eliminating $\Theta_{\rm W}$ between Eqs.~(\ref{RGEY},\ref{RGE2})
prior to reading off the `max' limit and noting that, by definition,
$\alpha_1^{\rm max} \equiv \alpha_5^{\rm max}$.
Reinsertion of Eq.~(\ref{al5gen}) facilitates the choice of reference to
one of $\alpha_5$ or $\alpha_5^{\rm max}$. 
\bea
\frac{1}{\alpha_1}
&=& \frac{1}{\alpha^{\rm max}_5} + \frac{b_{\rm Y} + 3/5(b_2-b_3)}{2\pi} \,\ln\frac{M^{\rm max}_{32}}{M_{32}}
\label{al1gen} \\
&=& \frac{1}{\alpha_5} + \frac{5 b_{\rm Y} + 3 b_2 - 8 b_3}{5\times 2\pi} \,\ln\frac{M^{\rm max}_{32}}{M_{32}}
\nonumber
\eea
The minimal scenario limits\footnote{The
coefficient $-\frac{3}{2\pi}$ in the $\left. {\rm limit} \right|^{\rm MSSM}_{\Xi_i=1}$
given in Eq.~(\ref{al5gen_min}) corrects the erroneous published value of $+\frac{33}{56\pi}$.}
of Eqs.~(\ref{al5max},\ref{al5gen},\ref{al1gen}) are shown below:
\begin{subequations}
\label{eqs:algen}
\bea
\left. \alpha^{\rm max}_5 \right|^{\rm MSSM}_{\Xi_i=1}
&=& \frac{20\, \alpha_{\rm em}\alpha_3 }{ 3 (\alpha_3 + 4 \alpha_{\rm em})}
\simeq 0.0412
\label{al5gen_max_min} \\
\left. \alpha_5 \right|^{\rm MSSM}_{\Xi_i=1} &=&
\left[
\frac{3}{5\alpha_3} + \frac{3}{20\alpha_{\rm em}}
- \frac{3}{2\pi}\,\ln \frac{M^{\rm max}_{32}}{M_{32}}
\right]^{-1}
\label{al5gen_min} \\
\left. \alpha_1 \right|^{\rm MSSM}_{\Xi_i=1} &=&
\left[
\frac{3}{5\alpha_3} + \frac{3}{20\alpha_{\rm em}}
+ \frac{9}{2\pi}\,\ln \frac{M^{\rm max}_{32}}{M_{32}}
\right]^{-1}
\label{al1gen_min}
\eea
\end{subequations}

The triple unification prediction for $\Theta_{\rm W}$ is likewise achieved from
Eq.~(\ref{RGE2}), using Eqs.~(\ref{m32max},\ref{al5max})
to eliminate reference to $M^{\rm max}_{32}$ and $\alpha^{\rm max}_5$.
\begin{subequations}
\label{thetamax}
\bea
\Theta^{\rm max}_{\rm W} \equiv
\Theta_{\rm W} \left(M^{\rm max}_{32}\right) &=&
%\sin^2\left(\theta_{\rm W}^{\rm\, eff} (M^{\rm max}_{32})\right) &=&
\frac{ 5 \alpha_{\rm em} \Xi_3 (b_{\rm Y} - b_2) + 3 \alpha_3 \Xi_{\rm Y} (b_2-b_3)
}{
\alpha_3 (5 b_{\rm Y} + 3 b_2 - 8 b_3)}
\label{thetamax_std}
\\
\left.\lim\right|^{\rm MSSM}_{\Xi_i=1} \Rightarrow
\sin^2(\theta_{\rm W}^{\rm\, eff}) &=&
\frac{1}{5} + \frac{7\, \alpha_{\rm em}}{15\, \alpha_3}
\simeq 0.2310
\label{thetamax_min}
\eea
\end{subequations}
The surprisingly fine numerical agreement of the minimal scenario limit with the experimental value quoted in Eq.~(\ref{lepdg})
suggests that the light and heavy thresholds, the second loop, and perhaps even
a reduction of the unification scale to $M_{\rm 32}$,
must effectively conspire to avoid undoing this baseline signal of unification. 

Considering $M_{32}$ for the moment to be an independent input, the correction to this expression
comes from solving Eq.~(\ref{lnmaxbym32}) for $\Theta_{\rm W}^{\rm DEP}$, with the superscript expressed to emphasize
that this dependent form of the (sine-squared of the) effective Weinberg angle may take a value either in deficit or surplus of that
prescribed by Eq.~(\ref{eq:sseff}) and the known thresholds.
The result reduces correctly in the triple unification
limit, and otherwise reveals the shift obtained at generic $M_{32}$.
\begin{subequations}
\label{eqs:thetagen}
\bea
\Theta_{\rm W}^{\rm DEP} &=& \Theta^{\rm max}_{\rm W}
- \frac{\alpha_{\rm em} (b_2-b_3)}{2\pi} \,\ln{\frac{M^{\rm max}_{32}}{M_{32}}}
\label{thetagen} \\
\left.\lim\right|^{\rm MSSM}_{\Xi_i=1} &\Rightarrow&
\frac{1}{5} + \frac{7\, \alpha_{\rm em}}{15\, \alpha_3}
- \frac{2\, \alpha_{\rm em}}{\pi} \,\ln{\frac{M^{\rm max}_{32}}{M_{32}}}
\eea
\end{subequations}
In such a case we may choose to attribute any deviation from the independently determined effective angle to the presence of an unknown threshold.
\bea
\Delta \xi_2^{\rm DEP} &\equiv& \frac{2\pi}{\alpha_{\rm em}} \left( \Theta^{\rm DEP}_{\rm W} - \Theta_{\rm W}^{\rm IND} \right) \label{eq:xi2dep} \\
&=& \frac{2\pi}{\alpha_{\rm em}} \left( \Theta^{\rm max}_{\rm W} - \Theta_{\rm W}^{\rm IND} \right) -
(b_2-b_3)\,\ln{\frac{M_{32}^{\rm max}}{M_{32}}} \nonumber
\eea
In particular, if one chooses to enforce the strict triple unification of Standard $SU(5)$ via $(\Theta_{\rm W}^{\rm DEP} = \Theta_{\rm W}^{\rm max})$,
then Eq.~(\ref{eq:xi2dep}) can yield a prediction for the shift $\xi_2^{\rm H}$ which must occur at the heavy scale relative to the
thresholds input at the better determined light scale.
\beq
\xi_2^{\rm H} \equiv \frac{2\pi}{\alpha_{\rm em}} \left( \Theta^{\rm max}_{\rm W} - \Theta_{\rm W}^{\rm light} \right) 
\label{heavyshift}
\eeq

The natural convention within the established formalism has been to cast the effect of any heavy thresholds onto renormalization of just the second gauge coupling as shown.
As hinted previously however, we will also have to make a compensatory shift
\bea
\xi_{\rm Y}^{\rm H} = - 3/5\, \xi_2^{\rm H} \label{eq:compshift}
\eea
to the hypercharge if $\Xi_{\rm Y}$ of Eq.~(\ref{eq:sseff}) is to be left undisturbed.
Whenever plotting, tabulating, or quoting results from the strict triple unification picture,
we will always implicitly assume,
without immediate concern for their origin,
the action of heavy thresholds
as specified for internal consistency
by Eqs.~(\ref{heavyshift},\ref{eq:compshift}).

Of course, enforcing $(\Delta \Xi_i = 0)$ can only be an approximation to the reality of simultaneous corrections to the running of all three gauge couplings.
Nevertheless, the fixing here of one degree of freedom argues for a parameterization of the cost in terms of a single threshold function.
Moreover, a suitably chosen linear combination of the actual $\xi_i$ can constitute under certain conditions a fair rendering of the true dependencies.
These issues will be taken up in detail in Section~(\ref{sct:heavythresholds}).

It can also be useful to isolate solutions for ($\alpha_5$,~$\alpha_1$,~$\Theta_{\rm W}$) as functions of $M_{32}$
without explicit reference to the `max' scale parameters from the triple unification.
This can be achieved by combination of Eqs.~(\ref{al5gen},\ref{al1gen},\ref{thetagen}) and Eqs.~(\ref{m32max},\ref{al5max},\ref{thetamax_std}),
or more simply, by turning directly to an appropriate remixing of the original Eqs.~(\ref{eqs:RGES}).
\begin{subequations}
\label{eqs:rgeofm32}
\bea
\frac{1}{\alpha_5} &=& \frac{\Xi_3}{\alpha_3} - \frac{b_3}{2\pi} \,\ln{\frac{M_{32}}{M_{\rm Z}}}
\label{eq:rgeofm32_1} \\
\frac{1}{\alpha_1} &=& \frac{3\,\Xi_{\rm Y}}{5\,\alpha_{\rm em}} - \frac{3\,\Xi_3}{5\,\alpha_3} -
\frac{ b_Y+3/5\,(b_2-b_3) }{2\pi} \,\ln{\frac{M_{32}}{M_{\rm Z}}}
\label{eq:rgeofm32_2} \\
\Theta_{\rm W} &=& \frac{\alpha_{\rm em}\Xi_3}{\alpha_3} + \frac{\alpha_{\rm em} (b_2-b_3)}{2\pi} \,\ln{\frac{M_{32}}{M_{\rm Z}}}
\label{eq:rgeofm32_3}
\eea
\end{subequations}
Or, going to the minimal limit:
\begin{subequations}
\label{eqs:rgeofm32_min}
\bea
\frac{1}{\alpha_5}
\left. \hspace{-2pt} \right|^{\rm MSSM}_{\Xi_i=1}
&=& \frac{1}{\alpha_3} + \frac{3}{2\pi} \,\ln{\frac{M_{32}}{M_{\rm Z}}}
\label{eq:rgeofm32_min_1} \\
\frac{1}{\alpha_1}
\left. \hspace{-2pt} \right|^{\rm MSSM}_{\Xi_i=1}
&=& \frac{3\, (\alpha_3 - \alpha_{\rm em})}{5\,\alpha_{\rm em}\alpha_3} - \frac{9}{2\pi} \,\ln{\frac{M_{32}}{M_{\rm Z}}}
\label{eq:rgeofm32_min_2} \\
\sin^2(\theta_{\rm W}^{\rm\, eff})
\left. \hspace{-2pt} \right|^{\rm MSSM}_{\Xi_i=1}
&=& \frac{\alpha_{\rm em}}{\alpha_3} + \frac{2\,\alpha_{\rm em}}{\pi} \,\ln{\frac{M_{32}}{M_{\rm Z}}}
\label{eq:rgeofm32_min_3}
\eea
\end{subequations}

We should also in turn consider the inversion of the relationship in Eq.~(\ref{thetagen}), where all thresholds are considered known,
and the discrepancy is attributed instead to the
down-shift\footnote{It is also possible to share the burden of low energy phenomenological
dependencies between the $\xi_i^{\rm H}$ and $M_{32}$, a freedom
which has previously~\cite{Ellis:1995at,Ellis:2002vk} distinguished the
flipped unification from its standard counterpart.} of $M_{32}$.
Indeed, at the end of the day, the quantity of primary practical interest is often the reduced mass
scale $M_{32}$ which plays a decisive role in establishing the rate of proton decay.
This value is extremely sensitive to the effects of thresholds, and one must
take proper care to employ reasonable estimates of their value.  Assuming such,
we will shift to the point of view with ($\Theta_{\rm W}^{\rm DEP}$ = $\Theta_{\rm W}^{\rm IND}$) as independent
input, and $M_{32}$ dependent output.
\begin{subequations}
\label{eqs:m32oftheta} \\
\bea
\ln\frac{M^{\rm max}_{32}}{M_{32}} &=&
\frac{2\pi \left( \Theta^{\rm max}_{\rm W} - \Theta_{\rm W} \right)
}{
\alpha_{\rm em} (b_2-b_3)}
\label{m32oftheta} \\
\left.\lim\right|^{\rm MSSM}_{\Xi_i=1}
&\Rightarrow&
\frac{\pi\left( \sin^2(\theta_{\rm W}^{\rm max}) - \sin^2(\theta_{\rm W}^{\rm\, eff}) \right)}{2\,\alpha_{\rm em}}
%\\
%&=&
%\frac{\pi}{10\,\alpha_{\rm em}} + \frac{7\pi}{30\,\alpha_3} - \frac{\pi \sin^2(\theta_{\rm W}^{\rm\, eff})}{2\,\alpha_{\rm em}} \nonumber
%\\
%&=&
%\frac{\pi\left( 3\,\alpha_3 - 8\,\alpha_{\rm em} \right)}{30\,\alpha_{\rm em}\alpha_3} - \ln{\frac{M_{32}}{M_{\rm Z}}} \nonumber
\eea
\end{subequations}
We return in the prior and henceforth to suppression of the (DEP/IND) superscripting,
emphasizing that the distinction was anyway interpretational rather than material.
The defining relation from Eq.~(\ref{eq:sseff}) applied to both views,
admitting that certain of the thresholds there employed
might originate, as in Eq.~(\ref{eq:xi2dep}), outside the scope of explicitly specified fields.

Eq.~(\ref{m32oftheta}) represents a solution for $M_{32}$ in terms of its extremal value $M_{32}^{\max}$ and the
corresponding prediction for the effective Weinberg angle at that scale.
It may be reinserted directly in the logarithmic form back into Eqs.~(\ref{al5gen},\ref{al1gen}) to realize
the analogous expressions for ($\alpha_5$,~$\alpha_1$).
Often however, no direct reference to the `max' triple unification parameters
is of interest. This being the case, the relations may be nicely reduced.
We will choose to forgo a direct printing of this equation set in the former style, in favor of the latter.
\begin{subequations}
\label{eqs:oftheta}
\bea
M_{32} &=&
M_{\rm Z} \times \exp \left\{
\frac{ 2\pi (\alpha_3 \Theta_{\rm W} - \alpha_{\rm em} \Xi_3)
}{
\alpha_{\rm em} \alpha_3 (b_2-b_3)}
\right\}
\label{m32ofthetasimple} \\
\frac{1}{\alpha_5} &=&
\frac{ \alpha_{\rm em} \Xi_3 b_2 - \alpha_3 \Theta_{\rm W} b_3
}{
\alpha_{\rm em} \alpha_3 (b_2-b_3) }
\label{alpha5oftheta} \\
\frac{1}{\alpha_1}
&=& \frac{ \frac{3}{5} \alpha_3 \Xi_{\rm Y} (b_2-b_3) - \alpha_3 \Theta_{\rm W} (b_{\rm Y} + \frac{3}{5}(b_2-b_3)) + \alpha_{\rm em} \Xi_3 b_{\rm Y}
}{ \alpha_{\rm em} \alpha_3 (b_2-b_3) }
\label{alpha1oftheta}
\eea
\end{subequations}
And again for the minimal limit:
\begin{subequations}
\label{eqs:oftheta_min}
\bea
M_{32}
\left. \hspace{-2pt} \right|^{\rm MSSM}_{\Xi_i=1}
&=&
M_{\rm Z} \times \exp \left\{
\frac{\pi}{2}
\left(
\frac{\sin^2(\theta_{\rm W}^{\rm\, eff})}{\alpha_{\rm em}}
-\frac{1}{\alpha_3}
\right)
\right\}
\\
\frac{1}{\alpha_5}
\left. \hspace{-2pt} \right|^{\rm MSSM}_{\Xi_i=1}
&=&
\frac{3\, \sin^2(\theta_{\rm W}^{\rm\, eff})}{4\, \alpha_{\rm em}} +
\frac{1}{4\,\alpha_3}
\\
\frac{1}{\alpha_1}
\left. \hspace{-2pt} \right|^{\rm MSSM}_{\Xi_i=1}
&=&
\frac{12 - 45\, \sin^2(\theta_{\rm W}^{\rm\, eff})}{20\,\alpha_{\rm em}} +
\frac{33}{20\,\alpha_3}
\eea
\end{subequations}

Returning to the strict triple unification picture, the extra constraint on the merger of $\alpha_1$ with $\alpha_5$
has meant that rather than choosing between a solution for either $\Theta_{\rm W}$ or $M_{32}$,
we have determined (the maximal limit of) both.
Specifically, we have been compelled to strictly impose some condition akin to Eq.~(\ref{eq:xi2dep})
to overcome any shortfall in the effective Weinberg angle.
There are however other possibilities if we will instead permit some previously fixed parameter to float.
In the spirit of Eq.~(\ref{alsmz}), we may now choose to isolate the strong coupling at the Z-mass as our third dependent function,
allowing the shift to an effective value $\alpha_3^{\rm max} \equiv \alpha_3 + \Delta\alpha_3^{\rm DEP}$ with $\Delta \xi_2^{\rm DEP} = 0$.
\begin{subequations}
\label{eqs:tripleofsstind}
\bea
\alpha_3^{\rm max} &=&
\left[ \frac{\Theta_{\rm W} (5 b_{\rm Y} + 3 b_2 - 8 b_3) - 3 \,\Xi_{\rm Y} (b_2-b_3)}{5\,\alpha_{\rm em}(b_{\rm Y}-b_2)} - \frac{\xi_3}{2\pi} \right]^{-1}
\label{eq:tripleofsstind_1} \\
&\approx&
\left[ \frac{\Theta_{\rm W} (5 b_{\rm Y} + 3 b_2 - 8 b_3) - 3 \,\Xi_{\rm Y} (b_2-b_3)}{5\,\alpha_{\rm em}\,\Xi_3(b_{\rm Y}-b_2)} \right]^{-1}
\nonumber \\
\alpha_5^{\rm max} &=&
\left[ \frac{\Theta_{\rm W}(5 b_{\rm Y}+3b_2) - 3\,\Xi_{\rm Y} b_2}{5\, \alpha_{\rm em} (b_{\rm Y}-b_2)} \right]^{-1}
\label{eq:tripleofsstind_2} \\
M_{32}^{\rm max} &=&
M_{\rm Z} \times \exp \left\{ \frac{2\pi(3\,\Xi_{\rm Y} - 8\,\Theta_{\rm W})}{5\,\alpha_{\rm em} (b_{\rm Y}-b_2)} \right\}
\label{eq:tripleofsstind_3}
\eea
\end{subequations}

Although the error bars have been reduced by an order of magnitude since the time when this
basic approach was advanced~\cite{Ellis:1995at}
as a way to {\it predict} $\alpha_3$, the long `lever arm' and still significant uncertainty of argue for its selection
as dependent parameter over any other outside choice.  Nevertheless, our interest in this formulation will primarily be indirect.
We will revisit it in Section~(\ref{sct:heavythresholds}) as a guide to our selection of the appropriate
linear combination of the $\delta \xi_i$ from which to construct $\xi_2^{\rm H}$ in Standard $SU(5)$.

We emphasize that Eqs.~(\ref{eq:tripleofsstind_2},\ref{eq:tripleofsstind_3}) are in general not numerically
equivalent to Eqs.~(\ref{al5max},\ref{m32max}), being that they place the burden of strict unification
onto two different proxies.  A correspondence is recovered however, by explicitly forcing evaluation
at the `maximal' (sine-squared) effective Weinberg angle $\Theta_{\rm W}^{\rm max}$ from Eq.~(\ref{thetamax_std}),
whereupon $\alpha_3^{\rm max}$ likewise reduces to its experimentally assigned value.
The minimal limits of the present equation set are:
\begin{subequations}
\label{eqs:oftheta_max_min}
\bea
\alpha_3^{\rm max}
\left. \hspace{-2pt} \right|^{\rm MSSM}_{\Xi_i=1}
&=&
\frac{7\, \alpha_{\rm em}}{15\, \sin^2(\theta_{\rm W}^{\rm\, eff}) - 3}
\\
\alpha_5^{\rm max}
\left. \hspace{-2pt} \right|^{\rm MSSM}_{\Xi_i=1}
&=&
\frac{28\, \alpha_{\rm em}}{36\, \sin^2(\theta_{\rm W}^{\rm\, eff}) - 3}
\\
M_{32}
\left. \hspace{-2pt} \right|^{\rm MSSM}_{\Xi_i=1}
&=&
M_{\rm Z} \times \exp \left\{
\frac{\pi \left(3 - 8 \sin^2(\theta_{\rm W}^{\rm\, eff})\right)}{14\, \alpha_{\rm em}}
\right\}
\eea
\end{subequations}

The underlying message of this section has been simple; we must clearly keep account of the available
constraints and matching number of dependent functions, choosing for their solution a properly orthogonalized
set of equations. We have presented a compendium of such solution sets,
applicable to various situations and needs,
for generic $\beta$-coefficients and thresholds, and also for the MSSM field limit with all thresholds
absorbed into $\sin^2(\theta_{\rm W}^{\rm\, eff})$. 

The triple, or Standard $SU(5)$, unification expressions for ($M_{32}^{\rm max}$,~$\alpha_5^{\rm max}$,~$\Theta_{\rm W}^{\rm max}$)
have been given in  Eqs.~(\ref{m32max},\ref{al5max},\ref{thetamax_std}).
The Flipped $SU(5)$ unification is treated either by taking the prior set in conjunction with
Eqs.~(\ref{al5gen},\ref{al1gen},\ref{thetagen}) to solve
for ($\alpha_5$,~$\alpha_1$,~$\Theta_{\rm W}^{\rm DEP}$), inserting $M_{32}$ as independent input,
or equivalently by using Eqs.~(\ref{eqs:rgeofm32})
which have been solved for the same variables without explicit reference to
parameters at the $M_{32}^{\rm max}$ scale, 
or finally, as has been our preference in practice, by using
Eqs.~(\ref{eqs:oftheta}) to solve
for ($M_{32}$,~$\alpha_5$,~$\alpha_1$), taking $\Theta_{\rm W}$ as independent input.
Eqs.~(\ref{eqs:tripleofsstind})
for ($\alpha_3^{\rm max}$,~$\alpha_5^{\rm max}$,~$M_{32}^{\rm max}$)
should not be mixed with the Flipped $SU(5)$ formulae, and indeed even for Standard $SU(5)$ we recommend
application of those expressions only in the abstract.

%%%%%%%%%%%%%%%%%%%%%%%%%%%%%%%%%%%%%%%%%%%%%%%%%%%%%%%%%%%%

\subsection{Grand Unification In The Second Loop\label{sct:2ndloop}}

The second order loop corrections are expected to be of comparable significance to the threshold effects.
As before, the sharp dependency of $\tau_p$ on the unification parameters can magnify even small shifts into the difference between potential detection and stalemate.
It is essential to have confidence in the numbers used, and we undertake here again a fresh analysis with full details provided.

We will begin with the renormalization group equations of the MSSM, written to the second loop in the gauge couplings $\alpha_i \equiv g_i^2/4\pi$, and to the first loop for the Yukawa couplings $\lambda_f \equiv y_f^{\dagger} y_f/4\pi$.
The Yukawa index $f$ takes three values (u,d,e) for the up- and down-type quark flavors and for the charged leptons.
Each $y_f$ itself is a $3\times 3$ matrix which spans the three generations.
Derivatives are taken with respect to $t = \textrm{ln}(\mu)$, the logarithm of the renormalization scale.
\beq
\frac{d \alpha_i}{dt} = \frac{b_i \alpha_i^2}{2\pi}
 +\frac{\alpha_i^2}{8\pi^2}
\left[~ \sum_{j=1}^3 B_{ij}  \alpha_j - \sum_{f=u,d,e}d_i^f
{\rm Tr}\left( \lambda_f \right) \right]
\label{mssmrge}
\eeq
\begin{subequations}
\label{eqs:yukawarge}
\bea
\frac{d \lambda_u}{dt} &=& \frac{\lambda_u}{2\pi}
\left[ 3 \lambda_u + \lambda_d 
+3{\rm Tr}( \lambda_u )
-\sum_{i=1}^3c_i^u \alpha_i \right] \\
\frac{d \lambda_d}{dt} &=& \frac{\lambda_d}{2\pi}
\left[ \lambda_u + 3 \lambda_d 
+{\rm Tr}(3 \lambda_d 
+ \lambda_e)
-\sum_{i=1}^3c_i^d \alpha_i \right] \\
\frac{d \lambda_e}{dt} &=& \frac{\lambda_e}{2\pi}
\left[ 3 \lambda_e +{\rm Tr}(3 \lambda_d 
+ \lambda_e)
-\sum_{i=1}^3c_i^e \alpha_i \right]
\eea
\end{subequations}
The relevant $\beta$-coefficients are:
\begin{subequations}
\label{eqs:mssmb}
\beq
b=\left(\frac{33}{5},1,-3\right)
\quad;\quad
B=\begin{pmatrix}
\frac{199}{25}& \frac{27}{5}&\frac{88}{5}\cr \frac{9}{5} & 25&24 \cr \frac{11}{5}&9&14
\end{pmatrix}
\label{mssmbB}
\eeq
\beq
d^u=\left(\frac{26}{5},6,4\right)
\quad;\quad d^d=\left(\frac{14}{5},6,4\right) \quad;\quad
d^e=\left(\frac{18}{5},2,0\right)
\label{mssmds}
\eeq
\beq
c^u=\left( \frac{13}{15}, 3, \frac{16}{3}\right)
\quad;\quad c^d=\left( \frac{7}{15}, 3, \frac{16}{3}\right) \quad;\quad
c^e=\left( \frac{9}{5}, 3, 0\right)
\label{mssmcs}
\eeq
\end{subequations}

The high entanglement and large dimensionality of these equations prohibits a direct attack.
However, a few basic physical approximations will ease the task of numerical computation and even allow for a usable closed form solution.
The tremendous falloff of mass between families suggests that the Yukawa couplings for generations one and two may be reasonably ignored here.
This reduces each $\lambda_f$ from a matrix to a simple number, and likewise eliminates the need for tracing over generations.
In fact, it will not be so unrealistic to disregard all Yukawa couplings except for the top quark itself, although we will specify explicitly when this second round of cuts is applied.

We can achieve some formal simplification by postulating an ansatz $\alpha_i^{-1} = -(b_i t + \zeta_i)/2\pi$ to extend
the one-loop indefinite solution through addition of an undetermined function $\zeta_i(t)$. 
Inserting this trial form into Eq.~(\ref{mssmrge}), cancellations leave a new differential equation for $\zeta_i$ which is first order in $\alpha_i$ and $\lambda_f$.
We emphasize that there is no approximation made in this step itself.
\beq
\frac{d \zeta_i}{dt} =
\frac{1}{4\pi}
\left[~ \sum_{j=1}^3 B_{ij}  \alpha_j - \sum_{f=t,b,\tau}d_i^f \lambda_f \right]
\label{zetarge}
\eeq 
If we recursively apply the first order definite solution without thresholds, $\alpha_i^{-1}(\mu) = \alpha_i^{-1}(M_{\rm Z}) - b_i/(2\pi)\times \textrm{ln}(\mu/M_{\rm Z})$, the shift in $\zeta_i$ from the gauge term of Eq.~(\ref{zetarge}) can be integrated directly without difficulty.
\beq
\left. \Delta \zeta_i \right|_{gauge} \approx \sum_{j=1}^3 \frac{B_{ij}}{4\pi} \times \frac{2\pi}{b_j}~
\textrm{ln}\left(1-\frac{\alpha_j(M_{\rm Z})~b_j}{2\pi}~ \textrm{ln}\frac{\mu}{M_{\rm Z}}\right)^{-1} 
\simeq \sum_{j=1}^3 \frac{B_{ij}}{4\pi} \times \frac{2\pi}{b_j}~ \textrm{ln}\left( \frac{\alpha_j(\mu)}{\alpha_j(M_{\rm Z})} \right)
\label{gaugeshift}
\eeq
For each $b_j = 0$, the limit $B_{ij}/(4\pi)\times\alpha_j(M_{\rm Z})~ \textrm{ln}(\mu/M_{\rm Z})$ replaces the prior.

To proceed analytically with the Yukawa contribution is somewhat more troublesome.
To achieve a suitably compact expression, one must sacrifice all but the top quark contribution as suggested previously, and also apply a simple boundary such as $y_t(M_{\rm Z})=1$.
Even less satisfactory however, is the need to further trim back to just a single gauge coupling $\alpha$ and the associated coefficients $b$ and $c^t$.
In these limits, the entire Yukawa sector is reduced to the equation shown following.
\beq
\frac{d \lambda_t}{dt} = \frac{\lambda_t}{2\pi}
\left[ 6 \lambda_t
- c^t \alpha \right]
\label{yukawasimple}
\eeq
This restriction can be made more palatable by the definition of composite constants which represent all three gauge groups simultaneously.
A difficulty emerges with this approach, in that the inverse couplings add simply, while the couplings themselves do not.
For a majority of the integration domain however, the Taylor expansion applies reasonably well.
\beq
\frac{\alpha_i(\mu)}{\alpha_i(M_{\rm Z})} \approx 1 + \frac{b_i \alpha_i(M_{\rm Z})}{2\pi}~\textrm{ln} \frac{\mu}{M_{\rm Z}} + \ldots
\label{tayloralpha}
\eeq
In this limit, the sum $\sum_i c_i^t~\alpha_i(\mu)$ from Eqs.~(\ref{eqs:yukawarge}) presents no obstacle to the needed reduction.
After using Eq.~(\ref{tayloralpha}) in reverse to restore the original functional form, the `averaged' constants shown following\footnote{This
choice is satisfactory but not unique, due to existence of a symmetry in the solution.} may be read off.
\beq
\alpha(M_{\rm Z}) \equiv \frac{\sum_i c_i^t~ \alpha_i(M_{\rm Z})}{\sum_i c_i^t}
~;~
b \equiv \frac{\sum_i c_i^t~ \sum_i b_i~ c_i^t~ \alpha_i^2(M_{\rm Z})}{\left(\sum_i c_i^t~\alpha_i(M_{\rm Z})\right)^2}
~;~
c^t \equiv \sum_{i=1}^3 c_i^t
\label{compcoef} 
\eeq 

Taking again the single loop expression for the recursion of $\alpha$ in Eq.~(\ref{yukawasimple}), a simple closed form solution for $\lambda_t$ is possible,
which in turn allows for integration of the Yukawa shift from Eq.~(\ref{zetarge}).
\begin{subequations}
\label{yukawashift}
\bea
\left. \Delta \zeta_i \right|_{Yukawa} &\approx&
- \frac{d_i^t}{4\pi} \times \frac{\pi}{3}~\textrm{ln}\left( 1 - \frac{3-3\left(\alpha(M_{\rm Z})/\alpha(\mu)\right)^{1+c^t/b}}{2\pi(b+c^t)\alpha(M_{\rm Z})} \right)^{-1}
\\
\lim_{b \rightarrow 0} &\Rightarrow&
- \frac{d_i^t}{4\pi} \times \frac{\pi}{3}~\textrm{ln}\left( 1 - \frac{3-3\left(M_{\rm Z}/\mu\right)^{c^t\alpha(M_{\rm Z})/2\pi}}{2\pi c^t\alpha(M_{\rm Z})} \right)^{-1}
\eea
\end{subequations}
All together, the effect of the second loop is summarized (consult the above for $b = 0$) by the following approximation.
\beq
\Delta \zeta_i \approx 
\sum_{j=1}^3 \frac{B_{ij}}{2 b_j}~ \textrm{ln}\left( \frac{\alpha_j(\mu)}{\alpha_j(M_{\rm Z})} \right)
- \frac{d_i^t}{12}~\textrm{ln}\left( 1 - \frac{3-3\left(\alpha(M_{\rm Z})/\alpha(\mu)\right)^{1+c^t/b}}{2\pi(b+c^t)\alpha(M_{\rm Z})} \right)^{-1}
\label{mssmzetas}
\eeq

The form of our trial solution is such that these shifts from the second loop may be applied in a manner directly parallel to the prior handling of the thresholds in Eq.~(\ref{eq:sseff}).
The below will extend and replace that form henceforth.
Having made this update, the subsequent expressions of Section~(\ref{sct:threshunif}) carry through without modification.
\begin{subequations}
\label{eq:sseff2}
\bea
\Xi_{\rm Y} &\equiv& 1 + \frac{\alpha_{\rm em}}{2\pi} \left\{ \frac{5}{3}\:(\xi_{\rm Y}-\zeta_{\rm Y}) + (\xi_2-\zeta_2) \right\} \\[4pt]
\Theta_{\rm W} &\equiv& \sin^2\theta_{\rm W} + \alpha_{\rm em} \:(\xi_2-\zeta_2) / 2\pi \\[8pt]
\Xi_3 &\equiv& 1 + \alpha_3 \:(\xi_3-\zeta_3) / 2\pi
\eea
\end{subequations}

For purposes of computation throughout this report however, we will instead prefer the improved accuracy of a purely numerical solution.
Unless otherwise stated, the level of detail used will be as follows.
The top {\it and} bottom quark Yukawa couplings will be considered from the third generation in the first loop.
The boundary conditions at $M_{\rm Z}$ will be given by
$y_t = \sqrt{2} m_t / (v_u \equiv v \sin \beta) \approx 0.999$, and
$y_b = \sqrt{2} m_b / (v_d \equiv v \cos \beta) \approx 0.242$, with the Higgs vacuum expectation value 
$v \equiv \sqrt{v_u^2 + v_d^2} = (\sqrt{2} G_F)^{-1/2} \approx 246~{\rm [GeV]}$, using
$G_F = 1.16637(1)\times10^{-5} {\rm [{GeV}^{-2}]}$ from the PDG~\cite{Amsler:2008zzb},
$m_t = 173.1(13)~{\rm [GeV]}$ from the TevEWWG~\cite{:2009ec},
$m_b = 4.20(17)~{\rm [GeV]}$ from the PDG, and
$\tan \beta \equiv v_u / v_d = 10$ from benchmark scenario~\cite{Battaglia:2003ab} $B'$.
All three gauge couplings will be used in the Yukawa renormalization {\it with} the second loop applied in the recursion.
The threshold correction terms, defaulting again to the MSSM spectrum of benchpoint $B'$, will also be applied wherever the gauge couplings $\alpha_i$ are
used\footnote{There is a minor inconsistency here, as the cumulative effect of the thresholds is applied in full at the outset rather than in steps across the integration.}.
Finally, recognizing that the second loop itself influences the upper limit $M_{32}$ of its own integrated contribution,
this feedback will be accounted for in the dynamic calculation of the unification scale.
The integrity of this construct is verified by comparing the computed value of $M_{32}$ to that produced by Eq.~(\ref{m32ofthetasimple}) with the use of $(\Theta_{\rm W},\Xi_3)$.

Our baseline second loop corrections for the MSSM are then calculated to take the values shown following for the Standard and Flipped $SU(5)$ GUTs.
For the triple unification of $SU(5)$, we are bound to also introduce the additional unknown heavy
threshold of Eq.~(\ref{heavyshift}) to supplement those light 
threshold factors held over from Eq.~(\ref{mssmthresh}). 
\begin{subequations}
\label{mssm2nd}
\bea
{\rm Standard} &:&
\zeta_{\rm Y} \approx 3.01
\quad;\quad
\zeta_2 \approx 5.32
\quad;\quad
\zeta_3 \approx 2.58
\quad;\quad
\xi_2^{\rm H} \approx 2.35
\\[8pt]
{\rm Flipped} &:&
\zeta_{\rm Y} \approx 2.98 
\quad;\quad
\zeta_2 \approx 5.27
\quad;\quad
\zeta_3 \approx
2.55
\eea
\end{subequations}

The standard unification occurs at $M_{32}^{\rm max} \approx 1.03 \times 10^{16}~{\rm [GeV]}$ with
$\alpha_5^{\rm max} \approx 0.04$, yielding a proton lifetime around $0.95 \times 10^{35}$ years. 
The flipped unification by comparison hits at $M_{32} \approx 0.58 \times 10^{16}~{\rm [GeV]}$, for a ratio $M_{32}/M_{32}^{\rm max} \approx 0.56$,
with $\alpha_5 \approx 0.04$, and a proton lifetime near $0.43 \times 10^{35}$ years.
This value is already substantially more rapid than that reported previously~\cite{Ellis:2002vk},
with the discrepancy attributable to the additional level of care taken here to calculation of the threshold and second loop effects.

The closed form approximation of Eq.~(\ref{mssmzetas}) evaluated at identical mass scales compares quite favorably here
and also in the extra TeV scale matter scenarios to be discussed shortly,
with an apparently systematic tendency to overstate the second loop effects by around ten to fifteen percent.
The majority of this difference is traced to numerical recursion of both the thresholds and the second loop,
without which the agreement improves to better than a couple of percent.
The proton lifetime tends to be underestimated in that approximation with just MSSM content,
but overestimated when additional TeV multiplets are included in the running.
The amount of disagreement varies, but averages under 20\% in magnitude for cases tested. 

In all cases, the negative Yukawa sector contributions to the second loop are subordinate
to the positive gauge sector term by a substantial multiplicative factor,
which ranges between about five and twenty.

%%%%%%%%%%%%%%%%%%%%%%%%%%%%%%%%%%%%%%%%%%%%%%%%%%%%%%%%%%%%

\subsection{Heavy Threshold Considerations\label{sct:heavythresholds}}

We turn now to evaluation of the super heavy GUT-scale threshold corrections from the residual Higgs and $SU(5)$ gauge fields.
The heavy lifting, so to speak, of this effort has been completed back in Section~(\ref{sct:multiplethresholds}),
where it was demonstrated, Eq.~(\ref{eq:lightandheavy}), that such effects can be considered retrospectively of the principal
analysis, without upgrading the existing cumulative $\beta$-functions, so long as the additional logarithms are
compared against the upper rather than the lower mass boundary.  Both prescriptions are custom made to order for the present calculation.
Indeed, the internal consistency of the view expressed in Eqs.~(\ref{eq:xi2dep},\ref{heavyshift}) which would attribute a shortfall in unification to the action
of an as yet undetermined threshold term is implicitly dependent on the $\beta$-decoupling.
Moreover, the ultra heavy masses under consideration are each naturally tethered by a common vacuum expectation value to the location of the
unification point itself, within variations in coupling strength which are typically judged to be of order unity.

From this point of view, we emphasize that the heavy thresholds are of a qualitatively different character than the light thresholds.
For the light fields the second order correction was in regards to at which point in the running the given degree of freedom,
already included in full to the first order, actually became active.
For the heavy fields the second order correction deals with whether the given field exists at all\footnote{This
does not necessarily imply that the heavy thresholds are comparatively smaller than the light.}. 
This distinction between an additive versus a subtractive compensation suggests that
GUT scale effects should accordingly be treated somewhat distinctly throughout our process.
In particular, we will neglect their second loop, and moreover discard their presence altogether,
even in the first loop, during calculation of the
second loop contributions of the light fields.

There is a level of mutual consistency to this logic.
In calculation of the second loop, Eqs.~(\ref{eqs:RGES}) were treated as continuous functions for all mass scales $\mu$.
This is not strictly correct, as the $\delta \xi_i$ are better considered to enter in step-function style after each threshold mass has been crossed.
Having passed the point of entry of all relevant thresholds however, the two representations afterward coalesce.
Absorption of these discontinuous processes into an adjustment in the `starting value' of the $M_{\rm Z}$ gauge couplings
precisely characterizes the approach taken with the effective Weinberg angle approach and its related constructs $\Xi_{({\rm Y},3)}$.
Since the thresholds considered have entered the running early\footnote{Even logarithmically,
even the TeV Vectors are still significantly smaller than $M_{32}$.},
the region of dispute likewise spans a narrow portion of the integration domain.
Allowing this error from the light thresholds to persist in the immediate neighborhood of the lower boundary is 
considered analogous in spirit to absolute exclusion of the heavy thresholds, whose presence in the continuous function would
skew the majority result for the benefit of only that region in closest proximity to the upper boundary.
Both surviving errors will be considered to represent comparatively small adjustments to an already small correction term.

The computationally pleasant consequence of all the above
is that we may introduce the heavy fields {\it \`{a} la carte}, holding the $b_i$ constant,
and calculating only shifts in the existing functions of Eqs.~(\ref{eq:sseff2}),
either each in turn or as some appropriately
crafted linear combination.  We will study both the Standard $SU(5)$
and Flipped $SU(5) \times U(1)_{\rm X}$ models, establishing the basic technology of analysis
and setting up a determination in Section~(\ref{sct:heavynumbers}) of the range
of possible consequences which inclusion of the canonical heavy fields may produce.

Let us first consider the Standard $SU(5)$ triple unification.
From the perspective of Eqs.~(\ref{eqs:tripleofsstind}),
we are usually faced, after inclusion of the known light thresholds,
with the task of `lowering $\alpha_{\rm s}$'~\cite{Ellis:1995at} into consistency with its experimental range.
We will generally in practice however, instead advocate use of
Eqs.~(\ref{m32max},\ref{al5max},\ref{thetamax_std}),
which treat $\Theta_{\rm W}^{\rm max}$
as a dependent variable, taking $\Delta\alpha_3^{\rm DEP} = 0$.
Unification {\it and} consistency with the central measured value for $\alpha_3(M_{\rm Z})$
are secured via invocation of unknown heavy thresholds given by Eq.~(\ref{heavyshift}).
However, as discussed in Section~(\ref{sct:threshunif}), the dilemma becomes determination of an effective 
linear combination $\xi_2^{\rm H}$ from the three $\xi_i$ which any physical threshold term may yield.
The merit of the solution for Eqs.~(\ref{eqs:tripleofsstind})
now becomes their use as a guide to this selection.
It is clear from inspection that no single consolidation of the three gauge thresholds can
mimic action of the whole simultaneously for each of ($\alpha_3^{\rm max}$,~$\alpha_5^{\rm max}$,~$M_{32}^{\rm max}$).
Specifically, the following applies in the limit of small changes.
\begin{subequations}
\label{eqs:varymax}
\bea
\frac{\delta \alpha_3^{\rm max}}{\alpha_3^{\rm max}}
&=& \alpha_3^{\rm max} \times \left\{ \frac{ (b_2-b_3) \,\delta \xi_{\rm Y} + (b_3-b_{\rm Y}) \,\delta \xi_2 + (b_{\rm Y}-b_2) \,\delta \xi_3}{2\pi (b_{\rm Y} - b_2)} \right\} \\
\frac{\delta \alpha_5^{\rm max}}{\alpha_5^{\rm max}}
&=& \alpha_5^{\rm max} \times \left\{ \frac{ b_2 \,\delta \xi_{\rm Y} - b_{\rm Y} \,\delta \xi_2}{2\pi (b_{\rm Y} - b_2)} \right\} \\
\frac{\delta M_{32}^{\rm max}}{M_{32}^{\rm max}}
&=& \frac{ \delta \xi_{\rm Y} - \,\delta \xi_2}{b_{\rm Y} - b_2}
\eea
\end{subequations}

Although the parameters ($\alpha_5^{\rm max},~M_{32}^{\rm max}$) at the unification scale are more directly relevant to our interest in proton lifetime, they are meaningless
if the unification itself is a failure.
We will thus direct focus on the variation of $\alpha_3^{\rm max}$ in Eqs.~(\ref{eqs:varymax}), ensuring that it is symmetric under
replacement of the actual $\xi_i$ with our choice for $\xi_2^{\rm H}$
and the book-matched hypercharge term $\xi_{\rm Y}^{\rm H}$ of Eq.~(\ref{eq:compshift}).
The point is that we can indeed then consistently lock down the strong coupling as we solve for $\Theta_{\rm W}^{\rm max}$,
used next to specify the `magnitude' of the missing heavy threshold as a single effective parameter.
The invariance of $\alpha_3^{\rm max}$ under this simplification ensures that
the coupling will not subsequently be displaced from its experimentally established strength.
Using the heavy field formulation of the $\xi_i$ from Eq.~(\ref{eq:lightandheavy}),
we will also wish to translate our result into the language of an effective $\beta$-coefficient $b_2^{\rm H}$.
In keeping with this discussion, the expression for $\xi_2^{\rm H}$, holding the $\Xi_i$ constant while
preserving the central value of $\alpha_3$  as a marker of successful unification,
and assuming $5 b_{\rm Y} + 3 b_2 - 8 b_3 \neq 0$,
is given following\footnote{We
emphasize again that the bulk $b_i$ reference the light fields {\it only}, while the $\delta b_i$
are the incremental additions to the baseline value of each $\beta$-coefficient attributable to the heavy fields.}.
We emphasize again that this formula is appropriate
only\footnote{The MSSM limit given in Eq.~(\ref{eq:xieff_min})
is consistent with published~\cite{Ellis:1991ri} results.
It has sometimes historically~\cite{Ellis:1995at,Ellis:2002vk} been
cross-applied to the flipped unification, a practice which we now discourage.}
for use in Standard $SU(5)$.
\begin{subequations}
\label{eqs:xieff}
\bea
\left.\xi_2^{\rm H}\right|^{\rm SU(5)}_{\Delta \Xi_i =\, 0}
&\equiv&
\frac{(b_2-b_3)\delta \xi_{\rm Y}+(b_3-b_{\rm Y})\delta \xi_2+(b_{\rm Y}-b_2)\delta \xi_3}{(b_3-b_{\rm Y})-\frac{3}{5}(b_2-b_3)}
\equiv
\sum_{\rm heavy} \delta b_2^{\rm H} \times \ln{\frac{M^{\rm max}_{32}}{M}}
\label{eq:xieff_std}
\\[2pt]
&=&
\sum_{\rm heavy} \left\{ \frac{ 5(b_2-b_3)\delta b_{\rm Y} + 5(b_3-b_{\rm Y})\delta b_2 +
5(b_{\rm Y}-b_2)\delta b_3}
{5 b_{\rm Y} + 3 b_2 - 8 b_3 }\right\}
\times \ln{\frac{M^{\rm max}_{32} }{M}}
\nonumber
\\[2pt]
\lim_{\rm MSSM} &\Rightarrow&
\sum_{\rm heavy} \left\{ \frac{1}{3} \,\delta b_{\rm Y} - \frac{4}{5} \,\delta b_2 + \frac{7}{15} \,\delta b_3 \right\}
\times \ln{\frac{M^{\rm max}_{32}}{M}}
\label{eq:xieff_min}
\eea
\end{subequations}

The practical effect of activating Standard $SU(5)$ heavy thresholds is, from this perspective, a
transverse rescaling of the strong coupling strength at $M_{\rm Z}$.  As for the first of Eqs.~(\ref{eqs:varymax}),
from which the below derives, we again stipulate the limit of small changes.
\begin{subequations}
\label{eqs:alpha3shift}
\bea
\alpha_3^{\rm max} &\Rightarrow&
\alpha_{3}^{\rm max} \times \left\{ 1 -
\xi_2^{\rm H}
\left( \frac{\alpha_3^{\rm max}}{2\pi} \right)
\left( \frac{5 b_{\rm Y} + 3 b_2 - 8 b_3}{5(b_{\rm Y}-b_2)} \right)
\right\} \label{alpha3shift_std} \\
\lim_{\rm MSSM} &\Rightarrow&
\alpha_{3}^{\rm max} \times \left\{ 1 -
\xi_2^{\rm H}
\left( \frac{15\, \alpha_3^{\rm max}}{14\pi} \right)
\right\} \label{alpha3shift_min} \\
\nonumber
\eea
\end{subequations}

Of course, the cost of favoring the strong coupling here is that predictions for updates to the
unified mass and coupling in terms of the effective threshold will be distorted. 
We strongly advocate deference to the more precise methods presented in Section~(\ref{sct:multiplethresholds})
whenever actual $\beta$-coefficients and mass scales are known, certainly for the light thresholds, and even for the heavy thresholds.
In fact, we most often suggest restricting direct application of even the concept of $\xi_2^{\rm H}$ to situations 
where it is important only to know that heavy thresholds do exist,
the nature of the fields involved being hypothetical, nonspecific or irrelevant.

In this view $\xi_2^{\rm H}$ represents a diagnostic tool for consolidating, abstracting and quantifying the failure of unification
rather than a literal implementation of actual thresholds by which a successful unification might be secured.
For this purpose and to this end we justify the underlying integrity of the construct, highlighting however
the generic limitation that only one of three dependent solutions may be accurately represented.
Nevertheless, we shall shortly demonstrate the effectiveness of Eq.~(\ref{alpha3shift_std}) to quickly and intuitively
determine what the bulk plausible effects of heavy thresholds could be, and as specific example,
whether they might be viable to salvage the triple unification of Standard $SU(5)$. 

We turn attention next to the partial unification at $M_{32}$ of Flipped $SU(5)$.
Since we are no longer strictly constrained by the triple unification, `lowering $\alpha_{\rm s}$' is no longer a principal concern.
To say it another way, there is no obstacle to the meeting of just two non-parallel lines at some distant point, and consequently
there is no fundamental reason to require the presence of an unknown heavy threshold $\xi^{\rm H}$.
Nevertheless, we do generally expect that there {\it will be} some heavy thresholds, and we may now by choice rather
than necessity stipulate the presence of such a term.  The result will be simply a relocation of the partial unification
$M_{32}$ point, and the values of the two couplings at that scale.

As before, we will need to confront the reality that only
one of these three parameters may be generically and accurately modeled by a single effective shift term.
With the integrity of our low energy phenomenology itself not in jeopardy, we may focus attention instead on fidelity
of the GUT scale rendering.  With $M_{32}$ being the most sensitive of our three choices to variation in the thresholds,
and considering in turn the great sensitivity of the proton lifetime to variation in $M_{32}$, it will become
our key to the definition of an analogue to Eq.~(\ref{eq:xieff_std}) for the Flipped $SU(5)$ context.
Taking our cue from the set of Eqs.~(\ref{eqs:oftheta}),
variations of the dependent parameters are as follows.
\begin{subequations}
\label{eq:flippedvariations}
\bea
\frac{\delta M_{32}}{M_{32}} &=& \frac{\delta \xi_2- \delta \xi_3}{b_2-b_3} \\
\frac{\delta \alpha_5}{\alpha_5} &=& \alpha_5 \times \left\{
\frac{b_3 \,\delta \xi_2 - b_2 \,\delta \xi_3}{2\pi (b_2-b_3)} \right\} \\
\frac{\delta \alpha_1}{\alpha_1} &=& \alpha_1 \times \left\{
\frac{b_{\rm Y} (\delta \xi_2 - \delta \xi_3 )}{2\pi (b_2-b_3)} -\frac{\delta \xi_{\rm Y}}{2\pi} \right\}
\eea
\end{subequations}
The effective single parameter shift which correctly mimics the dependency of the flipped unification scale may be simply read
off from the first member of the prior.
Since the meeting of the second and third couplings occurs independently of the hypercharge, there is no entanglement with $\xi_{\rm Y}^{\rm H}$.
\bea
\left.\xi_2^{\rm H}\right|^{\mathcal{F}\rm{-SU(5)}}_{\Delta \Xi_i =\, 0}
\quad &\equiv& \quad
\left( \delta \xi_2 - \delta \xi_3 \right)
\quad \equiv \quad
\sum_{\rm heavy} \delta b_2^{\rm H} \times \ln{\frac{M_{32}}{M}}
\label{eq:xieffflipped}
\\
\quad &=& \quad
\sum_{\rm heavy} \left\{ \delta b_3 - \delta b_2\right\}
\times \ln{\frac{M_{32}}{M}}
\nonumber
\eea

The practical effect of activating heavy thresholds in Flipped $SU(5)$ is expressed most clearly as a lateral scaling of the unification point.
\begin{subequations}
\label{m32shift}
\bea
M_{32} &\Rightarrow& M_{32} \times \exp{\left\{\frac{\xi_2^{\rm H}}{(b_2-b_3)}\right\}}
\quad = \quad M_{32} \times \prod_{\rm heavy} {\left( \frac{M}{M_{32}} \right)}^{\delta (b_2-b_3) \,/\, (b_2-b_3)} \label{m32shift_std} \\
\lim_{\rm MSSM} &\Rightarrow&
M_{32} \times \exp{\left\{\frac{\xi_2^{\rm H}}{4}\right\}}
\label{m32shift_min}
\eea
\end{subequations}
The exponential amplification of the unification point's dependency
on the heavy thresholds brings a potentially large measure of
uncertainty into our calculation.
Detailed effects near the GUT scale being largely detached from low energy phenomenological constraint, we
have precious little data to guide the selection of a value for this term. 
Indeed, the plausible range of variation remains sufficient to loom as a lingering threat
to swamp out any detailed consideration of better known elements. 

The correction presented in Eq.~(\ref{alsmz}) manifests itself also in Eqs.~(\ref{m32shift}),
with the coefficient $1/4$ from the MSSM limit given in Eq.~(\ref{m32shift_min})
superseding the previously published value of $5/11$.
The new value being smaller by nearly a half, the overall uncertainty
from the heavy thresholds is trimmed back to roughly its own square-root.
Unfortunately, this benefit is immediately countered by substitution of the flipped
effective threshold from Eq.~(\ref{eq:xieffflipped}) for the
inapplicable Standard $SU(5)$ form of Eq.~(\ref{eq:xieff_min}).
The net result is essentially a wash in comparison to prior calculations.

%%%%%%%%%%%%%%%%%%%%%%%%%%%%%%%%%%%%%%%%%%%%%%%%%%%%%%%%%%%%

\section{$\mathcal{F}$-ast Proton Decay\label{main:fast}}

%%%%%%%%%%%%%%%%%%%%%%%%%%%%%%%%%%%%%%%%%%%%%%%%%%%%%%%%%%%%

\subsection{Principal Proton Lifetime Predictions}

We present in this section the key tabular and graphical results of our present effort.

\begin{table}[htbp]
\begin{center}
\begin{tiny}
\setlength{\extrarowheight}{4pt}
$
\begin{array}{|c||c|c|c||c|c|c||c|c|c||c|c||c|c||c|}
\hline
\textrm{Scenario} &
\xi_{\rm Y} & \xi_2 & \xi_3 &
\zeta_{\rm Y} & \zeta_2 & \zeta_3 &
\Xi_{\rm Y} & \Theta_{\rm W} & \Xi_3 &
\Delta\xi_2^{\rm DEP} &
\Delta \alpha_3^{\rm DEP} &
\alpha^{\rm max}_5 & M^{\rm max}_{32} &
{\textrm{\normalsize $\tau$ }}_{\!\!p}^{\!\!{(e\vert\mu)}^{\!+}\! \pi^0}
\\[3pt]
\hline
\hline
{\textrm{SU(5)}}_0 &
3.20 & 5.48 & 8.05 & 3.01 & 5.32 & 2.58 & 1.001 & 0.231 & 1.102 & 2.35 & 0.012 & 0.040 & 1.03 & 0.95
\\[3pt]
\hline
{\textrm{SU(5)}}_{\rm I} &
10.4 & 12.7 & 15.2 & 8.84 & 16.9 & 14.9 & 0.998 & 0.226 & 1.006 & 3.94 & 0.022 & 0.117 & 1.85 & 1.15
\\[3pt]
\hline
{\textrm{SU(5)}}_{\rm II} &
10.4 & 12.7 & 15.2 & 8.35 & 12.8 & 14.0 & 1.004 & 0.231 & 1.023 & 1.26 & 0.006 & 0.115 & 2.08 & 1.92
\\[3pt]
\hline
\end{array}
$
\setlength{\extrarowheight}{0pt}
\end{tiny}
\end{center}
\caption{\label{tb:resultsstd}
Principal results for Standard $SU(5)$, with and without heavy vector multiplets.
Mass is in units of $[10^{16}~{\rm GeV}]$.
Lifetimes are given in $[10^{35}~{\rm Y}]$.}
\end{table}

The principal numerical results for our study of Standard $SU(5)$, for the MSSM, and also
for F-theory models which include extra TeV Scale vector-like multiplets,
are accumulated in Table~(\ref{tb:resultsstd}).  For each scenario under consideration,
we present first the $\xi_i$ and $\zeta_i$ of Sections~(\ref{sct:multiplethresholds},\ref{sct:2ndloop}),
which account respectively for the presence of light (including TeV-scale) thresholds and the second loop.
These are combined according to Eqs.~(\ref{eq:sseff2}) into an effective sine-squared Weinberg angle
and the corresponding factors for the hypercharge and strong coupling.
As discussed around Eq.~(\ref{eq:xi2dep}), strict triple-unification will not generally be
consistent with a given specification of physical thresholds, and one may posit the
existence of an unknown (likely heavy) compensating threshold $\Delta\xi_2^{\rm DEP}$ to close the gap.
An alternative mechanism for quantifying the failure of unification, as discussed around
Eq.~(\ref{eq:tripleofsstind_1}), is to lock the thresholds and push any residual discrepancy instead
onto a deviation of the strong coupling
$\left( \Delta \alpha_3^{\rm DEP} \equiv \alpha_3^{\max} - \alpha_{\rm s}(M_{\rm Z}) \right)$,
now viewed as a dependent parameter.
This perspective\footnote{We have opted to simply carry over the second loop factor
from the original perspective with $\alpha_3$ independent.
} has an advantage of intuitive immediacy by comparison to 
experimental uncertainty (0.002) of $\alpha_{\rm s}$.  We see in the table that even the best
case studied (with Type~II vectors) suffers a mismatch of three standard deviations. 
As discussed in Section~(\ref{sct:heavythresholds}), heavy thresholds are unable to reconcile this disagreement.
Finally, we present the unified coupling and mass, and based on these results, the predicted proton lifetime.
Second order effects are responsible for reducing the Type~0 unification scale down from its baseline
value of $M_{32}^{\rm max} \simeq 2\times10^{16}~{\rm [GeV]}$ in Eq.~(\ref{m32max_min}).
Although the F-theory scenarios are more strongly coupled at unification, this effect is more than
offset by something like a relative doubling of their unification mass points,
resulting in (slightly) longer overall lifetime predictions.
We must emphasize again however, that the central message from
the Standard $SU(5)$ scenarios is a failure of the unification process itself, rendering any other
conclusions in practice academic.

\begin{table}[htbp]
\begin{center}
\begin{tiny}
\setlength{\extrarowheight}{4pt}
$
\begin{array}{|c||c|c|c||c|c|c||c|c|c||c|c|c||c||c|}
\hline
\textrm{Scenario} &
\xi_{\rm Y} & \xi_2 & \xi_3 &
\zeta_{\rm Y} & \zeta_2 & \zeta_3 &
\Xi_{\rm Y} & \Theta_{\rm W} & \Xi_3 &
\alpha_1 & \alpha_5 & M_{32} & \%\, M_{32}^{\rm max} &
{\textrm{\normalsize $\tau$ }}_{\!\!p}^{\!\!{(e\vert\mu)}^{\!+}\! \pi^0}\!
\\[3pt]
\hline
\hline
{\mathcal{F}\textrm{-SU(5)}}_0 &
3.20 & 5.48 & 8.05 & 2.98 & 5.27 & 2.55 & 1.001 & 0.231 & 1.103 & 0.039 & 0.041 & 0.58 & 55.8 & 0.43
\\[3pt]
\hline
{\mathcal{F}\textrm{-SU(5)}}_{\rm I} &
4.64 & 12.7 & 15.2 & 6.21 & 16.4 & 14.3 & 0.992 & 0.227 & 1.018 & 0.045 & 0.116 & 0.68 & 0.01 & 0.10
\\[3pt]
\hline
{\mathcal{F}\textrm{-SU(5)}}_{\rm II} &
7.51 & 12.7 & 15.2 & 6.50 & 16.4 & 14.3 & 0.998 & 0.227 & 1.017 & 0.061 & 0.116 & 0.68 & 1.06 & 0.10
\\[3pt]
\hline
\end{array}
$
\setlength{\extrarowheight}{0pt}
\end{tiny}
\end{center}
\caption{\label{tb:resultsflipped}
Principal results for Flipped $SU(5)$, with and without heavy vector multiplets.
Mass is in units of $[10^{16}~{\rm GeV}]$.
Lifetimes are given in $[10^{35}~{\rm Y}]$.}
\end{table}

The collected results for our study of Flipped $SU(5)$ are likewise presented in
Table~(\ref{tb:resultsflipped}).  An essential point of distinction from the prior
is that this unification of two couplings is intrinsically self consistent\footnote{Whether
this be virtue (avoidance of fine tuning) or vice (triple unification is more
special and thus more suggestive) is open to interpretation.}, and thus has 
no need for the compensation of unknown thresholds.
As before, we see the F-theoretic
constructions dramatically lifting the $SU(5)$ coupling at (partial) unification.
Since the scale $M_{32}$ now remains relatively stable between the three scenarios,
this enhancement of $\alpha_5$ is able to net something like a four-fold
comparative reduction in the overall lifetime.
We provide also in this table the relative percentage of the ratio $M_{32}/M_{32}^{\rm max}$.
As will be discussed in Section~(\ref{sct:superunification}), the putative point of triple unification
$M_{32}^{\rm max}$ for a given Flipped $SU(5) \times U(1)_{\rm X}$ model appears,
as a rule of thumb, to target the eventual point of full unification
between $\alpha_{\rm X}$ and $\alpha_5$.
As generically expected, the mass scale reductions relative to $M_{32}^{\rm max}$ are more than
sufficient to offset the five-fold flipped rate suppression mentioned in Section~(\ref{sct:plife}),
producing a significant downward shift of the proton lifetime for all three flipped models. 
In the bare MSSM Type~0 scenario, we see a very conventional `mild' flipping ratio of about $56\%$.
In the models with TeV scale vectors, the flipping becomes extreme\footnote{It
is important to recognize that the limit $M_{32} \rightarrow M_{32}^{\rm max}$
does not in this case actually take you into a continuum with Standard $SU(5)$.
This is because the vector multiplets have been assigned specifically
for compatibility with flipped charge assignments.}, dropping to just one
percent for Type~II, and significantly lower yet for Type~I.
This corresponds to a significant extension of the (full) unification point, while keeping the
initial $SU(5)$ merger, that which is relevant to proton decay, pleasingly low.
This constitutes a natural and stable distinction between what might
usually be called the GUT scale, and a point of super unification
around the reduced Planck mass.

\begin{figure}[htp]
\begin{center}
\includegraphics[width=\plotwidth,angle=0]{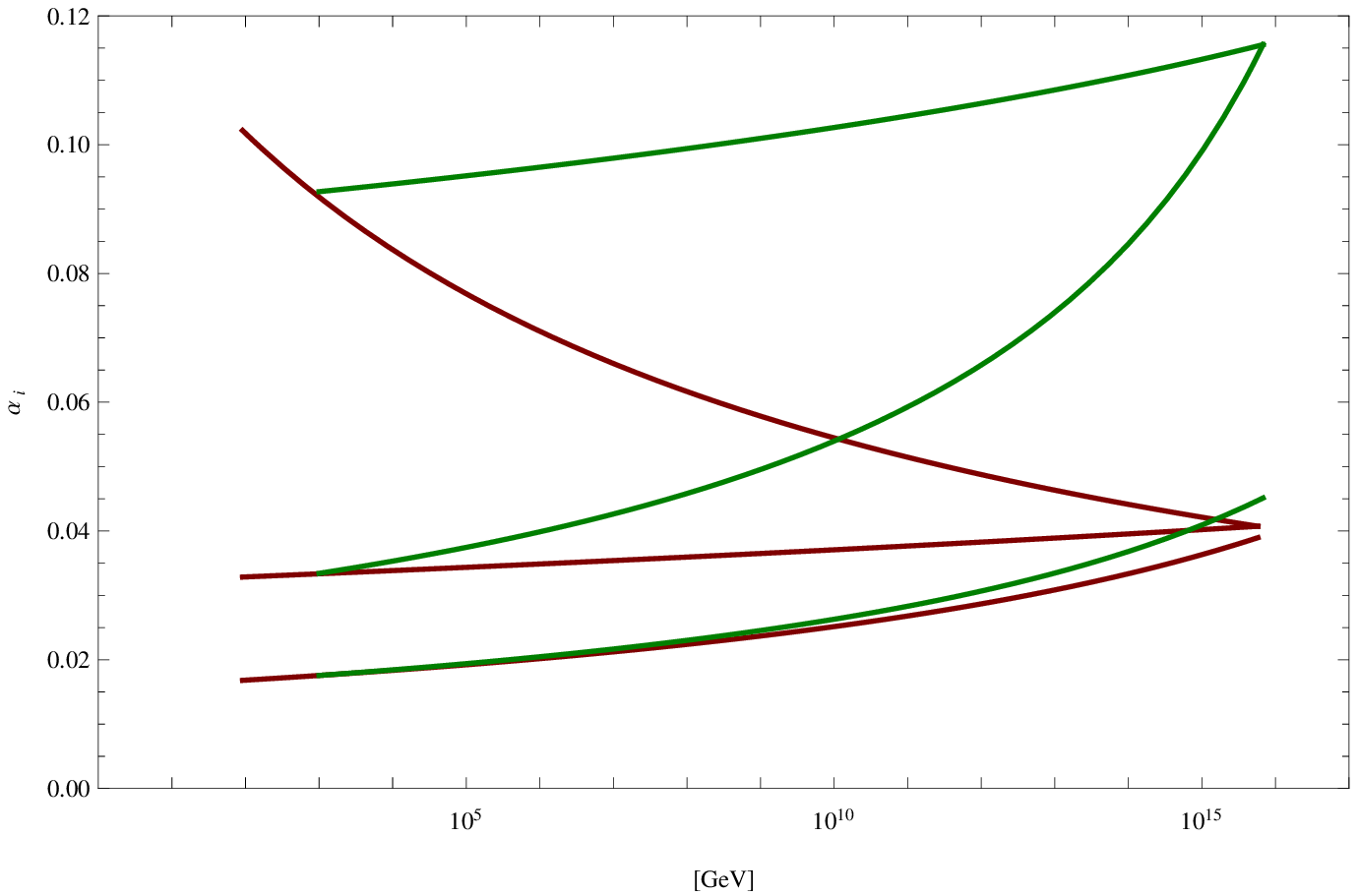} \\
\includegraphics[width=\plotwidth,angle=0]{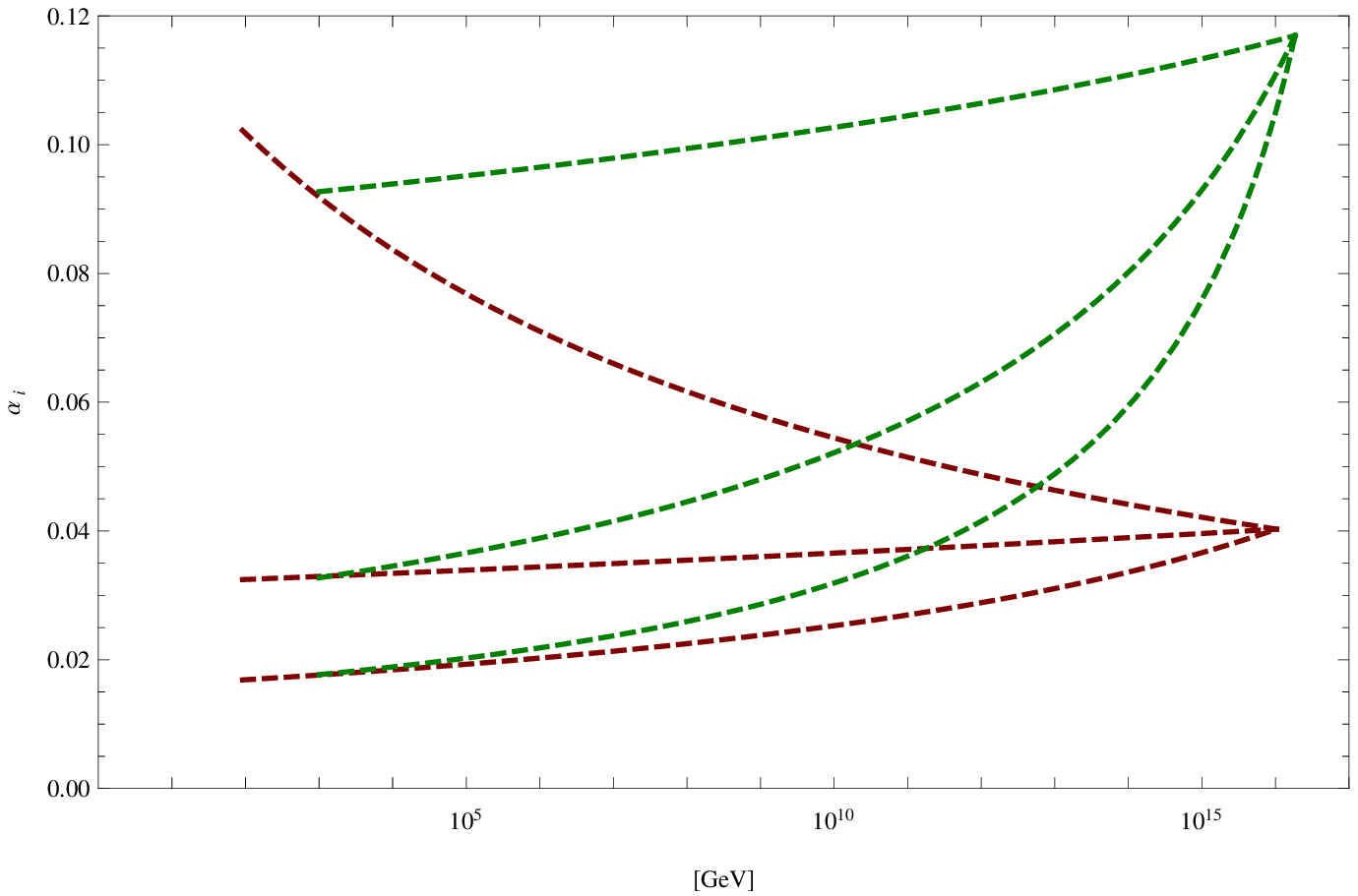}
\end{center}
\caption{\label{fig:running}
Running of the gauge couplings is compared for 
${\mathcal{F}\textrm{-SU(5)}}_0$ (solid red) Vs. ${\mathcal{F}\textrm{-SU(5)}}_{\rm I}$ (solid green) and
${\textrm{SU(5)}}_0$ (dashed red) Vs. ${\textrm{SU(5)}}_{\rm I}$ (dashed green).
The upper line of each set depicts $\alpha_3$, with $\alpha_2$ beneath, and then $\alpha_{\rm Y}$.}
\end{figure}

\begin{figure}[htp]
\begin{center}
\includegraphics[width=\plotwidth,angle=0]{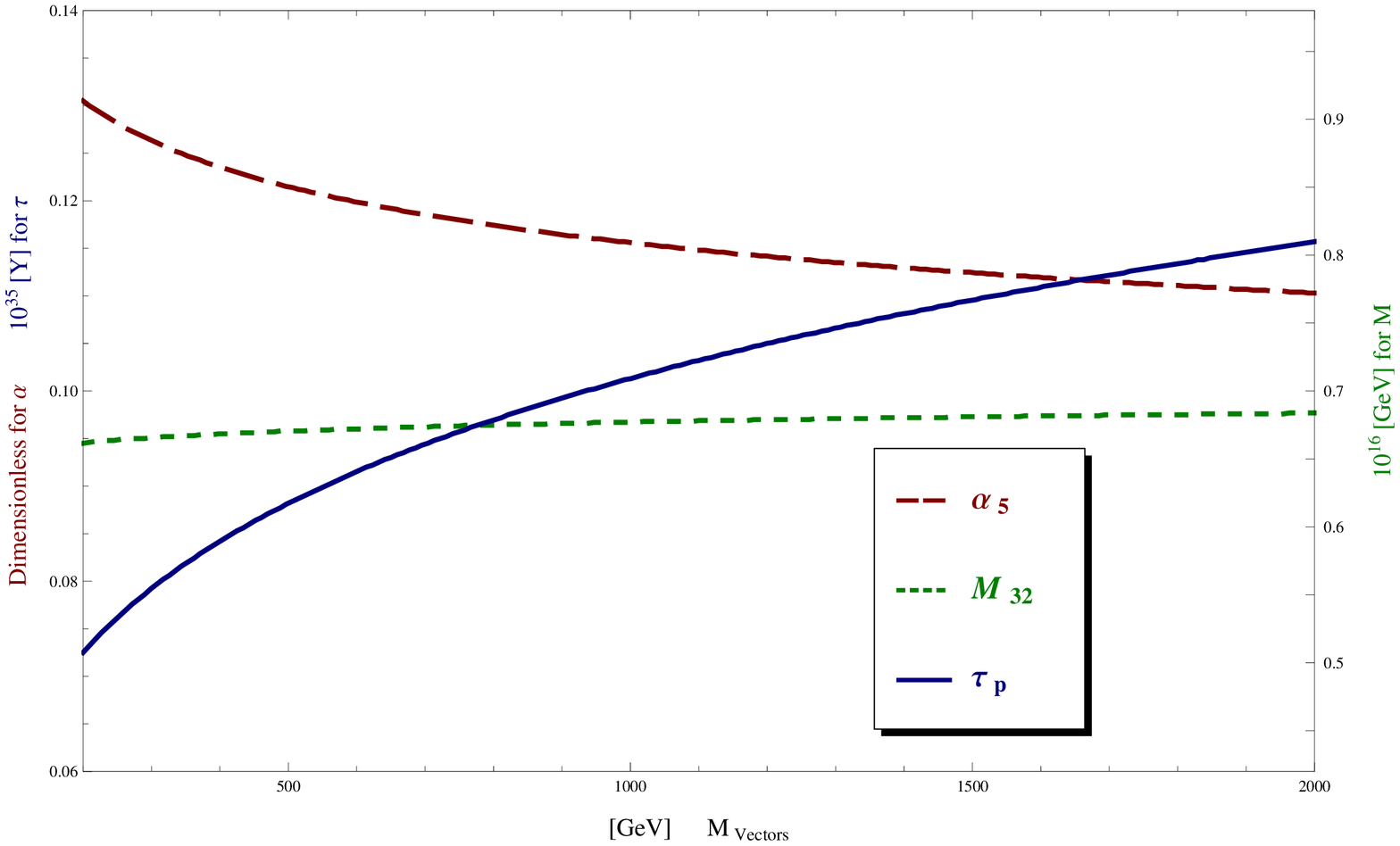} \\
\includegraphics[width=\plotwidth,angle=0]{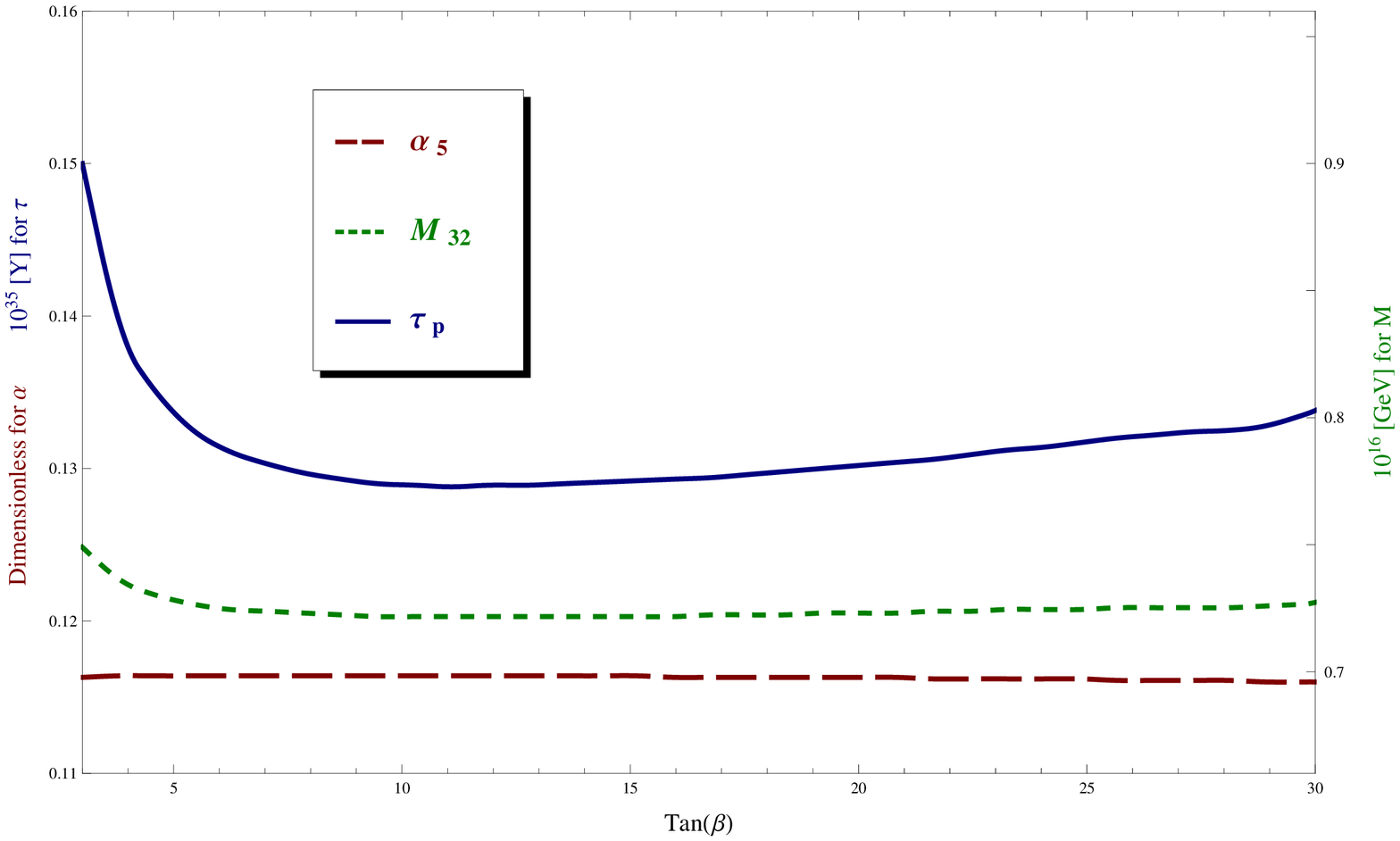}
\end{center}
\caption{\label{fig:variations}
The left-hand numerical scales here perform double duty, read dimensionlessly
for $\alpha_5$, and in units of $[10^{35}\, \rm{Y}]$ for the proton partial lifetime
${\textrm{\normalsize $\tau$ }}_{\!\!p}^{\!\!{(e\vert\mu)}^{\!+}\! \pi^0}$.  $M_{32}$ is
read on the right-hand scale in $[10^{16}~\rm{GeV}]$.
The first example visually explores variation of these parameters
with respect to changes in the mass of the F-theory vector-like multiplets of Flipped
$SU(5)$ Type~II.  The vector mass is run from $200~{\rm [GeV]}$ up to $2~{\rm [TeV]}$, bookending the
value of $1~{\rm [TeV]}$ to which we otherwise default.  The SUSY spectrum is taken
at benchmark B' as usual.  We see a decline in the coupling strength
as the vector mass increases, with the (partial) unification mass holding essentially steady.
The result is a modest gradual rise in proton lifetime from left to right, starting some $25\%$
below, and ending some $15\%$ above our baseline value. 
In the second plot we instead vary the SUSY Higgs parameter $\tan \beta$ from 3 to 30,
bounding again our default value of 10 for benchmark B'.
We regenerate the SUSY spectrum along this span with ISASUGRA, otherwise leaving the
B' input parameters intact.  This change of approach with respect to the spectrum is responsible
for some overall up-tick in the lifetime.
The model again is Flipped $SU(5)$ Type~II, with the vector mass locked at $1~{\rm [TeV]}$.
In this example we instead see $\alpha_5$ flatline, while the unification mass declines
steeply to a minimum near our default region, thereafter rebounding slowly.
The proton lifetime tracks with and amplifies changes in the unification
mass, although the overall variation is still mild, the peak within about $15\%$ of the valley
for the range considered.  The slope approaching the left edge however is steep,
and continuing too far outside the graph in either direction we experience pathologies
in application of the Yukawa boundary condition for calculation of the second loop.}
\end{figure}

\begin{figure}[htp]
\begin{center}
\includegraphics[width=.65\plotwidth,angle=0]{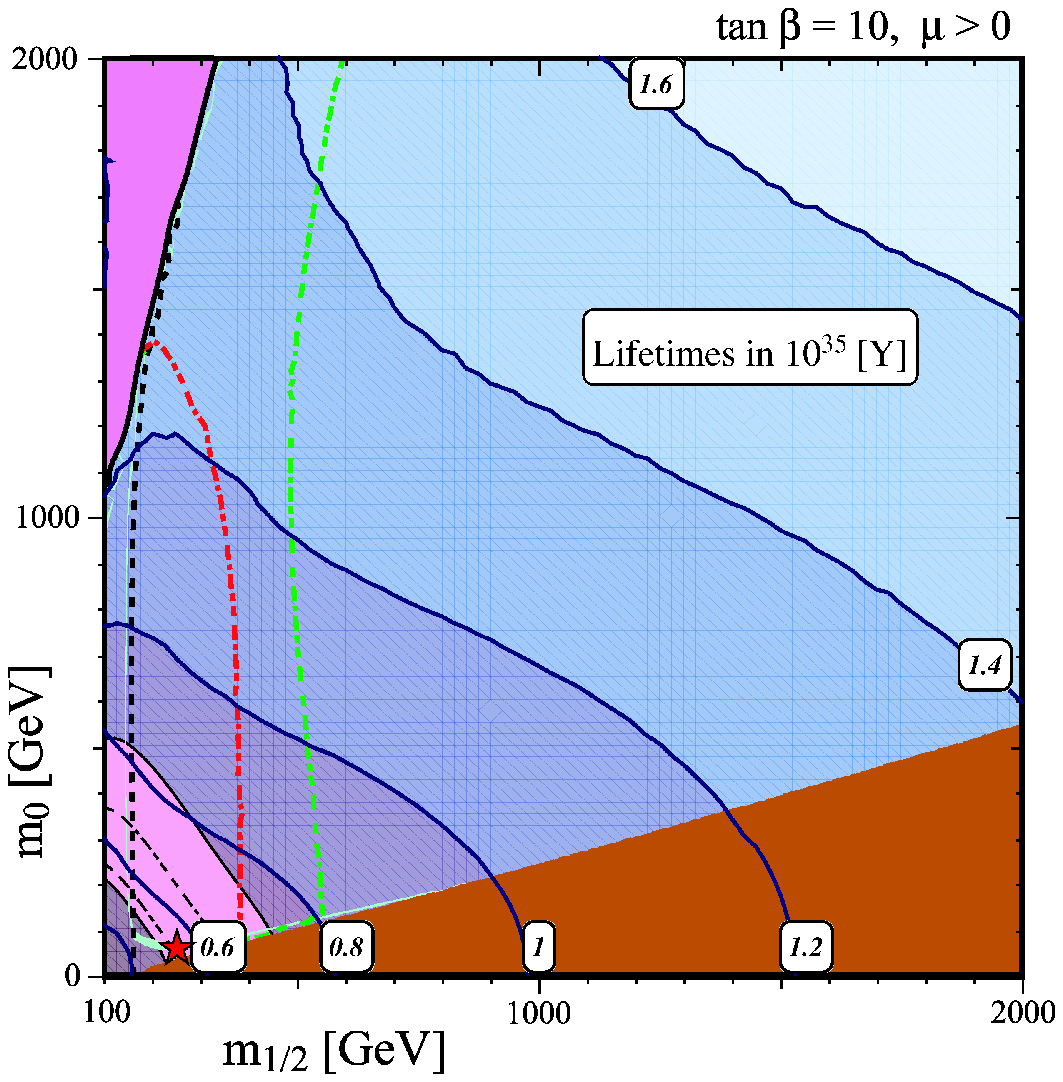}
\hspace{-25pt}
\includegraphics[width=.65\plotwidth,angle=0]{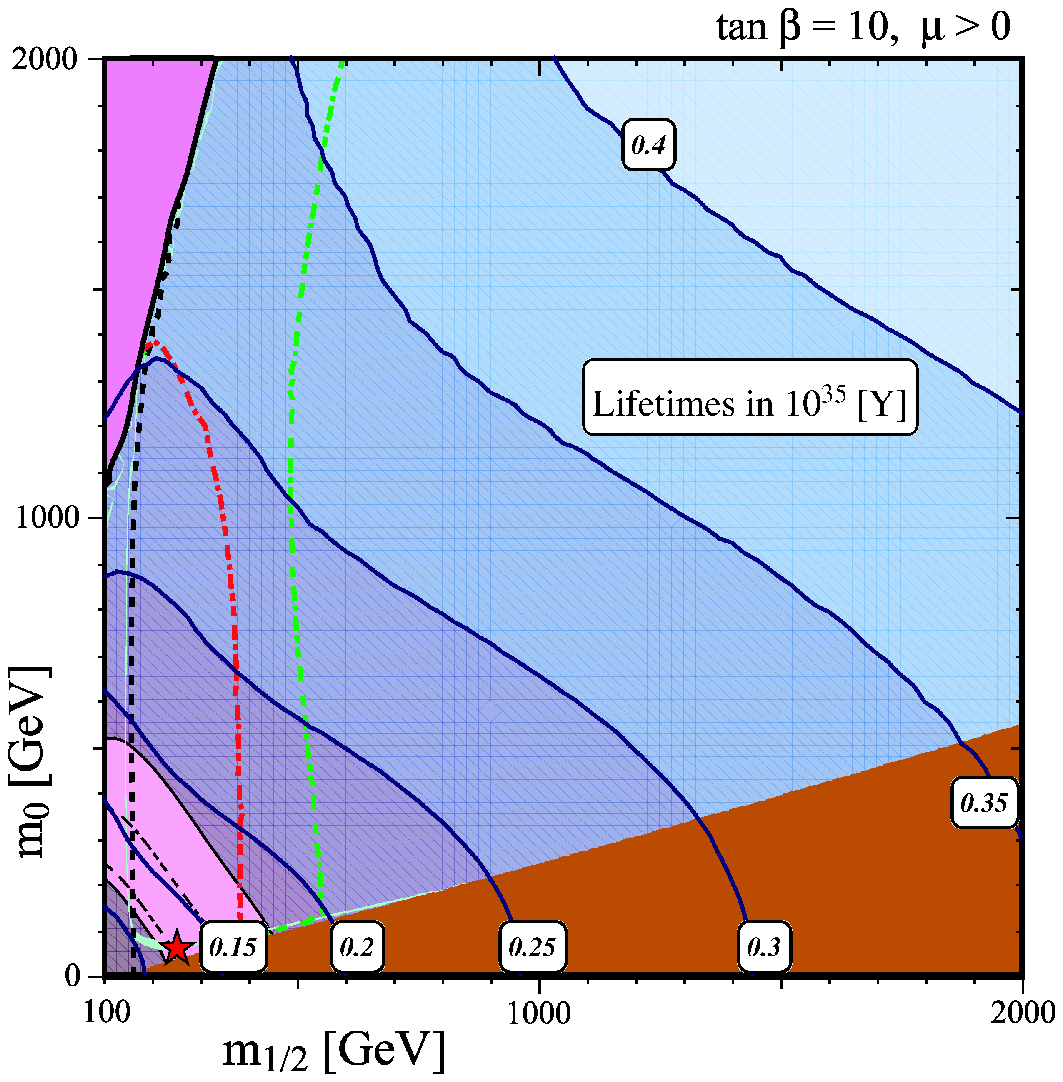}
\end{center}
\caption{\label{fig:contours}
Proton ${\textrm{\normalsize $\tau$ }}_{\!\!p}^{\!\!{(e\vert\mu)}^{\!+}\! \pi^0}$
lifetime contours (dark blue), for Flipped $SU(5)$ first with only MSSM field content,
and subsequently including Type~II vector-like multiplets,
are overlaid onto a background map of phenomenological limits 
on the CMSSM for $\tan \beta = 10$, and $\mu >0$,
that courtesy of Olive et~al.~{\rm\cite{Ellis:2009ai}}.
The orange-brown regions in the lower right of each graphic
suffer a charged LSP, while the dark pink
at upper left fail electroweak symmetry breaking.
The dotted green, red and black lines mark the scalar neutralino-nucleon cross section
at $10^{-9}~\rm{[pB]}$, the Higgs at $114~{\rm [GeV]}$, and a chargino mass of
$104~{\rm [GeV]}$.
The light pink band in the lower left is consistent with measurements of
${(g-2)}_\mu$, and the narrow turquoise strip marks agreement between the
cosmological CDM limits from WMAP and the relic neutralino prediction
after coannihilation.  Our baseline benchmark B' (red star)
sits in the $\mathcal{F}$-ast lower left corner of each diagram.}
\end{figure}

In Figure Set~(\ref{fig:running}), we comparatively depict the renormalization of gauge couplings
for both Standard and Flipped $SU(5)$ models, both with and without the inclusion of F-theory
TeV-scale vector multiplets.
We expect $\alpha_3$ to diminish with increasing energy in the MSSM,
but the presence of the TeV-scale vectors intervenes to reverse this trend
and start the line back upwards in the F-theory related models\footnote{
A very mild effective shift in the $M_{\rm Z}$ scale couplings may be observed
graphically, as was discussed in the note following Eqs.~(\ref{eqs:RGES}).  With inclusion of
the TeV scale vector multiplets however, the effective coupling shift, particularly for $\alpha_3$,
would be quite severe.  To alleviate this visual artifact and produce a more realistic plot,
we splice the F-theory curves onto the MSSM curves once the 1~TeV boundary has been crossed.}.
The essential flatness of the strong coupling is consistent with
our first order expectations for the F-theoretic models with $b_3 = 0$.
Prior analysis~\cite{Ellis:1995at,Ellis:2002vk} of thresholds and the
`$M_{32}$' effect tied variation in proton lifetime strongly to the
partial unification scale, with never much change in the coupling between different scenarios.
Here however, such a severe discontinuity to all three couplings so early in the running can change things.
In fact, it is the leveling out of the strong coupling which forces the flavor coupling to make
a greater accommodation, potentially tripling or quadrupling the value of $\alpha_{32}$,
who's effect on the decay is itself to be squared.  The consequence is a significant reduction in the
${\textrm{\normalsize $\tau$ }}_{\!\!p}^{\!\!{(e\vert\mu)}^{\!+}\! \pi^0}$ mean lifetime.

Figure Set~(\ref{fig:variations}) is a graphical study of variations in the proton lifetime with respect to
changes first in the mass of the F-theory vector multiplets, near the TeV order but departing from the value 1,
and second in $\tan(\beta)$, the ratio of vacuum expectation values for the up- and down-type Higgs.
Figure Set~(\ref{fig:contours}) depicts an overlay of proton lifetime contours onto a background of phenomenological
constraints on the CMSSM parameter space.

%%%%%%%%%%%%%%%%%%%%%%%%%%%%%%%%%%%%%%%%%%%%%%%%%%%%%%%%%%%%

\subsection{Deconstruction of Updates}

We undertake in this section a comparative forensic deconstruction
of prior scenarios and methods of investigation
into unification and proton lifetime, isolating the specific updates in data,
context, and technique which have culminated in the presently more rapid predictions for decay.

\begin{table}[htbp]
\begin{center}
\begin{footnotesize}
\setlength{\extrarowheight}{4pt}
$
\begin{array}{|c|c|c|}
\hline
\textrm{Computational Scenario and Method} &
{\textrm{\large $\tau$ }}_{\!\!p}^{\!\!{(e\vert\mu)}^{\!+}\! \pi^0} &
\Delta \%
\\[3pt]
\hline
\hline
{\mathcal{F}\textrm{-SU(5)}}_0 \textrm{ at pt. B with Legacy Threshold and $2^{nd}$ Loop (2002)}&
1.44 & -
\\[3pt]
\hline
\textrm{with Corrections to the Isolation of Dependent Parameters} &
11.1 &
\uparrow 670 \%
\\[3pt]
\hline
\textrm{with Revisions from the Particle Data Group \& pt. B $\rightarrow$ B'} &
5.95 &
\downarrow 46 \%
\\[3pt]
\hline
\textrm{with Light Thresholds Applied to Each Coupling} &
0.61 &
\downarrow 90 \%
\\[3pt]
\hline
\textrm{with Detailed Evaluation of the Second Loop} &
0.43 &
\downarrow 29 \%
\\[3pt]
\hline
\textrm{with Vector-like Multiplets at 1~[TeV] from }
{\mathcal{F}\textrm{-SU(5)}}_{\rm II} &
0.10 &
\downarrow 77 \%
\\[3pt]
\hline
\hline
\multicolumn{3}{|c|}{\textrm{The net change is $\downarrow 70 \%$ for ${\mathcal{F}\textrm{-SU(5)}}_0$
OR $\downarrow 93 \%$ including ${\mathcal{F}\textrm{-SU(5)}}_{\rm II}$ fields}}
\\[3pt]
\hline
\end{array}
$
\setlength{\extrarowheight}{0pt}
\end{footnotesize}
\end{center}
\caption{\label{tb:pctchanges}
A comparative deconstruction of
various model scenarios and computational methods for the determination of proton lifetime.
Lifetimes are given in $[10^{35}\, {\rm Y}]$.}
\end{table}

Table~(\ref{tb:pctchanges}) takes Flipped $SU(5)$ with MSSM field content as its baseline,
applying the legacy procedures and input data of~\cite{Ellis:2002vk}.
Percentage changes\footnote{Note for example that: $\,11.1 \div 1.44 \approx 7.7$; $\,(7.7-1)\times 100\% = +\, 670\%$.}
are determined sequentially from this reference, with each methodological
adjustment combined with those described preceding.
The published benchmark mass spectra from~\cite{Battaglia:2001zp,Battaglia:2003ab}, computed by SSARD and FeynHiggs at points (B,B')
respectively, are applied at face value.

The corrections described in Section~(\ref{sct:threshunif}), specifically as manifest in
Eqs.~(\ref{m32max_min},\ref{eq:maxto32_min},\ref{al5gen_max_min},\ref{al5gen_min}),
taking also the effective shift in $\alpha_{\rm Y}$ of Eq.~(\ref{defcoups}) from Eq.~(\ref{NLO}),
make up the first transition, fully eliminating errors related to the incorrect
isolation of dependent parameters.  Taken in isolation, this step actually
{\it lengthens} the proton lifetime by a large multiplicative factor of nearly eight times.
It will be all downhill from however from this inauspicious start, starting with simple revisions
to the central values of $M_{\rm Z}$-scale parameters reported by the PDG~\cite{Amsler:2008zzb}, 
accumulated over the last several years.
Together with the transition to the updated benchmark $B'$, this accounts for
a not insignificant downward shift by a factor of almost two.
The methods, described and applied in Sections~(\ref{sct:multiplethresholds},\ref{sct:mssmthresh}) respectively,
for the correct individual application of light thresholds to each of the three running couplings
take the lion's credit for reduction in predicted lifetime, a full order of magnitude.
The detailed evaluation of the second loop\footnote{The relative percentage reductions hold basically steady
if the order of application of the modern treatment of the light thresholds and of the second loop is reversed.}
from Section~(\ref{sct:2ndloop}),
applied again individually to each coupling, yields a modest yet still significant reduction of about a third.
The cumulative effect on proton lifetime for Flipped $SU(5)$ with MSSM field content is 
a cut of seventy percent, due solely to refinements of technique and parameter input.

Including also the F-theory vector-like multiplets of Type II from Section~(\ref{sct:vectors}),
proton lifetime drops substantially yet again, now to roughly a quarter of the preceding value.
This change is attributed primarily to the strengthening of the $\alpha_5$ coupling at the
partial unification point.  At just $1.0 \times 10^{34}$ Years, this prediction sits
coyly on the upper lip of current experimental bounds.
The net downward percentage shift from our original baseline number is a rather astonishing $93\%$. 
Although experimental uncertainty in parameters at $M_{\rm Z}$ can rescale this result
either up or down by a factor of about two, the potentially more significant heavy thresholds
appear capable only of elongating the result, upward into the yet unexplored territories.
This fact is elaborated upon in the next section, and visually summarized in Figure Set~(\ref{fig:heavy}),
wherein it is furthermore clear that departure from our phenomenologically
preferred neighborhood of benchpoint $B'$ should likewise almost certainly slow the decay.

%%%%%%%%%%%%%%%%%%%%%%%%%%%%%%%%%%%%%%%%%%%%%%%%%%%%%%%%%%%%

\subsection{Computation of Heavy Thresholds\label{sct:heavynumbers}}

We will now make concrete the results and discussion of Section~(\ref{sct:heavythresholds})
by specifically computing the range of possible consequences which may arise
from introduction of the canonical GUT scale fields of Standard and Flipped $SU(5)$.
Table~(\ref{tb:heavy_betas}) summarizes the supersymmetric field content for each case~\cite{Ellis:1991ri}.

\begin{table}[htbp]
\begin{center}
\setlength{\extrarowheight}{8pt}
$
\begin{array}{c|c|c|c||c|c|c|}
\cline{2-7}
& \multicolumn{3}{c||}{\textrm{SU(5)}} & \multicolumn{3}{c|}{\mathcal{F}\textrm{-SU(5)}}
\\[4pt] 
\cline{2-7}
& M_{H_3} & M_\Sigma & M_V & M_{H_3} & M_{{\overline{H}}_{3}} & M_V
\\[4pt] 
\hline
\hline
\multicolumn{1}{|c||}{\delta b_{\rm Y}} &
\frac{2}{5} & 0 & -10 & \frac{2}{5} & \frac{2}{5} & -10
\\[4pt]
\hline
\multicolumn{1}{|c||}{\delta b_2} &
0 & 2 & -6 & 0 & 0 & -6
\\[4pt]
\hline
\multicolumn{1}{|c||}{\delta b_3} &
1 & 3 & -4 & 1 & 1 & -4
\\[4pt]
\hline
\hline
\multicolumn{1}{|c||}{\delta b_2^{\rm H}} &
\frac{3}{5} & -\frac{1}{5} & -\frac{2}{5} & 1 & 1 & 2 
\\[4pt]
\hline
\end{array}
$
\setlength{\extrarowheight}{0pt}
\end{center}
\caption{\label{tb:heavy_betas}
$\beta$-function coefficients~\cite{Ellis:1991ri}
for the GUT-scale fields of Standard and Flipped $SU(5)$.
The Effective $\beta$-function coefficients $b_2^{\rm H}$ are listed in the final row.}
\end{table}

For Standard $SU(5)$, the three relevant mass scales are $M_{H_3}$, the heavy down-quark type
triplet from the $({\bf 5},\overline{\bf 5})$ electroweak-breaking Higgs,
$M_\Sigma$, representing the octet and triplet from the GUT scale adjoint Higgs ${\bf 24}$, and $M_V$,
representing the (X,Y) gauge fields and the corresponding residual elements of the adjoint Higgs.

For Flipped $SU(5)\times U(1)_{\rm X}$, $M_{(H_3,{\overline{H}}_3)}$ are the down-quark
type elements of the GUT scale $({\bf 10},\overline{\bf {10}})$ Higgs,
and $M_V$ again represents the (X,Y) gauge fields, now in combination with the quark-type
doublet-triplets remaining from the fundamental Higgs.

Having established the preferred forms of our effective threshold factors in
Eqs.~(\ref{eq:xieff_std},\ref{eq:xieffflipped}), the final row of Table~(\ref{tb:heavy_betas})
presents a tally of the relevant $\delta b_2^{\rm H}$ for each of the heavy field groupings.
It is here the case, both for Standard and Flipped $SU(5)$,
that each of the models listed in Table~(\ref{tb:modelcontent})
produce numerically identical $\delta b_2^{\rm H}$.
Although this seems to occur only by simple coincidence,
we thus make no effort to distinguish between Types~(0,I,II) in the tabulation.

We will adopt the conventional wisdom~\cite{Ellis:1991ri,Ellis:1995at}
with regards to mass generation at the heavy scale,
\ie that mass $M_i$ is proportionally linked to a common vacuum expectation
value ${\langle v \rangle}_{\rm GUT}$ by multiplication of the relevant coupling
strength $\lambda_i$, either gauge or Yukawa.
Moreover the relative coupling strengths, combined with any dimensionless numerical factors,
should be of a single order in magnitude, usually defined to be within about a factor of three.
The largest of the set of heavy masses present should condense at the onset of the breaking
of the gauge symmetry, and thus correspond to $M_{\rm GUT}$, or here $M_{32}$ itself,
and specifically $M_{32}^{\rm max}$ for the case of Standard $SU(5)$.  All together:
\beq
M_{32}^{\rm (max)} \div 3 \quad\leq\quad
M_i \simeq \lambda_i {\langle v \rangle}_{\rm GUT}
\quad\leq\quad M_{32}^{\rm (max)}
\label{vevscale}
\eeq

For Standard $SU(5)$, the $\delta b_2^{\rm H}$ of Table~(\ref{tb:heavy_betas}) in hand,
we may proceed to determine a plausible
range for the factor $\xi_2^{\rm H}$ according to the prescription of Eq.~(\ref{eq:xieff_std}).
However, there is a most important and well known complication.
Namely, Standard $SU(5)$ notoriously suffers from dangerously fast dimension five proton decay
mediated by supersymmetric graphs which mix the Higgs $({\bf 5},\overline{\bf 5})$ 
as required for tuning against the adjoint Higgs $\mathbf{24}$ to enable doublet-triplet splitting.
Suppression of this catastrophic decay places a strong lower bound~\cite{Murayama:2001ur}
on the color-triplet Higgs mass $M_{H_3}$, implying that we must surely take it
to be the heaviest member of our set.

We see however, that the triplet Higgs is the only heavy
field available to Standard $SU(5)$ with a positive
value for the effective $\beta$-coefficient $b_2^{\rm H}$.  The negative coefficients
for $(M_\Sigma,M_V)$, along with their positive logarithms, imply that the
net contribution to $\xi_2^{\rm H}$ can only itself be negative.
This is incompatible with the summary data of Table~(\ref{tb:resultsstd}),
which specify a positive value between about one and four
for the dependent parameter $\Delta \xi_2^{\rm DEP}$ in order to salvage 
unification in the various models considered.

As an alternate perspective, taking explicit heavy thresholds contributions from
$(M_\Sigma,M_V)$, each at a third of $M_{32}^{\rm max}$, with no
additional implicit heavy fields,
the deviation $\Delta \alpha_3^{\rm max}$ in the `predicted' strong coupling increases
to $(0.016,0.027,0.010)$ for the Standard $SU(5)$ models of Type~(0,I,II) respectively.
These results, calculated here by direct reevaluation of Eq.~(\ref{eq:tripleofsstind_1})
with all three $\delta b_i$ of Table~(\ref{tb:heavy_betas}),
agree very nicely with $\delta \alpha_3^{\rm max}$ from Eq.~(\ref{alpha3shift_std}),
applying the effective $\beta$-coefficient combinations $\delta b_2^{\rm H}$
which are common to all three scenarios studied.

We thus hereby reaffirm the previously advertised~\cite{Murayama:2001ur,Ellis:2002vk} conclusion that
the triple unification of Standard $SU(5)$ is incapable
of simultaneously surviving proton decay limits while maintaining agreement with
precision electroweak experiments.
This result remains intact with equal force for the Standard $SU(5)$ F-theory variations
considered directly herein.
We do note carefully however that modern embeddings of supersymmetric $SU(5)$ into M-theory~\cite{Acharya:2001gy,Witten:2001bf}
and F-theory~\cite{Donagi:2008ca,Beasley:2008dc,Beasley:2008kw,Donagi:2008kj} model building are to be distinguished
from the historical Standard $SU(5)$ picture by the presence of alternative scenarios for doublet-triplet splitting,
namely the isolation of Higgs doublet/triplet elements on distinct matter curves~\cite{Beasley:2008kw}, and
suppression of dimension five proton decay.

For Flipped $SU(5) \times U(1)_{\rm X}$, doublet-triplet splitting is achieved naturally
via the missing partner mechanism~\cite{Barr:1981qv,Derendinger:1983aj,Antoniadis:1987dx},
so that no threat from dimension five decay exists.
Interestingly, the available $\delta b_2^{\rm H}$ from Table~(\ref{tb:heavy_betas})
are all positive for the flipped models, suggesting that $\xi_2^{\rm H}$ can
likewise only be positive, and thus by Eq.~(\ref{m32shift_std}) that $M_{32}$ can
only be rescaled to a larger value, slowing proton decay.
We will quickly check the numbers in order to estimate the size of this effect,
taking $(M_{H_3} = M_{{\overline{H}}_{3}})$ for simplicity\footnote{Since
$\delta b_i^H = \delta b_i^{\overline{H}}$, there is no loss of generality
from the alternate condition sometimes quoted~\cite{Ellis:1995at,Ellis:2002vk} that
$M_V \div 3 \leq \sqrt{M_{H_3} \times M_{{\overline{H}}_{3}}} \leq 3 \times M_V $
}.

\begin{figure}[htp]
\begin{center}
\includegraphics[width=.8\plotwidth,angle=0]{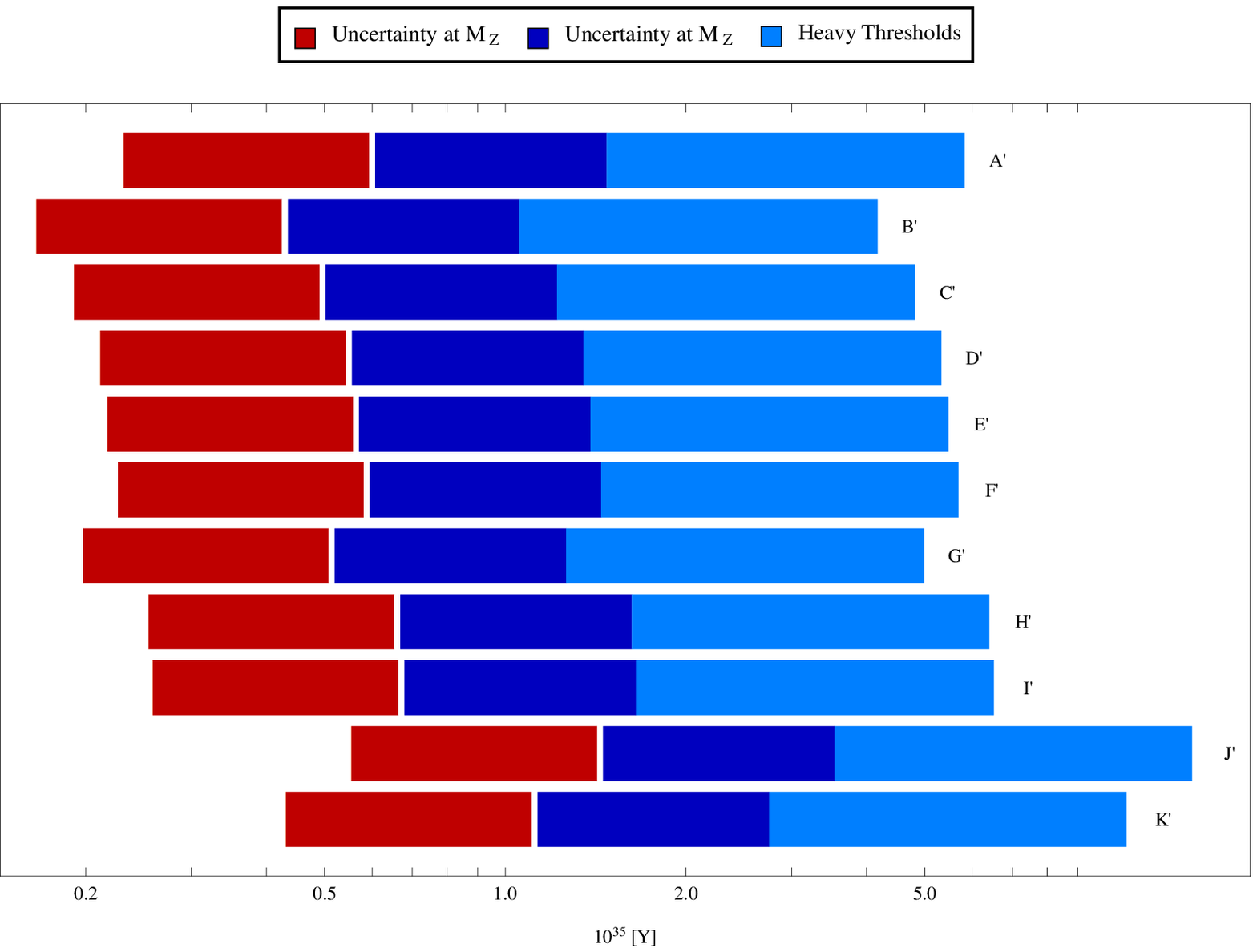} \\
\includegraphics[width=.8\plotwidth,angle=0]{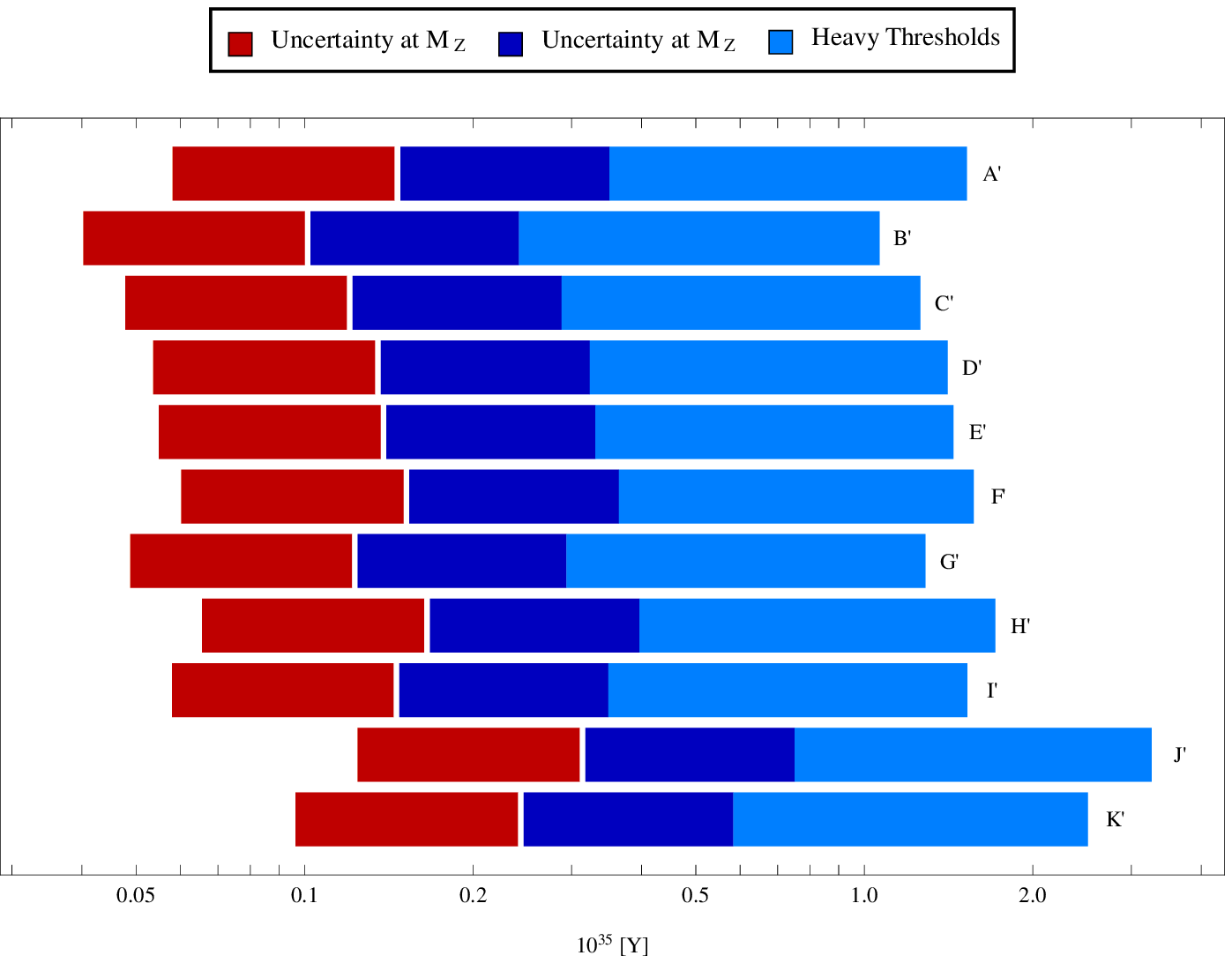}
\end{center}
\caption{\label{fig:heavy}
The central value and key uncertainties of the predicted proton lifetime 
${\textrm{\normalsize $\tau$ }}_{\!\!p}^{\!\!{(e\vert\mu)}^{\!+}\! \pi^0}$
are depicted for each of the benchmark scenarios (A'-K') 
for Flipped $SU(5)$ with MSSM field content only (upper) and
adding Type~II vector-like multiplets (lower), respectively.
The central prediction lies at the thin white line dividing
the dark red and blue bands which themselves represent
the cumulative uncertainty, combined in quadrature, in the electroweak
mass scale $M_{\rm z}$ and the coupling parameters
$(\alpha_{\rm em},\sin^2  \theta^{\overline{\rm MS}}_{\rm W},\alpha_{\rm s})$
measured at that scale, cf. Eq.~(\ref{lepdg}).
The most significant variation is by far attributed to the strong coupling.
The light blue bar represents plausible action of the unknown heavy thresholds
as described in Sections~(\ref{sct:heavythresholds},\ref{sct:heavynumbers}),
measured from the central value, and acting only to slow decay in all cases.
The second loop is computed once for each benchmark at the central evaluation point.
The overall left to right motion of the bar sets gives some idea of the variation
attributable to motion within the supersymmetric parameter space.  It is a
factor somewhat larger than two, comparable to the $M_{\rm Z}$-scale uncertainties.
Other potential sources of error, such as that from the hadronic matrix elements, are expected
to be smaller in magnitude.}
\end{figure}

There are now two options for evaluating the sum from Eq.~(\ref{eq:xieffflipped}).
Either $M_V$ is placed as the heavy member at $M_{32}$, eliminating its logarithm,
or the same happens instead to the pair of triplet Higgs.  In the former case, the
Higgs each contribute $\delta b_2^{\rm H} = 1$ for a total of $2$, and in the latter
case, $M_V$ replicates the result identically, yielding a factor of $2$ by itself.
In either case, to realize the boundary scenario of Eq.~(\ref{vevscale}),
we will apply the full factor of three in the mass ratio of the surviving logarithm.
The result is then $(\xi_2^{\rm H} \leq 2 \ln 3 \approx 2.2)$, corresponding to an
enlargement\footnote{Note that since $(b_2-b_3) = 4$ applies for each of the flipped scenarios
under consideration, the mass rescaling from Eq.~(\ref{m32shift_std}) will likewise be universal.}
of $M_{32}$ by a factor of up to about $(e^{2.2/4} \approx 1.7)$.  Since the proton
lifetime is proportional to the fourth power of this mass, the decay rate can be reduced 
by a factor as large as nine.  This result is fully consistent with direct numerical
reevaluation of Eq.~(\ref{m32ofthetasimple}), taking the $\delta b_i$ individually.
Its significance is demonstrated graphically in Figure Set~(\ref{fig:heavy}), first for baseline
Flipped $SU(5)$ in the MSSM, and next with TeV-scale vector multiplets of Type~II included.

The heavy thresholds represent by far the largest single source of uncertainty
in determination of the flipped proton lifetime
${\textrm{\normalsize $\tau$ }}_{\!\!p}^{\!\!{(e\vert\mu)}^{\!+}\! \pi^0}$.
In Section~(\ref{sct:superunification}),
we will explore the possibility that `downward pressure' from the true grand unification
of $SU(5)$ with $U(1)_{\rm X}$ might suppress excessive upward motion of the scale
$M_{32}$, thereby also constraining runaway uncertainty in the rate of proton decay. 

%%%%%%%%%%%%%%%%%%%%%%%%%%%%%%%%%%%%%%%%%%%%%%%%%%%%%%%%%%%%

\section{$\mathcal{F}$-inale\label{main:finale}}

%%%%%%%%%%%%%%%%%%%%%%%%%%%%%%%%%%%%%%%%%%%%%%%%%%%%%%%%%%%%

\subsection{The $\mathcal{F}$-inal Unification of $SU(5)\times U(1)_{\rm X}$\label{sct:superunification}}

The most immediately distinctive characteristics of the Flipped $SU(5)$ `GUT'
are the factorization into the direct product $SU(5) \times U(1)_{\rm X}$,
and the existence of a partial unification between ($\alpha_3$,$\alpha_2$)
at the scale $M_{32}$, leaving $\alpha_{\rm Y}$ external to the majority field content. 
$M_{32}$ is shifted downward from its maximal value in conventional $SU(5)$,
creating further separation from the fundamental regime of the reduced Planck mass
$M_{\rm Pl} \equiv \sqrt{\hbar c / 8\pi G_N} \simeq 2.4\times10^{18}~{\rm [GeV/c^2]}$.

However, this is not the whole story, as we are free to consider that
($\alpha_1,\alpha_5$) further continue their own running upward toward a meeting at
some true grand unified scale $M_{51}$.
In the crossing of a boundary at which the gauge group enlarges (or breaks down depending on direction traversed),
there may be residual group elements released which are free to perform a remixing analogous to that of 
$SU(2)_{\rm L} \times U(1)_{{\rm Y}'}$ at the electroweak scale, as elaborated in Section~(\ref{sct:mssmthresh}).
Here, in crossing the scale $M_{32}$, the breakdown of $SU(5)$ (rank 4) into $SU(3)\times SU(2)$
ranks (2,1), releases a $U(1)$ (rank 1) symmetry, which remixes with
the coupling $\left( \alpha_1 \equiv \alpha_{\rm Y}(M_{32}) \right)$ of $U(1)_{\rm Y}$.
\beq
\frac{25}{\alpha_1}=\frac{1}{\alpha_5}+\frac{24}{\alpha_{\rm X}}
\label{u1disc}
\eeq

The $U(1)$ symmetry which emerges from this collision, which we will label $U(1)_{\rm X}$,
and the $SU(5)$ coupling are driven upward by renormalization group equations of the familiar form.
This process was depicted previously in Figure~\ref{fig:running}, where
the Eq.~(\ref{u1disc}) discontinuity in the $U(1)$ transition
across $M_{32}$ is clearly visible.
In the spirit of Section~(\ref{sct:2ndloop}), we will extend renormalization to the the second loop
as with Eq.~(\ref{mssmrge}), but without the Yukawa sector.
Derivatives are taken with respect to $t = \textrm{ln}(\mu)$, the logarithm of the renormalization scale.
\beq
\frac{d \alpha_i}{dt} = \frac{b_i \alpha_i^2}{2\pi}
 +\frac{\alpha_i^2}{8\pi^2}
\left[~ \sum_{j=1}^3 B_{ij}  \alpha_j 
\right]
\label{superrge}
\eeq

The one- and two-loop $\beta$-coefficients for the the field content of each of the three
Flipped $SU(5)$ scenarios under consideration are given following. 
\begin{subequations}
\label{eqs:superbB}
\setlength{\extrarowheight}{4pt}
\bea
{\mathcal{F}\textrm{-SU(5)}}_0&:\quad
b = \left(\frac{15}{2},-5\right) ~;&~
B = \begin{pmatrix}
\frac{33}{4} & 60 \cr
\frac{5}{2} & 82 
\end{pmatrix}
\\[4pt]
{\mathcal{F}\textrm{-SU(5)}}_{\rm I}&:\quad
b = \left(8,-2\right) ~;&~ 
B = \begin{pmatrix}
\frac{83}{10} & \frac{336}{5} \cr
\frac{14}{5} & \frac{776}{5}
\end{pmatrix}
\\[4pt]
{\mathcal{F}\textrm{-SU(5)}}_{\rm II}&:\quad
b = \left(\frac{37}{4},-2\right) ~;&~
B = \begin{pmatrix}
\frac{457}{40} & \frac{336}{5} \cr
\frac{14}{5} & \frac{776}{5}
\end{pmatrix}
\eea
\end{subequations}
\setlength{\extrarowheight}{0pt}

Postulating again the ansatz $\alpha_i^{-1} = -(b_i t + \zeta_i)/2\pi$ to extend
the one-loop indefinite solution, the undetermined function $\zeta_i(t)$ will represent
the totality of the second loop contribution as before, either via the approximate
closed form gauge sector solution of Eq.~(\ref{gaugeshift}), or by direct numerical integration.
Although we will not anticipate additional use of thresholds during this final span, we will
allow for that generality as well via the natural analogs to Eqs.~(\ref{eq:sseff2}). 
\begin{subequations}
\label{eqs:superxieff}
\bea
\Xi_{\rm X} &\equiv& 1 + \alpha_{\rm X} \:(\xi_{\rm X}-\zeta_{\rm X}) / 2\pi
\label{superxieff_1} \\[4pt]
\Xi_5 &\equiv& 1 + \alpha_5 \:(\xi_5-\zeta_5) / 2\pi
\label{superxieff_2}
\eea
\end{subequations}

The renormalization group solutions, accurate to the second loop,
then mirror Eqs.~(\ref{eqs:RGES}).
\begin{subequations}
\label{eqs:RGESU}
\bea
\frac{\Xi_{\rm X}}{\alpha_{\rm X}}-\frac{1}{\alpha_{51}}&=&\frac{b_{\rm X}}{2\pi}\, 
\ln{\frac{M_{51}}{M_{32}}}
\label{RGE1} \\[4pt]
\frac{\Xi_5}{\alpha_5}-\frac{1}{\alpha_{51}}&=&\frac{b_5}{2\pi}\,
\ln{\frac{M_{51}}{M_{32}}}
\label{RGE5}
\eea
\end{subequations}
We solve Eqs.~(\ref{eqs:RGESU}) for ($\alpha_{51}$,$M_{51}$),
the grand unified coupling and mass scale.
\begin{subequations}
\label{eqs:masu}
\bea
\alpha_{51} &=& \left[
\frac{\Xi_5}{ (1-b_5/b_{\rm X})\,\alpha_5} +
\frac{\Xi_{\rm X}}{(1-b_{\rm X}/b_5)\,\alpha_{\rm X}}
\right]^{-1}
\label{alpha51} \\[4pt] 
M_{51} &=& M_{32} \times
\exp \left\{
\frac{2\pi}{b_{\rm X} - b_5}\left(
\frac{\Xi_{\rm X}}{\alpha_{\rm X}} - \frac{\Xi_5}{\alpha_5}
\right)
\right\}
\label{m51}
\eea
\end{subequations}

\begin{table}[htbp]
\begin{center}
\begin{footnotesize}
\setlength{\extrarowheight}{3pt}
$
\begin{array}{|c||c|c||c|c||c|c|c|c||c|c|}
\hline
\textrm{Scenario} &
\zeta_{\rm X} & \zeta_5 &
\Xi_{\rm X} & \Xi_5 &
\alpha_1 & \alpha_5 & \alpha_{\rm X} & M_{32} & \alpha_{51} & M_{51}
\\[2pt]
\hline
\hline
{\mathcal{F}\textrm{-SU(5)}}_0 &
0.13 & 0.16 & 0.999 & 0.999 & 0.039 & 0.041 & 0.039 & 0.58 & 0.040 & 1.04 
\\[2pt]
\hline
{\mathcal{F}\textrm{-SU(5)}}_{\rm I} &
6.01 & 13.1 & 0.958 & 0.758 & 0.045 & 0.116 & 0.044 & 0.68 & 0.104 & 9470
\\[2pt]
\hline
{\mathcal{F}\textrm{-SU(5)}}_{\rm II} &
3.29 & 6.86 & 0.969 & 0.873 & 0.061 & 0.116 & 0.060 & 0.68 & 0.110 & 88.1 
\\[2pt]
\hline
\end{array}
$
\setlength{\extrarowheight}{0pt}
\end{footnotesize}
\end{center}
\caption{\label{tb:resultssuper}
Principal results for the grand unification of Flipped $SU(5) \times U(1)_{\rm X}$,
with and without heavy vector multiplets.
Mass is in units of $[10^{16}~{\rm GeV}]$.}
\end{table}

The collected results for our study of the grand unification of Flipped $SU(5) \times U(1)_{\rm X}$
are presented in Table~(\ref{tb:resultssuper}).  The $\zeta_i$ account for the effects of the second
loop.  Although our calculations
are numerical, we again find excellent agreement with the closed form approximation, valid here to
about $10 \%$.  The second loop is found here to be important, but not decisive.
Without the $\zeta_i$, $M_{51}$ would be reduced
by about a quarter for model II, and by about a half for model I.
These factors are assembled
according to Eqs.~{\ref{eqs:superxieff}} to form the $\Xi_i$, in terms of which our solutions to
the renormalization group equations are conveniently expressed.  We have not considered
the possibility of crossing any mass thresholds in the transit from $M_{32}$ to $M_{51}$,
and the $\xi_i$ are thus zero.  We provide the prior solutions for ($\alpha_1$,$\alpha_5$,$M_{32}$)
for reference, and compute the $U(1)_{\rm X}$ coupling according to Eq.~(\ref{u1disc}).
Finally, we tabulate the grand unified coupling and mass from Eqs.~(\ref{eqs:masu}). 

It is of great interest that string theory generically predicts that any lower
energy gauge structures emerge from their `Super-Unification' at a string
scale $M_{\rm str}$ which is significantly depressed from the Planck mass.
In weakly coupled heterotic theory, $M_{\rm str} \simeq 5\times10^{17}~{\rm [GeV]}$ is
typical.  The presence of some (slightly) enlarged extra dimension~\cite{Dienes:1998vg}
might indeed be capable of bringing the Super-Unification well down into the
$10^{(15-16)}~{\rm [GeV]}$ range, although we favor extreme moderation in the application
of this scenario.  The broader point is that the assent of the string
to make some extension in our direction, encourages our prospects for likewise
`reaching up' to make the identifications\footnote{Moreover, in the context of
heterotic string model building, the string mass and coupling are deeply entangled:
$M_{\rm str} \propto \sqrt{\alpha_{\rm str}} M_{\rm str}$.}:
\beq
\alpha_{51} \Leftrightarrow \alpha_{\rm str} ~~;~~
M_{51} \Leftrightarrow M_{\rm str}
\label{assocsu}
\eeq

\begin{figure}[htp]
\begin{center}
\includegraphics[width=\plotwidth,angle=0]{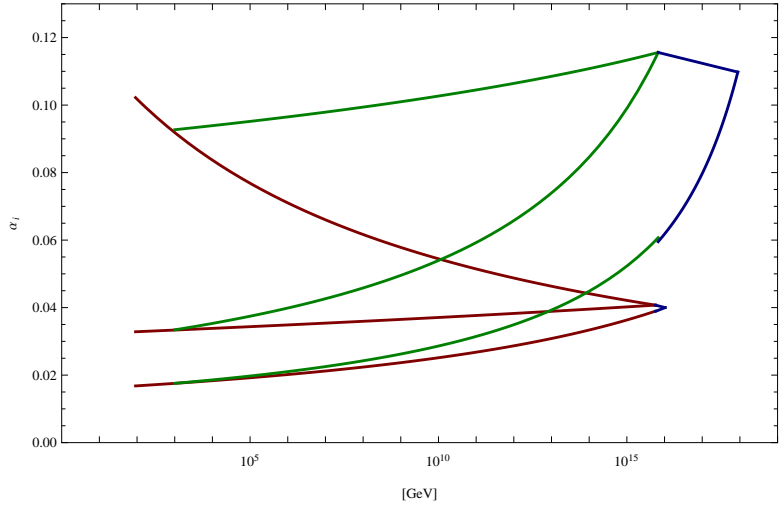}
\end{center}
\caption{\label{fig:superunification}
In this plot we follow the Flipped $SU(5) \times U(1)_{\rm X}$ Type~II model (solid green)
further upward to a possible super unification (solid blue) just shy of $10^{18}~[\rm{GeV}]$.
This is nearly two orders of magnitude heavier than the corresponding extension of the Type~0 
model (solid red) with MSSM content only.  The running is computed to two loops.}
\end{figure}

Examining $M_{51}$ from Table~(\ref{tb:resultssuper}), Flipped $SU(5)$ in the bare MSSM shows little
promise of realizing this merger.  The problem here is the overly close proximity of the
couplings ($\alpha_1$,$\alpha_5$), so that no `breathing room' is reserved to allow for
a significant second phase of the running.  Model Type~I with vector multiplets suffers
from an opposite problem, catastrophically overshooting the string scale. 
This problem may be remedied by judicious application of
heavy threshold corrections near the $M_{32}$ scale.
Model Type~II, on the other hand, represents a most appealing result,
demonstrated graphically in Figure~(\ref{fig:superunification}), striking at about
$9 \times 10^{17}~{\rm [GeV]}$, less than a factor of three shy of the reduced Planck mass.

Since the grand unification mass $M_{51}$ from
Eq.~(\ref{m51}) is directly proportional to the partial unification
$M_{32}$, we expect that the multiplicative transfer described in Eq.~(\ref{m32shift_std})
from the presence of heavy thresholds should be carried over
directly to the upper mass scale\footnote{Variation in the coupling $\alpha_{51}$ of Eq.~(\ref{alpha51})
is both relatively more sedate, and of less practical consequence.},
neglecting corrections to the second loop as manifest in the $\Xi_i$.
Ideally, with the baseline prediction already approaching the upper mass boundary, full exhaustion
of the heavy thresholds allocated in Section(\ref{sct:heavynumbers}) would be precluded
by `downward pressure' from the higher scale.  However, the allowed rescaling of $M_{51}$ by
a factor of around $1.7$ does not here present any definite incompatibility with unification
around $M_{\rm Pl}$.  Nonetheless, the tendency discussed for $M_{\rm str}$ to itself significantly
descend from $M_{\rm Pl}$ suggests that the mechanism is in principle viable, although it is
difficult to be more concrete in this case. 
Reminding ourselves that the heavy thresholds never threatened to lower $M_{32}$
the result could be a significantly tighter constraint from both directions
on unification, and in particular the proton lifetime,
than we might otherwise have imagined or feared. 

It is typical, especially of string derived models, that the constant $b_{\rm X}$
be conspicuously large, causing $\alpha_{\rm X}$ to climb much more rapidly than did the hypercharge.
This quite general result~\cite{Lopez:1995pr,Lopez:1997hq,Nanopoulos:1997ip}, exists because
negative $\beta$-function contributions, as would be required to help cancel the large
number of positively contributing loops arise, only from non-Abelian self-interactions,
of which the group $U(1)$ has, by definition, none.
Driven by the large value of $b_{\rm X}$, the rapid ascent of $\alpha_{\rm X}$
runs the danger of squandering an initial coupling
separation, rejoining prematurely with $\alpha_5$.

It has been natural in the context of Flipped $SU(5)$ to expect that the split
couplings continue upward to a true unification.
It is in fact only in this way that we can hope to recover the beneficial GUT
properties such as a correlated charge quantization and successful
prediction\footnote{Although divergence from this relation for the lighter generations
is sometimes mentioned as a failure of Grand Unification, we take the view
that competing ${\cal O}$(MeV) corrections must in fact be expected to wash
out masses in that same order.} of $m_b/m_\tau$.
Such memories of a simpler past are naturally inherited if $SU(5)\times U(1)_{\rm X}$
is descendant\footnote{It is of great interest that the $\mathbf{16}$ spinor of $S0(10)$
breaks to the quantum numbers of only the flipped variety $SU(5)$, while the
($\mathbf{5},\mathbf{\bar{5}}$) of SM Higgs with conventional charges readily
emerge from a fundamental~$\mathbf{10}$.} from a structure such as $SO(10)$.
That the string is willing to provide suitable representations for this purpose
in vicinity of the necessarily escalated GUT scale is quite a strong enticement
for the union of these pictures.

Being that we are now well approaching energy scales thought to be associated
with the onset of large quantum gravitational effects, it is natural to inquire
what the superstring may have to say, if anything, on the topic of proton decay.
The possibility of modifications to the relevant decay amplitudes
by the presence of compact extra dimensions
is addressed in Ref.~\cite{Alciati:2005ur}.
Interestingly, certain studies~\cite{Klebanov:2003my,Friedmann:2002ty} suggest
overall corrections of order one.  In other words, it may likely be correct,
as has been our perspective in this report,
to view proton decay as a pure particle physics topic.
We mention also with interest the suggestion~\cite{Donagi:2008kj,Wijnholt:2008db}
that there may be a logarithmic enhancement to the dimension six amplitude
associated with the breaking of GUT groups in F-theoretic constructions.
%We mention with also with interest the suggestion~\cite{Donagi:2008kj} that there may be 
%an enhancement of $\sqrt{\alpha_{\rm GUT}}$ to the dimension six amplitude
%associated with the breaking of GUT groups in F-theoretic constructions.
%Proton lifetime being inversely proportional to the amplitude squared, this would
%constitute a further order of magnitude speeding of the decay in our models,
%wherein $\alpha_5 \simeq \frac{1}{10}$.
%The problem is that we do not know the normalization of internal Green function, \ie,
%the uncertainties can be at the same order, preventing clean predictions.
%Moreover, these effects appear to only apply to $SU(5)$
%and $SO(10)$ models that are broken at the usual GUT scale.
%logarithmic enhancement ~\cite{Wijnholt:2008db}

We see a number of tantalizing conspiracies at work in the models under consideration,
particularly model Type~II.
The dramatic elevation of the GUT coupling, in itself already speeding proton decay,
is just right to allow space for a second stage running of some significance.  The more traditional
mild flipping ratios $M_{32}/M_{32}^{\rm max}$ in the neighborhood of $50\%$ seem by
comparison to be almost `not worth the effort'.  In the TeV-scale vector multiplet models however,
we see a dynamically emergent rationale for the closely adjacent yet stably disparate GUT and Planck scales.
Any GUT which cannot cast light on the origin and sustenance of this hierarchy
seems in comparison phenomenologically pale.
In fact, if we stipulate the existence of an upper Super Unification as something akin to
`experimental constraint', the possibility emerges of selecting between string derived models
by merit of the supplemental $\beta$-functions which they carry, and their effects on the string
mass scale and coupling.  The loss of grand unification at $10^{16}~{\rm [GeV]}$ seems a light price
for Flipped $SU(5)$ to pay, as we are anyways waiting for the union with gravity itself, already on its doorstep.

%%%%%%%%%%%%%%%%%%%%%%%%%%%%%%%%%%%%%%%%%%%%%%%%%%%%%%%%%%%%

\subsection{Summary and Conclusion}

Proton decay is one of the most unique
yet ubiquitous predictions of Grand Unification.
Its study draws upon and bears timely relevance to a
demonstrably broad palette of topics.

We have considered the proton decay process
$p \!\rightarrow\! {(e\vert\mu)}^{\!+}\! \pi^0$ via dimension six operators for 
heavy gauge boson exchange.
Including uncertainties for light and heavy
thresholds, we conclude that a majority of the
parameter space for proton decay is indeed within the reach of proposed
next-generation experiments such as Hyper-Kamiokande and DUSEL for the
Type~I and Type~II extensions to Flipped $SU(5)\times U(1)_X$.
The minimal Flipped $SU(5)\times U(1)_X$ model is also
testable if the heavy threshold corrections are small.
In particular, detectability of TeV scale vector supermultiplets
at the LHC presents an opportunity for cross correlation of results
between the most exciting particle physics experiments of the coming decade.

We have significantly upgraded the analysis of gauge coupling unification~\cite{Ellis:1995at,Ellis:2002vk},
correcting a subtle inconsistency in usage of the effective Weinberg angle,
improving resolution of the light threshold corrections,
and undertaking a proprietary determination of the second loop,
starting fresh from the standard RGEs, {\it cf.}~\cite{Jiang:2006hf}.
The step-wise entrance of the top quark and supersymmetric particles into the RGE
running is now properly accounted to all three gauge couplings individually 
rather than to a single composite term for the effective shift.
In addition to the light $M_Z$-scale threshold corrections from the superpartner's entry into the RGEs,
there may also be shifts occurring near the
$M_{23}$ unification point due to heavy Higgs fields and the
broken gauge generators of $SU(5)$.
The light fields carry strong correlations to cosmology and low energy
phenomenology, so that we are guided toward plausible estimates of their mass distribution.

The two-loop contribution has likewise been individually numerically determined
for each gauge coupling, including the top and bottom quark Yukawa couplings
from the third generation, taken themselves in the first loop.  All three gauge couplings are 
integrated recursively with the second loop into the Yukawa coupling renormalization, with
the boundary conditions at $M_{\rm Z}$ treated correctly for various values of $\tan \beta$.
The light threshold correction terms, defaulting to a phenomenologically favored benchmark point
$B'$~\cite{Battaglia:2003ab, DeRoeck:2005bw} of the constrained MSSM spectrum space,
are included wherever the gauge couplings $\alpha_i$ are used.
Recognizing that the second loop itself influences the upper limit $M_{32}$
of its own integrated contribution, this feedback is accounted for in the dynamic
calculation of the unification scale. 
We have also provided a closed form solution which is highly consistent with
our numerical results, allowing for transparent `low tech' precision analysis.

We have provided a comprehensive dictionary of solutions for the relevant
unification parameters of both Flipped and Standard $SU(5)$,
allowing for generic $\beta$-function coefficients,
light and heavy threshold factors, and corrections from the second loop.
We find that the conjunction of this detailed $\mathcal{F}$-resh analysis with the context of
$\mathcal{F}$-theory constructions of $\mathcal{F}$-lipped $SU(5)$ GUTs
to result in comparatively $\mathcal{F}$-ast proton decay.
Large portions of the lifetime prediction only narrowly evade existing detection limits,
and a vast majority of the most plausible range is
within the scope of the next generation detectors.
Moreover, inclusion of TeV scale vector multiplets in the renormalization significantly magnifies
the separation of the flipped scale $M_{32}$, at which $SU(5)$ breaks and proton decay is
established, from $M_{51}$, the point of $\mathcal{F}$-inal Grand Unification,
which can be extended to the order of the reduced Planck mass.

%%%%%%%%%%%%%%%%%%%%%%%%%%%%%%%%%%%%%%%%%%%%%%%%%%%%%%%%%%%%

\section*{Acknowledgments}
We thank D. B. Cline for
helpful private communication with regards in particular to
plans for the DUSEL project.
We thank James Maxin for helpful comments regarding constraints on the
MSSM spectrum, and for heading up the ``Flipped \texttt{SUSPECT} / \texttt{micrOMEGAs}''
programming efforts, which will feature in a future publication.
This research was supported in part 
by  the DOE grant DE-FG03-95-Er-40917 (TL and DVN),
by the Natural Science Foundation of China 
under grant No. 10821504 (TL),
and by the Mitchell-Heep Chair in High Energy Physics (TL).

%%%%%%%%%%%%%%%%%%%%%%%%%%%%%%%%%%%%%%%%%%%%%%%%%%%%%%%%%%%%

\appendix

\section{Brief Review of del Pezzo Surfaces}

The del Pezzo surfaces $dP_n$, where $n=1,~2,~...,~8$, are
defined by blowing up $n$ generic points of 
$\mathbb{P}^{1}\times\mathbb{P}^{1}$ or 
$\mathbb{P}^2$. The homological group
$H_2(dP_n, Z)$ has the generators
\beq
H,~E_1, ~E_2,~...,~E_n~,~\,
\eeq
where $H$ is the hyperplane class for $P^2$, and $E_i$, the
exceptional divisors at the divergent points, are
isomorphic to $\mathbb{P}^{1}$.
The intersection numbers of the generators are:
\beq
H\cdot H=1
\quad;\quad
E_{i}\cdot E_{j}=-\delta_{ij}
\quad;\quad
H\cdot E_{i}=0
\eeq
The canonical bundle on $dP_{n}$ is given by:
\beq
K_{dP_{n}}=-c_{1}(dP_{n})=-3H+\sum_{i=1}^{n}E_{i}
\eeq

For $n\geq3$,  we can define the generators as follows, where $i=(1,2,\ldots,n-1)$.
\beq
\alpha_i=E_i-E_{i+1} \quad;\quad \alpha_n=H-E_1-E_2-E_3
\eeq
Thus, all the generators $\alpha_i$ are perpendicular to the canonical
class $K_{dP_{n}}$, and the intersection products are equal to the negative Cartan
matrix of the Lie algebra $E_n$, and can be considered as simple roots. 
The curves $\Sigma_i$ in $dP_n$ where the particles are localized 
must be divisors of $S$, and the genus for the curve $\Sigma_i$ is:
\beq
2 g_i-2~=~[\Sigma_i]\cdot ([\Sigma_i]+K_{dP_{k}})
\eeq

For a line bundle $L$ on the surface $dP_{n}$ with
\beq
c_{1}(L)=\sum_{i=1}^{n}a_{i}E_{i}~,
\eeq
where $a_{i}a_{j}<0$ for some $i\neq j $, the K\"ahler form
$J_{dP_{n}}$ can be constructed as shown following~\cite{Beasley:2008dc},
where $\sum_{i=1}^k a_{i}b_{i}=0$ and $b_0 \gg b_{i}>0$.
By this construction, it is easy to see that the line bundle $L$ solves
the BPS equation $J_{dP_k}\wedge c_{1}(L)=0$.
\beq
J_{dP_{k}}=b_0H-\sum_{i=1}^{n}b_{i}E_{i},
\eeq

%%%%%%%%%%%%%%%%%%%%%%%%%%%%%%%%%%%%%%%%%%%%%%%%%%%%%%%%%%%%

\bibliography{bibdata}

\end{document}